\documentclass[12pt]{article}
\usepackage{amssymb}
\usepackage{epsf}
\usepackage{rotate}
\usepackage{graphics}

\newcommand{\insertfig}[2]{\mbox{\epsfxsize=#1cm
\epsfbox{#2.eps}}}

\newcommand{\Bx}{x_{\rm B}}
\newcommand{\Ax}{x_{\rm A}}
\newcommand{\AM}{M_{\rm A}}

\newcommand{\cQ}{{\cal Q}}
\newcommand{\re}{\Re\mbox{e}}
\newcommand{\im}{\Im\mbox{m}}
\newcommand{\GeV}{\mbox{GeV}}

\begin{document}

\begin{titlepage}

\centerline{\large\bf  Deeply virtual Compton scattering off nuclei}

\vspace{10mm}

\centerline{\bf A. Kirchner$^a$, D. M\"uller$^{b}$}
\vspace{10mm}

\centerline{\it $^a$Institut f\"ur Theoretische Physik,
                Universit\"at Regensburg}
\centerline{\it D-93040 Regensburg, Germany}

\vspace{5mm}

\centerline{\it $^b$Fachbereich Physik, Universit\"at  Wuppertal}
\centerline{\it D-42097 Wuppertal, Germany}

\vspace{10mm}

\centerline{\bf Abstract}

\vspace{2cm}

\noindent
We consider the hard leptoproduction of a photon off nuclei up to
spin-1. As a new result we present here the general azimuthal angular
dependence of the differential cross section for a spin-1 target. Its
twist-two Fourier coefficients of the interference and squared deeply
virtual Compton scattering amplitude are evaluated in leading order
approximation of perturbation theory in terms of generalized parton
distributions, while the pure Bethe--Heitler cross section is exactly
calculated in terms of electromagnetic form factors. Relying on a simple
model for the nucleon generalized parton distribution $H$, which
describes the existing DVCS data for a proton target, we estimate the
size of unpolarized cross sections, beam and longitudinal target spin as
well as unpolarized charge asymmetries for present fixed target
experiments with nuclei. These estimates are confronted with
preliminary HERMES data for deuterium and neon.
\vspace{6cm}


\vspace{1cm}


\end{titlepage}

\tableofcontents

\section{Introduction}

Exclusive two-photon processes in the light--cone dominated region, i.e.,
in the generalized Bjorken limit, are most suitable for the exploration
of the partonic content of hadrons. This comes from the fact that the
dominant contribution to the amplitude of such processes arises from
Feynman diagrams in which both photons directly couple to one quark line
\cite{MueRobGeyDitHor94}. The processes of interest are different
photon-to-meson transition form factors, i.e., $\gamma^\ast
\gamma^{(\ast)} \to M$, the production of hadron pairs by photon fusion
and the crossed processes like $\gamma^\ast H \to B \gamma $ or
$\gamma^{(\ast)} H \to B l^+ l^- $. The latter class might be denoted as
deeply virtual Compton scattering (DVCS). It contains different
processes: DVCS on a hadron target without and with excitation of the
final state due to the leptoproduction of a real photon
\cite{Rad96,Ji96a,DieGouPirRal97,BelMueNieSch00,BelMueKirSch00,
BelMueKir01}, the photoproduction \cite{BerDiePir01} or leptoproduction
\cite{GuiVan02,BelMue02a} of a lepton pair.

The factorization of short- and long-range dynamics is formally given by
the operator product expansion (OPE) of the time ordered product of two
electromagnetic currents, which has been worked out at leading twist-two
in next-to-leading order (NLO) and at twist-three level in leading order
(LO) of perturbation theory (for references see \cite{BelMueKir01}).
However, one should be aware that the partonic hard-scattering part,
i.e., the Wilson coefficients, contains collinear singularities, which
are absorbed in the non-perturbative distributions by a factorization
procedure, which has been proven at twist-two level
\cite{ColFre98,JiOsb98}.

The non-perturbative distributions are defined in terms of light--ray
operators with definite twist sandwiched between the corresponding
hadronic states. These distributions, depending on the two-photon
process under consideration, are sensitive to different aspects of
hadronic physics. Especially, in DVCS one can access the so-called
generalized parton distributions (GPDs). The second moment of the flavor
singlet GPD is related to the expectation value of the energy momentum
tensor. Thus, it contains information on the angular orbital momentum
fraction of the nucleon spin carried by quarks \cite{Ji96}. Moreover, in
contrast to ordinary parton distributions, measurable in inclusive
reactions, GPDs carry also information about the parton distribution in
transverse direction and might so provide us with a hologramatic picture
of the nucleon \cite{Bur00,RalPir01,Die02,BelMue02}.

Although GPDs are accessible in the leptoproduction of mesons
\cite{ColFraStr96}, DVCS is the theoretically cleanest tool to probe the
partonic content of the nucleon on the level of amplitudes. The DVCS
process in the leptoproduction of a photon on a proton target has been
measured by the H1 collaboration \cite{Adletal01} in the small $\Bx$
region (see also \cite{Sau00}), due to the single beam spin asymmetry by
the HERMES \cite{Airetal01} and CLAS \cite{Steetal01} collaborations as
well as due to the charge asymmetry at HERMES \cite{Ell02}.
Unfortunately, in this DVCS process a deconvolution of GPDs from a
measured amplitude is practically impossible. Thus, a model for GPDs is
required that satisfies the constraints coming from first principles
\cite{MueRobGeyDitHor94,Rad96,Ji96a}, namely, support properties, sum
rules and the reduction to the parton densities in the forward
kinematics. Moreover, different authors derived positivity constraints
(see Ref.\ \cite{Pob02a} and references therein). Although these
derivations are intuitively understandable, a number of questions can be
raised that are hardly to answer. At LO in perturbation theory all
experimental data are consistent with an oversimplified GPD model
\cite{BelMueKir01} that fulfills the first principle and also the
positivity constraints, derived in Ref.\ \cite{PirSofTer98,Rad98a}.

Recently, the measurements of beam spin asymmetries in the
leptoproduction of a photon on neon and deuteron targets have been
reported by the HERMES collaboration \cite{EllShaVol02}. Certainly, it
is appealing to employ DVCS for the investigation of the internal
structure of nuclei. Although the binding energy is negligibly small
compared to the virtuality of the exchanged photon, one might expect
that DVCS observables are affected by the binding forces. While for
spin-0 and -1/2 targets one might assume that the nucleus GPDs can be
expressed in terms of slightly modified proton GPDs, in the case of a
spin-1 target new GPDs enter the DVCS observables, in which bound state
effects are dominantly manifested. An appropriate candidate for studying
such effects is the deuteron, which has been widely used as a target in
lepton-scattering experiments. This nucleus has been extensively studied
in both deep-inelastic \cite{HooJafMan89} and elastic
\cite{BroJiLep83,GilGro01,GarOrd01} scattering. From the theoretical
point of view, it would be desirable to have complementary information,
which would give us a deeper understanding of the binding forces and
could hopefully shed some light of its effective description and the
fundamental degree of freedom in QCD.

On the theoretical side it is time to study DVCS on nuclei in more
detail. For deuterium a parameterization of leading twist-two operator
matrix elements in terms of GPDs has been proposed in Ref.\
\cite{BerCanDiePir01}. First estimates for the beam spin asymmetry have
been presented in Ref.\ \cite{KirMue02}, based on a drastic
simplification of the theoretical prediction, and in Ref.\
\cite{CanPir02,CanPir02a} by means of a convolution model. In Ref.
\cite{Pol02} it has been argued that nuclei GPDs can deliver information
about the spatial distribution of the strong forces in terms of the
fundamental degrees of freedom in QCD. In this paper we complete the
twist-two sector for the theoretical predictions for a spin-1 target,
generalize our oversimplified model, and estimate different asymmetries
for fixed target experiments.

The outline of the paper is the following: In Section \ref{Sec-ObsDVCS} the
OPE approach is applied to DVCS on a target with arbitrary spin.
Employing these general results, in Section \ref{Sec-FouCoe} we evaluate
the Fourier coefficients appearing in the cross section for a spin-1
target at leading twist and at leading order in perturbation theory. In
Section \ref{Sec-ModGPD} we present oversimplified GPD models for
nuclei, expressed by the proton ones, where bound state effects for
longitudinal degrees of freedom are neglected. Relying on qualitative
properties of proton GPDs, consistent with the DVCS data for the proton
target, we estimate in Section \ref{Sec-EstObs} by a rough kinematical
approximation beam spin, charge, and target spin asymmetries for fixed
target experiments at HERMES and JLAB kinematics. We then provide
numerical results for these asymmetries and the unpolarized cross
section. Finally, we summarize and give conclusions. Appendices are
devoted to the parameterization of the deuteron electromagnetic form
factors, the results for the Fourier coefficients of a spin-1 target,
and the target mass corrections.

\section{Azimuthal angular dependence of the cross section}
\label{Sec-ObsDVCS}

In this Section we evaluate the differential cross section for the
leptoproduction of a photon
\begin{eqnarray}
l^\pm (k) {\rm A} (P_1) \to l^\pm  (k^\prime) {\rm A} (P_2) \gamma (q_2)
\end{eqnarray}
off a nucleus target A with atomic mass number $A$ and mass $\AM$. 
The goal is to express the azimuthal angular harmonics in terms of the
electromagnetic current and Compton amplitudes to twist-two accuracy at
LO in $\alpha_s$. The considerations are valid for any target with
arbitrary spin content. 

\begin{figure}[t]
\vspace{-2.5cm}
\begin{center}
\mbox{
\begin{picture}(0,250)(250,0)
\put(70,0){\insertfig{13}{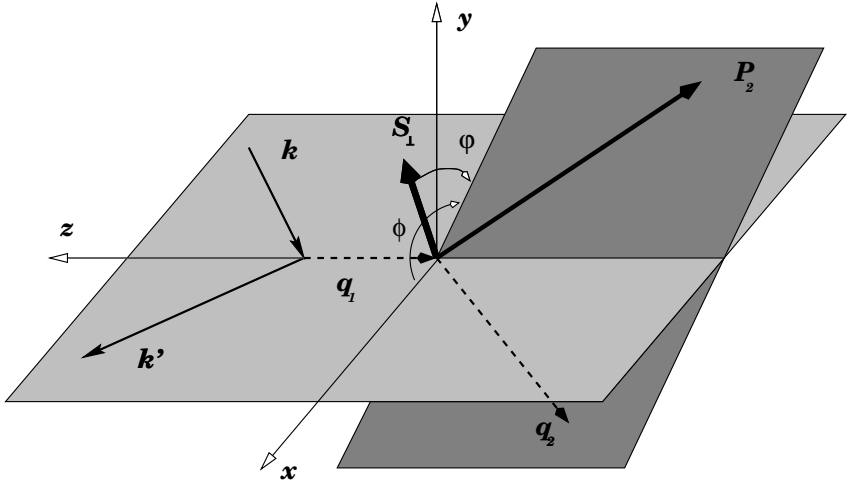}}
\end{picture}
}
\end{center}
\caption{\label{Fig-Kin}
The kinematics of the leptoproduction in the target rest frame. The
$z$-direction is chosen counter-along the three-momentum of the incoming
virtual photon. The lepton three-momenta form the lepton scattering plane,
while the recoiled nucleus and outgoing real photon define the nucleus
scattering plane. In this reference system the azimuthal angle of the
scattered lepton is $\phi_l = 0$, while the azimuthal angle between the
lepton plane and the recoiled nucleus momentum is $\phi_N = \phi$. When the
nucleus is transversely polarized (in this reference frame) $S_\perp = (0,
\cos {\mit\Phi}, \sin {\mit\Phi}, 0)$, the angle between the spin
vector and the scattered nucleus is denoted as $\varphi = {\mit\Phi} -
\phi_N$. }
\end{figure}

The five-fold differential cross section 
\begin{eqnarray}
\label{WQ}
\frac{d\sigma}{d\Ax dy d|\Delta^2| d\phi d\varphi}
=
\frac{\alpha^3  \Ax y } {16 \, \pi^2 \,  {\cal Q}^2 \sqrt{1 + \epsilon^2}}
\left| \frac{\cal T}{e^3} \right|^2 , \quad
\epsilon \equiv 2 \Ax \frac{\AM}{{\cal Q}}\, 
\end{eqnarray}
depends on the scaling variable $\Ax = {\cal Q}^2 / 2 P_1\cdot q_1 $,
where ${\cal Q}^2 = -q_1^2$ with $q_1 = k - k'$, the momentum transfer
square $\Delta^2 = (P_2 - P_1)^2$, the photon energy fraction $y=
P_1\cdot q_1/P_1\cdot k$ and, in general, two azimuthal angles. In the
following we refer to the rest frame, shown in Fig.\ \ref{Fig-Kin}. The
scaling variable $\Ax$ is related to the Bjorken  variable $\Bx$ by
\begin{eqnarray}
\Bx = \frac{\cQ^2}{2 M_N E y } \approx  A\, \Ax\, ,
\end{eqnarray}
where $M_N$ is the nucleon mass and $E$ is the lepton beam energy. The
photon energy fraction $y$ and the Bjorken variable satisfy the well
know relation
\begin{eqnarray}
\Bx y = \frac{\cQ^2}{s_N-M_N^2} \, , 
\end{eqnarray}
where $s_N$ is the center--of--mass energy squared for the lepton
scattering off a nucleon. Below it is sometimes more convenient to use
 the scaling variable 
\begin{eqnarray}
\label{Def-KinVarBj}
\xi \approx - \eta \approx \frac{\Ax}{2-\Ax}  \approx  \frac{\Bx}{2 A -\Bx }
\, 
\end{eqnarray}
instead of $\Ax$,
where $\Delta^2/\cQ^2$ corrections have been neglected. Here $\eta$ is
called the skewedness parameter, giving the longitudinal momentum
fraction in the $t$-channel. As we see, in the DVCS kinematics it is
simply related to the Bjorken like scaling variable $\xi$. To define the
azimuthal angles we point the virtual photon three-momentum towards the
negative $z$-direction. $\phi=\phi_l-\phi_N$ is the angle between the
lepton and nucleus scattering planes and $\varphi = {\mit\Phi} - \phi_N$
is the difference of the azimuthal angle $\mit\Phi$ of the spin vector
\begin{eqnarray}
\label{Def-SpiVec}
S^\mu = (0, \cos{\mit\Phi}
\sin{\mit\Theta}, \sin{\mit\Phi} \sin{\mit\Theta},
\cos{\mit\Theta}),
\end{eqnarray}
giving the magnetic quantization direction for the initial nucleus, and
the azimuthal angle $\phi_N$ of the recoiled nucleus as depicted in
Fig.\ \ref{Fig-Kin}. 

We consider this process in the (generalized) Bjorken limit, i.e.,
$\cQ^2 \sim P_1\cdot q_1$ should be large compared to $\AM^2$ and
$\Delta^2$. Obviously, increasing the atomic mass number $A$ one will
violate the condition $\cQ^2 \gg \AM^2$, since $\cQ^2$ is in reality
restricted by the experimental settings. It is instructive to consider
the situation in deeply inelastic scattering. Here the hadronic tensor
is given by the absorptive part of the forward scattering amplitude. In
the kinematical forward case the target mass corrections are given by
$\epsilon^2$, which is independent on the mass number $A$. In our
kinematics the situation is more complex \cite{BelMue01}. Fortunately,
it is shown in Appendix \ref{App-TarMasCor} that within our GPD model,
specified below, in the DVCS kinematics the target mass
corrections\footnote{We distinguish between power suppressed corrections that
arise from multi--parton correlation GPDs and  target mass
corrections. The former ones contain new dynamical information and
not much is known about them. The latter ones arise from trace
subtraction of twist-two operators. They have been elaborated in Ref.\
\cite{BelMue01} and are given as convolution with so-called double distributions
or alternatively with GPDs.} possess the same $A$  scaling as the twist-two 
contributions. Thus, they have nearly the same size as for a nucleon target and 
we may employ the OPE approach also for realistic $\cQ^2$ and larger $A$.

\begin{figure}[t]
\begin{center}
\mbox{
\begin{picture}(550,110)(0,0)
\put(0,-20){\insertfig{14}{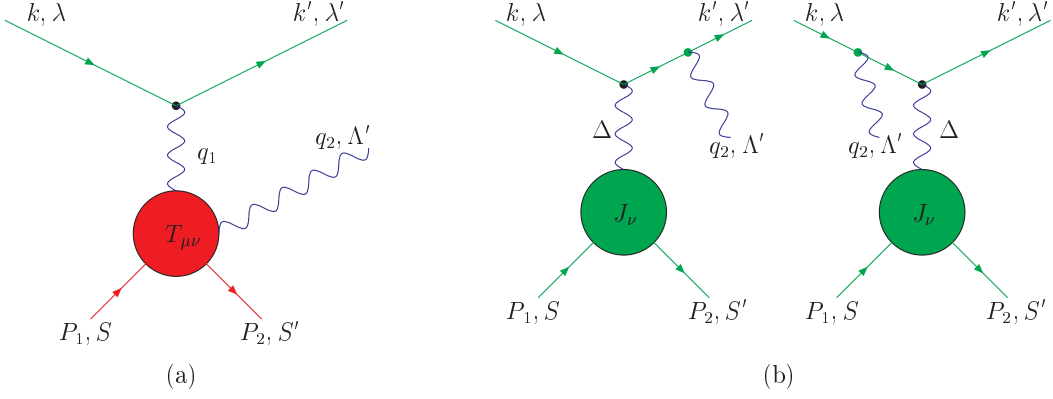}}
\end{picture}
}
\end{center}
\caption{\label{Fig-DVCSBH}
Both DVCS  (a) and BH (b) amplitudes contribute to the hard leptoproduction
amplitude of a photon. 
}
\end{figure}

In the one virtual photon exchange approximation the amplitude ${\cal
T}$ is the sum of the DVCS ${\cal T}_{\rm DVCS}$ and Bethe-Heitler (BH)
${\cal T}_{\rm BH}$ ones, displayed in Fig.\ \ref{Fig-DVCSBH}:
\begin{equation}
\label{Def-T2}
{\cal T}^2
=  \left\{
|{\cal T}_{\rm BH}|^2 + |{\cal T}_{\rm DVCS}|^2 + {\cal I} \right\},
\end{equation}
with the interference term
\begin{equation}
{\cal I} =
{\cal T}_{\rm DVCS} {\cal T}_{\rm BH}^\ast
+ {\cal T}_{\rm DVCS}^\ast {\cal T}_{\rm BH},
\end{equation}
where the recoiled lepton $(\lambda^\prime)$, nucleus $(S^\prime)$ as
well as photon $(\Lambda^\prime)$ polarization will usually not be observed.  
Each of these three terms in Eq.\ (\ref{Def-T2}) is given by the
contraction of DVCS tensor
\begin{eqnarray} 
T_{\mu\nu}(P,\Delta,q|S,S^\prime) = \frac{i}{e^2} \int dx {\rm e}^{i x \cdot q}
\langle P_2, S^\prime | T j_\mu (x/2) j_\nu (-x/2) | P_1,S\rangle 
\end{eqnarray}
 or/and electromagnetic current
\begin{eqnarray} 
J_\alpha(P,\Delta|S,S^\prime)=
\frac{1}{e} \langle P_2, S^\prime | j_\alpha (0)| P_1,S \rangle ,
\qquad
j_\alpha = \sum_{i=u,d,s} e_i \bar\psi\gamma_\alpha\psi ,
\end{eqnarray}
with a corresponding leptonic tensor 
\begin{eqnarray}
|{\cal T}_{\rm DVCS}|^2 \!\!\! &=&\!\!\!
\frac{e^6}{q_1^4} (-g^{\alpha\beta})  L^{\mu\nu}_{\rm DVCS}\, 
\sum_{S^\prime} T_{\alpha\mu}
\left(T_{\beta\nu}\right)^{\dagger},
\\
|{\cal T}_{\rm BH}|^2 \!\!\! &=&\!\!\!
\frac{e^6}{\Delta^4}
L^{\mu\nu}_{\rm BH}\, \sum_{S^\prime}  J_{\mu} J_{\nu}^\dagger,
\\
{\cal I} \!\!\! &=&\!\!\!
\frac{\pm e^6}{q_1^2 \Delta^2}  L^{\alpha\mu\nu} 
\sum_{S^\prime}  \left[J_\mu  \left(T_{\alpha\nu}\right)^{\dagger} + {\rm h.c.}\right]
\qquad \left\{ {+ \mbox{\ for\ } e^- \atop - \mbox{\ for\ } e^+} \right. . 
\end{eqnarray}
Here $P = P_{1} +P_{2}$ and $q=(q_1+q_2)/2$ and we summed over the  
photon  $(\Lambda^\prime)$ and lepton $(\lambda^\prime)$ polarizations, where the second
summation is included in the leptonic tensors.
For any nucleus (or hadron) target a general parameterization of the DVCS tensor
at twist-three level and in LO approximation of perturbation theory
has been given in Ref.\ \cite{BelMue00a}. To NLO accuracy a new Lorentz structure
occurs due to the flip of the photon helicity
 by two units, which is caused by the so-called twist-two gluonic transversity
\cite{HooJi98,BelMue00,Die01}, describing an angular orbital momentum flip 
by two units, as well as twist-four effects \cite{KivMan01}.
In the following we only consider its twist-two part, where it is expected
that for the typical kinematics of present fixed target experiments the twist-three
contributions are small for different observables, but not for all of them.
In this approximation the DVCS tensor is parameterized by a vector $V_1^\mu$ and an
axial--vector $A_1^\mu$ part \cite{BelMueNieSch00}:
\begin{eqnarray}
\label{Par-HadTen}
T_{\mu\nu} = - {\cal P}_{\mu}^{\ \,\sigma} g_{\sigma\tau} {\cal P}^\tau_{\ \nu}
\frac{q \cdot V_1}{P \cdot q}
- {\cal P}_{\mu}^{\ \,\sigma} i\epsilon_{\sigma \tau q \rho} {\cal P}^\tau_{\ \nu}
\frac{A_{1}^\rho}{P \cdot q}
, \end{eqnarray}
where the projector
$
{\cal P}_{\mu\nu} = g_{\mu\nu} - q_{1 \mu} q_{2 \nu}/{q_1 \cdot q_2}
$
ensures current conservation.
The leptonic tensors are evaluated by means of the standard projector
technique and read after summation over the helicity of the final state
lepton (see Fig.\ \ref{Fig-DVCSBH})
\begin{eqnarray}
\label{Con-LepHadTen-DVCS}
L_{\rm DVCS}^{\mu\nu}\!\!\! &=&\!\!\!
\sum_{\lambda^\prime} 
\bar{u}(k^\prime,\lambda^\prime) \gamma^{\mu} u(k,\lambda)
\bar{u}(k,\lambda) \gamma^{\nu}
u(k^\prime,\lambda^\prime)  =
{\rm Tr}\,  \gamma^{\mu}  \slash\!\!\! k \frac{1-\lambda \gamma_5}{2}
\gamma^{\nu} \slash\!\!\! k^\prime
\nonumber\\
\!\!\! &=& \!\!\!
2\left(k^{\mu} k^{\prime \nu} + k^{\nu} k^{\prime \mu} -
k\cdot k^\prime\, g^{\mu\nu} - i \lambda\, \epsilon^{\mu \nu \alpha\beta}\,
k_\alpha k^\prime_\beta \right),
\\
\label{Con-LepHadTen-BH}
L^{\mu\nu}_{\rm BH}\!\!\! &=&\!\!\!
(- g_{\alpha\beta}) {\rm Tr}\,  \Gamma^{\alpha\mu}(k,k^\prime,\Delta)
\slash\!\!\! k \frac{1-\lambda \gamma_5}{2}
\Gamma^{\nu\beta}(k,k^\prime,\Delta)
\slash\!\!\! k^\prime,
\\
\label{Con-LepHadTen-Int}
L^{\alpha\mu\nu}\!\!\! &=&\!\!\!
(- g^{\alpha}_{\ \,\beta}) {\rm Tr}\,  \Gamma^{\beta\mu}(k,k^\prime,\Delta)
\slash\!\!\! k\frac{1-\lambda \gamma_5}{2}
\gamma^{\nu} \slash\!\!\! k^\prime,
\end{eqnarray}
where
\begin{eqnarray}
\Gamma_{\alpha \mu}(k,k^\prime,\Delta) =
	\frac{\gamma_\alpha (\slash\!\!\! k - \slash\!\!\!\! \Delta) \gamma_\mu}
		{(k-\Delta)^2}	+
	\frac{\gamma_\mu (\slash\!\!\! k^\prime +\slash\!\!\!\! \Delta) \gamma_\alpha}
		{(k^\prime +\Delta)^2}
\end{eqnarray}
is the leptonic part of the BH amplitude.

Employing the parameterizations (\ref{Def-EM-Cur-1}),
(\ref{Def-V1-Sp1}) and (\ref{Def-A1-Sp1}), the contractions of
leptonic and DVCS tensors result in the kinematically exact
expression for the squared BH term (of course, in tree
approximation)
\begin{eqnarray}
|{\cal T}_{\rm BH}|^2 \!\!\! &=&\!\!\!
\frac{-8 e^6}{{\cal P}_1(\phi){\cal P}_2(\phi) \cQ^4 \Delta^2} \sum_{S^\prime}\Bigg\{
q\cdot J\ k \cdot J^\dagger + k\cdot J\ q\cdot J^\dagger - 
q\cdot J\ q\cdot J^\dagger - 2 k\cdot J\ k\cdot J^\dagger
\nonumber\\
& &\hspace{3.2cm}-\frac{({\cal Q}^2- \Delta^2 + 2  k\cdot\Delta)^2 +
4 (k\cdot\Delta)^2 }{4 \Delta^2} J\cdot J^\dagger
\\
& &\hspace{3.2cm} +\frac{i\lambda
\left[\left({\cal Q}^2 + 4 k\cdot \Delta -\Delta^2 \right)
\epsilon_{q \Delta J J^\dagger}
+ \Delta^2 \epsilon_{k (2 q + \Delta) J J^\dagger} \right]}
{2\Delta^2}
\Bigg\},
\nonumber
\end{eqnarray}
while the interference term\footnote{Note that the sign in front of the Levi-Civita
tensors in Eqs.\ (20) and (23) of Ref.\ \cite{BelMueNieSch00} is erroneously.} 
(here for negatively charged lepton scattering) 
\begin{eqnarray}
\label{Def-T2-Int}
{\cal I}
\!\!\! &=&\!\!\!
\frac{2 - 2y + y^2}{y^2 {\cal P}_1(\phi){\cal P}_2(\phi) }
 \frac{ 4 e^6 \xi}{\Delta^2 {\cal Q}^4}
 \left( k^\sigma -\frac{q^\sigma}{y}  \right)
\sum_{S^\prime} \left[
\left( J_\sigma + 2 \Delta_\sigma \frac{q\!\cdot\!J}{{\cal Q}^2} \right)
q\!\cdot\!V_1^\dagger
+ 2 i \epsilon_{\sigma q \Delta J} \frac{q\!\cdot\!A_1^\dagger}{{\cal Q}^2} \right]
\\
&&\!\!\!\! +\frac{\lambda(2 - y) y}{y^2 {\cal P}_1(\phi){\cal P}_2(\phi)}
 \frac{4 e^6 \xi}{\Delta^2 {\cal Q}^4}
\left( k^\sigma -\frac{q^\sigma}{y}  \right)
\sum_{S^\prime} \left[
\left( J_\sigma + 2 \Delta_\sigma \frac{q\!\cdot\!J}{{\cal Q}^2} \right)
q\!\cdot\! A_1^\dagger
+ 2 i \epsilon_{\sigma q \Delta J} \frac{q\!\cdot\!V_1^\dagger}{{\cal Q}^2}
\right] + {\rm h.c.},
\nonumber
\end{eqnarray}
and the squared DVCS amplitude
\begin{eqnarray}
\label{Def-T2-DVCS}
|{\cal T}_{\rm DVCS}|^2 &=&
8 e^6 \frac{2 - 2y + y^2}{y^2} \frac{\xi^2}{{\cal Q}^6}
\sum_{S^\prime} \left(q\!\cdot\! V_1\ q\!\cdot\! V_1^\dagger +
q \!\cdot\! A_1\ q \!\cdot\! A_1^\dagger \right)
\nonumber \\
&&+ 8 e^6 \frac{\lambda (2-y)}{y} \frac{\xi^2}{{\cal Q}^6}
\sum_{S^\prime} \left(q\!\cdot\! V_1\ q\!\cdot\! A_1^\dagger +
q\!\cdot\! A_1\ q\!\cdot\! V_1^\dagger  \right)
\end{eqnarray}
are expanded with respect to $1/\cQ$. In contrast to the
squared DVCS amplitude the interference as well as the squared BH terms
have an additional $\phi$-dependence due to the (scaled) BH propagators
\begin{eqnarray}
\label{Par-BH-Pro}
{\cal P}_1 \!\!\! &\equiv & \!\!\! \frac{(k-q_2)^2}{\cQ^2}
= - \frac{1}{y (1 + \epsilon^2)} \left\{ J + 2 K \cos(\phi) \right\},\\
{\cal P}_2 \!\!\! &\equiv & \!\!\! \frac{(k-\Delta)^2}{\cQ^2}
=
1 + \frac{\Delta^2}{\cQ^2} +
\frac{1}{y (1 + \epsilon^2)}\left\{J   + 2 K \cos(\phi)
\right\}
\nonumber ,
\end{eqnarray}
where
\begin{eqnarray*}
J =
\left( 1 - y - \frac{y \epsilon^2}{2} \right)
\left( 1 + \frac{\Delta^2}{\cQ^2} \right)
-
(1 - x) (2 - y) \frac{\Delta^2}{\cQ^2} \, 
\end{eqnarray*}
and 
\begin{equation}
\label{Def-K}
K^2 = -\frac{\Delta^2}{{\cal Q}^2} (1 - \Ax)
\left( 1 - y - \frac{y^2\epsilon^2}{4} \right)
\left( 1 - \frac{\Delta^2_{\rm min}}{\Delta^2} \right)
\left\{
\sqrt{1 + \epsilon^2}
+ \frac{4\Ax (1 - \Ax) + \epsilon^2}{4(1 - \Ax)}
\frac{\Delta^2 - \Delta^2_{\rm min}}{{\cal Q}^2}
\right\} 
\end{equation}
with the plus sign taken for the square root in Eq.\ (\ref{Par-BH-Pro}).
It vanishes at the kinematical boundary $\Delta^2 = \Delta_{\rm min}^2$,
determined by the minimal value
\begin{eqnarray}
\label{Def-tmin}
-\Delta_{\rm min}^2
=  {\cal Q}^2
\frac{2(1 - \Ax) \left(1 - \sqrt{1 + \epsilon^2}\right) + \epsilon^2}
{4\Ax (1 - \Ax) + \epsilon^2}
\approx \frac{\AM^2 \Ax^2}{1 - \Ax + \Ax \AM^2/{\cQ}^2}
\, ,
\end{eqnarray}
as well as at
\begin{eqnarray*}
y(x,\cQ^2)
=
y_{\rm max}
\equiv
2 \frac{\sqrt{1 + \epsilon^2} - 1}{\epsilon^2}
\approx
1 - \frac{\AM^2 \Ax^2}{\cQ^2} \, .
\end{eqnarray*}

In the frame chosen the contractions of leptonic, depending on $k,\
q_1,$ and $\Delta$, and DVCS, a function of $q,\ P,\ \Delta$ and
the spin vector $S$, tensors in Eqs.\
(\ref{Con-LepHadTen-DVCS}--\ref{Con-LepHadTen-Int}), yield finite sums
of Fourier harmonics \cite{DieGouPirRal97,BelMueKir01}:
\begin{eqnarray}
\label{Par-BH}
&&\hspace{-0.8cm}|{\cal T}_{\rm BH}|^2
= \frac{e^6 (1 + \epsilon^2)^{-2}}
{\Ax^2 y^2  \Delta^2\, {\cal P}_1 (\phi) {\cal P}_2 (\phi)}
\left\{
c^{\rm BH}_0
+  \sum_{n = 1}^2
\left[ c^{\rm BH}_n \, \cos{(n\phi)}
 + s^{\rm BH}_n \, \sin{(\phi)}\right]
\right\} \, ,
\\
\label{AmplitudesSquared}
&&\hspace{-0.8cm} |{\cal T}_{\rm DVCS}|^2
=
\frac{e^6}{y^2 {\cal Q}^2}\left\{
c^{\rm DVCS}_0
+ \sum_{n=1}^2
\left[
c^{\rm DVCS}_n \cos (n\phi) + s^{\rm DVCS}_n \sin (n \phi)
\right]
\right\} \, ,
\\
\label{InterferenceTerm}
&&\hspace{-0.8cm}{\cal I}
= \frac{\pm e^6}{\Ax y^3 \Delta^2 {\cal P}_1 (\phi) {\cal P}_2 (\phi) }
\left\{
c_0^{\cal I}
+ \sum_{n = 1}^3
\left[
c_n^{\cal I} \cos(n \phi) +  s_n^{\cal I} \sin(n \phi)
\right]
\right\} \, ,
\end{eqnarray}
where the $+$ ($-$) sign in the interference term stands for the negatively
(positively) charged lepton beam. In the twist-two sector the $\phi$ 
independent contribution of the squared DVCS amplitude is 
\begin{eqnarray}
c_0^{\rm DVCS} \!\!\! &=&\!\!\! 8\xi^2 \sum_{S^\prime}\Bigg[
(2 - 2y + y^2)\frac{q\!\cdot\! V_1\ q\!\cdot\! V_1^\dagger +
q \!\cdot\! A_1\ q \!\cdot\! A_1^\dagger}{{\cal Q}^4}
\nonumber\\
& &\hspace{1.3cm}
+ \lambda (2-y)y
\frac{q\!\cdot\! V_1\ q\!\cdot\! A_1^\dagger +
q\!\cdot\! A_1\ q\!\cdot\! V_1^\dagger}{{\cal Q}^4} \Bigg]\, ,
\end{eqnarray}
while the Fourier coefficients of the interference term are
\begin{eqnarray}
\left\{ {c_1^{\cal I} \atop s_1^{\cal I}}
\right\} \!\!\! &=& \!\!\!
8 \xi \Ax \left\{ {\Re\mbox{e} \atop \Im\mbox{m}}\right\}
 \frac{y k^\sigma -q^\sigma}{\cQ^2} 
\nonumber\\
& &\times\sum_{S^\prime}
\Bigg\{ 
(2 - 2y + y^2)
\left[
\left( J_\sigma + 2 \Delta_\sigma \frac{q\!\cdot\!J}{{\cal Q}^2} \right)
\frac{q\!\cdot\!V_1^\dagger}{\cQ^2}
+ \frac{ 2 i\epsilon_{\sigma q \Delta J}}{\cQ^2}
 \frac{q\!\cdot\!A_1^\dagger}{{\cal Q}^2} \right]
\\
&&\!\!\!\!\hspace{1.8cm} +\lambda(2 - y) y
\left[
\left( J_\sigma + 2 \Delta_\sigma \frac{q\!\cdot\!J}{{\cal Q}^2} \right)
\frac{q\!\cdot\! A_1^\dagger}{\cQ^2}
+ \frac{2 i \epsilon_{\sigma q \Delta J}}{\cQ^2} \frac{q\!\cdot\!V_1^\dagger}{{\cal Q}^2}
\right] \Bigg\}
\nonumber,
\end{eqnarray}
expanded in leading order of $1/{\cQ}$. Note that the $\cos(3\phi)$ and
$\cos(2\phi)$ harmonics in the interference and squared DVCS term,
respectively, also appear  at twist-two level. Since they arise from the
gluon transversity, they are suppressed by $\alpha_s/\pi$.

\section{Fourier coefficients for a spin-one target}
\label{Sec-FouCoe}

So far the formalism, presented in the previous section, is rather
general and at LO of perturbation theory it can be easily extended to
the twist-three sector. Introducing an appropriate form factor
decomposition of the electromagnetic current and the DVCS amplitudes, it
can be employed for any target. The resulting Fourier coefficients for
the spin-0 and -1/2 targets are presented at twist-three level in Refs.\
\cite{BelMueKirSch00,BelMueKir01} for a pion\footnote{Here the variable
$\Bx$ has be defined in the rest frame with respect to the target mass
rather than to the nucleon mass.} and proton target, respectively. The
results for a nucleus target follow from the replacements of kinematical
variables, form factors $F_i(\Delta^2)$ and (set) of GPDs:
\begin{eqnarray}
\label{RepRul2Nuc}
&&\!\!\!\!\!\Bx \rightarrow \Ax\, ,\quad \xi \rightarrow \xi\, , 
\quad M\rightarrow  M_A\, , 
\qquad\qquad
F_i(\Delta^2)\rightarrow F^{\rm A}_i(\Delta^2)\, , 
\\
&&\!\!\!\!\!F= \left\{H,E,\widetilde H,\widetilde E,\right\}(x,\eta,\Delta^2)
\rightarrow
F^{\rm A}= 
\left\{H^{\rm A},E^{\rm A},\widetilde H^{\rm A},\widetilde E^{\rm A},\right\}(x,\eta,\Delta^2)\, .
\nonumber
\end{eqnarray}

Let us now derive the Fourier coefficients for a spin-1 target in terms
of GPDs. The electromagnetic current
\begin{eqnarray}
\label{Def-EM-Cur-1}
J_\mu = - \epsilon_2^\ast\! \cdot\! \epsilon_1 P_\mu\, G_1
+ \left(\epsilon_2^\ast\! \cdot\! P \epsilon_{1\mu} +
 \epsilon_1\! \cdot\! P \epsilon_{2\mu}^\ast \right) G_2  -
\epsilon_2^\ast\! \cdot P\,  \epsilon_1\! \cdot\! P
\frac{P_\mu}{2 \AM^2}\,  G_3
\end{eqnarray}
is given by three form factors $G_i(\Delta^2)$ with $i=\{1,2,3\}$, where
$\epsilon_{1\mu}$ ($\epsilon_{2\mu}$) denote the three polarization
vectors for the initial (final) nucleus. The form factors
$G_i(\Delta^2)$ can be measured due to the elastic scattering of a
lepton on a nucleus. For the deuteron their parameterizations are
available in the literature, see Ref.\ \cite{GilGro01,GarOrd01} and
references therein.

At twist-two level the amplitudes $V_1$ and $A_1$, appearing in the
parameterization (\ref{Par-HadTen}) of the DVCS tensor, can be
decomposed in a complete basis of nine Compton form factors (CFFs)
\begin{eqnarray}
{\cal F}(\xi,\Delta^2,\cQ^2)=
\left\{{\cal H}_1, {\cal H}_2, {\cal H}_3, {\cal H}_4, {\cal H}_5,
\widetilde {\cal H}_1, \widetilde {\cal H}_2, \widetilde {\cal H}_3, 
\widetilde {\cal H}_4 \right\}(\xi,\Delta^2,\cQ^2). 
\end{eqnarray} 
For convenience we employed here the scaling variable $\xi$, c.f., Eq.\
(\ref{Def-KinVarBj}). Adopting the notation of Ref.\
\cite{BerCanDiePir01}, the CFFs read for the vector
\begin{eqnarray}
\label{Def-V1-Sp1}
V_{1\,\mu} &=& - \epsilon_2^\ast\! \cdot\! \epsilon_1 P_\mu\, {\cal H}_1
+ \left(\epsilon_2^\ast\! \cdot\! P \epsilon_{1\mu} +
 \epsilon_1\! \cdot\! P \epsilon_{2\mu}^\ast \right) {\cal H}_2
-\epsilon_2^\ast\! \cdot P\,  \epsilon_1\! \cdot\! P
\frac{P_\mu}{2 \AM^2}\,  {\cal H}_3 \\
&&\!\!\!
+ \left(\epsilon_2^\ast\! \cdot\! P \epsilon_{1\mu} -
 \epsilon_1\! \cdot\! P \epsilon_{2\mu}^\ast \right) {\cal H}_4
+
\left(\frac{2 \AM^2 \left\{
\epsilon_2^\ast\! \cdot\! q \epsilon_{1\mu} +
 \epsilon_1\! \cdot\! q \epsilon_{2\mu}^\ast \right\}}{P\cdot q}
 + \frac{\epsilon_2^\ast\! \cdot\! \epsilon_{1}}{3}
 P_\mu \right) {\cal H}_5,
\nonumber
\end{eqnarray}
and  axial--vector
\begin{eqnarray}
\label{Def-A1-Sp1}
A_{1\,_\mu} &=&
i \epsilon_{\mu \epsilon_2^\ast  \epsilon_1 P } \widetilde {\cal H}_1
- \frac{i\epsilon_{\mu \Delta P \epsilon_1}\, \epsilon_2^\ast\! \cdot\! P+
i \epsilon_{\mu \Delta P \epsilon_2^\ast}\, \epsilon_1\! \cdot\! P}{\AM^2}
 \widetilde {\cal H}_2
\\
 & &\!\!\!
- \frac{i\epsilon_{\mu \Delta P \epsilon_1}\, \epsilon_2^\ast\! \cdot\! P-
i \epsilon_{\mu \Delta P \epsilon_2^\ast}\, \epsilon_1\! \cdot\! P}{\AM^2}
 \widetilde {\cal H}_3
- \frac{i\epsilon_{\mu \Delta P \epsilon_1}\, \epsilon_2^\ast\! \cdot\! q+
i \epsilon_{\mu \Delta P \epsilon_2^\ast}\, \epsilon_1\! \cdot\! q}{q\! \cdot\! P}
\widetilde {\cal H}_4,
\nonumber
\end{eqnarray}
where $1/\cQ$-power suppressed effects have been neglected.

The CFFs in Eqs. (\ref{Def-V1-Sp1}) and (\ref{Def-A1-Sp1}) are given by
a convolution of perturbatively calculable coefficient functions
$C^{(\pm)}$ and twist-two GPDs via
\begin{eqnarray}
\label{DefTw2}
{\cal H}_k
&\!\!\! =\!\!\! & \sum_{i=u,\dots}
\int_{- 1}^{1} \! dx \, C_i^{(-)} (\xi, x,\cQ^2,\mu^2) 
H_k^i(x, \eta, \Delta^2,\mu^2)_{|\eta=-\xi},
\quad \mbox{for}\quad k=\{1,\dots,5\},
 \\
\widetilde {\cal H}_k 
&\!\!\! =\!\!\! &\sum_{i=u,\dots}
\int_{- 1}^{1} \! dx \, C_i^{(+)} (\xi, x,\cQ^2,\mu^2)
\widetilde H^i_k(x, \eta, \Delta^2,\mu^2)_{|\eta=-\xi},
\quad \mbox{for}\quad k=\{1,\dots,4\} ,
\end{eqnarray}
where $\mu$ denotes the factorization scale. For each quark species $i$
we have nine GPDs. The two sets $\{H^i_1,\dots,H^i_5\}$ and
$\{\widetilde{H}^i_1,\dots, \widetilde{H}_4^i\}$ are defined by
off-forward matrix elements of vector and axial--vector light--ray
operators. All GPDs of a given set satisfy the same evolution equations,
which govern the logarithmical dependence on the factorization scale
$\mu$. The coefficient functions $C^{(\mp)}$ have been perturbatively
expanded. In LO they read for the even ($-$) and odd ($+$) parity
sectors
\begin{equation}
\label{CoeffFunction}
\xi\, {C_{(0)i}^{(\mp)}} \left( \xi, x \right)
= \frac{Q^2_i}{1 - x/\xi - i 0}
\mp
\frac{Q^2_i}{1 + x/\xi - i 0},
\end{equation}
where $Q_i$ is the fractional quark charge.

The Fourier coefficients $c/s_i$ can be calculated from Eqs.\
(\ref{Def-T2-Int}), (\ref{Def-T2-DVCS}), and analogous ones for the
squared BH amplitude by summing over the polarization $\Lambda^\prime$,
where we can employ the common projector technique. For a massive spin-1
particle we have (see for instance \cite{Pas83})
\begin{eqnarray}
\label{Def-PolSum}
\epsilon_{1\mu}^\ast(\Lambda=0) \epsilon_{1\nu} (\Lambda=0) &=& S_\mu S_\nu,
\\
\epsilon_{1\mu}^\ast(\Lambda=\pm 1) \epsilon_{1\nu} (\Lambda=\pm 1) &=&
\frac{1}{2} \left(-g_{\mu\nu} + \frac{P_{1\mu}P_{1\nu}}{\AM^2}
- S_\mu S_\nu + \frac{i \Lambda}{\AM} \epsilon_{\mu\nu P_1 S}  \right),
\nonumber
\end{eqnarray}
where $\Lambda=\{+1,0,-1\}$ denotes the magnetic quantum number with
respect to the quantization direction given by the spin vector $S_\mu$,
defined in Eq.\ (\ref{Def-SpiVec}). Obviously, the spin sum of the
recoiled nucleus is
\begin{eqnarray}
\sum_{\Lambda^\prime=-1}^1
\epsilon_{2\mu}^\ast(\Lambda^\prime) \epsilon_{2\nu} (\Lambda^\prime) =
-g_{\mu\nu} + \frac{P_{2\mu}P_{2\nu}}{\AM^2} .
\end{eqnarray}

As we see, the Fourier coefficients  for a spin-1 target quadratically
depend on the spin vector $S_\mu$ and, thus, we introduce the following
decomposition
\begin{eqnarray}
\label{Def-FouDec}
c_n^{T} &=& \frac{3}{2} \Lambda^2 c_{n,{\rm unp}}^{T} +
\Lambda\left\{ c_{n,{\rm LP}}^{T} \cos({\mit\Theta}) +
c_{n,{\rm TP}}^{T}(\varphi) \sin({\mit\Theta}) \right\}
+
\left(1-\frac{3}{2}\Lambda^2\right)
\nonumber\\
&&\times \left\{
c_{n,{\rm LTP}}^{T}(\varphi)  \sin(2{\mit\Theta}) +
c_{n,{\rm LLP}}^{T} \cos^2({\mit\Theta})  +
c_{n,{\rm TTP}}^{T}(\varphi) \sin^2 ({\mit\Theta})
\right\}
\end{eqnarray}
for $ T=\{{\rm BH}, {\cal I}, {\rm DVCS}\}$.
An analogous decomposition holds true for the odd harmonics $s_n^{T}$.
The unpolarized and the longitudinally polarized coefficients
$c/s_{n,{\rm unp}}^{T}$, $c/s_{n,{\rm LP}}^{T}$, and $c/s_{n,{\rm
LLP}}^{T}$, respectively, are independent of $\varphi$. The transverse
coefficients $c/s_{n,{\rm TP}}^{T}$ and the transverse-longitudinal
interference terms may be decomposed with respect to the first harmonics
in the azimuthal angle $\varphi$
\begin{eqnarray}
c_{n,{\rm TP}}^{T}(\varphi)  &=& c_{n,{\rm TP}+}^{T} \cos(\varphi) +
s_{n,{\rm TP}-}^{T} \sin(\varphi),\\
c_{n,{\rm LTP}}^{T}(\varphi) &=& c_{n,{\rm LTP}+}^{T} \cos(\varphi) +
s_{n,{\rm LTP}-}^{T} \sin(\varphi),
\end{eqnarray}
while $c/s_{n,{\rm TTP}}^{T}$ may be written as
\begin{eqnarray}
c_{n,{\rm TTP}}^{T}(\varphi)  &=&
c_{n,{\rm TTP}\Sigma}^{T}+
 c_{n,{\rm TTP}\Delta}^{T} \cos(2\varphi) +
s_{n,{\rm TTP}\pm}^{T} \sin(2\varphi).
\end{eqnarray}
Analogous definitions are used for the odd harmonics (just replace $c
\leftrightarrow s$). Obviously, $c_{n,{\rm TTP}\Sigma}^{T}$ does not
belong to an independent frequency, rather it can be included in the
constant and $\cos^2(\theta)$ terms of Eq.\ (\ref{Def-FouDec}). Indeed,
the calculation of the Fourier coefficients establishes the equality:
\begin{eqnarray}\label{TTPAbh}
 c_{0,{\rm LLP}}^{T}=  3 c_{0,{\rm unp}}^{T} -
 2 c^{T}_{0,{\rm TTP}\Sigma} , \qquad T = \{{\rm BH}, {\cal I},
{\rm DVCS}\},
\end{eqnarray}
and an analogous one for $s_{1,{\rm TTP}\Sigma}^{{\cal I}}$ coefficients.

We will now present the results for the dominant harmonics $c/s_1^{\cal I}$ of
the interference term. 
\begin{table}[t]
\vspace{-0.5cm}
\begin{center}
\begin{tabular}{|c|c|c|c|}
\hline
$\Bigg.$ i
& ${\cal L}_{1,i}^{{\cal I}c}$ & ${\cal L}^{{\cal I}s}_{1,i}$ &
${\cal L}_{0,i}^{\rm DVCS}$
\\
\hline\hline $\Bigg.$
unp &   $-8 K (2 - 2y + y^2)$   &
        $8 \lambda K y(2-y)$    &
        $  2 - 2 y + y^2 $
\\ \hline $\Bigg.$
LP &    $-8 \lambda K  y(2-y)$  &
        $8 K (2 - 2y + y^2)$    &
        $\lambda  y(2-y)$
\\ \hline  $\Bigg.$
TP$+$ & $-8 \lambda \frac{\sqrt{1-y} \AM}{\cal Q}  y (2 - y)$ &
        $8 \frac{\sqrt{1-y} \AM}{\cal Q} (2 - 2y + y^2) $     &
        $\lambda \frac{{\cal Q} K}{\sqrt{1-y} \AM}  y (2 - y)$
\\ \hline $\Bigg.$
TP$-$ & $8\lambda \frac{\sqrt{1-y} \AM}{\cal Q}  y (2 - y)$   &
        $8\frac{\sqrt{1-y}\AM}{\cal Q} (2 - 2y + y^2) $       &
        $-\frac{{\cal Q} K}{\AM\sqrt{1-y}} (2 - 2y + y^2) $
\\ \hline $\Bigg.$
LLP &   $-8 K (2 - 2y + y^2)$ &
        $8\lambda  K y(2-y) $ &
        $2 - 2y + y^2$
\\ \hline $\Bigg.$
LTP$+$ &$-8\frac{\sqrt{1-y} \AM}{\cal Q} (2 - 2y + y^2) $ &
        $8\lambda \frac{\sqrt{1-y} \AM}{\cal Q} y (2 - y)$     &
        $\frac{{\cal Q} K}{\AM\sqrt{1-y}} (2 - 2y + y^2)$
\\ \hline $\Bigg.$
LTP$-$ &$8\frac{\sqrt{1-y} \AM}{\cal Q} (2 - 2y + y^2) $  &
        $8\frac{\sqrt{1-y} \AM}{\cal Q} \lambda y (2 - y)$    &
        $\lambda \frac{{\cal Q} K}{\sqrt{1-y} \AM}  y (2 - y)$
\\ \hline $\Bigg.$
TTP$\Sigma$ &   $-8 K (2 - 2y + y^2)$   &
                $8 \lambda K y(2-y)$    &
                $2 - 2y + y^2$
\\ \hline $\Bigg.$
TTP$\Delta$ & $-8 K (2 - 2y + y^2)$ &
              $8 \lambda K  y(2-y)$ &
              $2 - 2y + y^2$
\\ \hline $\Bigg.$
TTP$\pm$ &  $8 K (2 - 2y + y^2)$    &
            $8 \lambda K y(2-y)$    &
            $- \lambda y(2-y) $
\\ \hline
\end{tabular}
\end{center}
\caption{Kinematical prefactors that appear in the Fourier coefficients of the
interference term (\ref{Def-FC-Int}) and squared DVCS amplitude (\ref{Def-FC-DVCS}).}
\label{Tab-LepFac}
\end{table}
We write them as a product of leptonic prefactors $\cal L$, defined in
Tab.\ \ref{Tab-LepFac}, with the real/imaginary part of `universal'
functions ${\cal C}_{i}^{\cal I}$:
\begin{eqnarray}
\label{Def-FC-Int}
\left\{ c_{1,i}^{\cal I} \atop s_{1,i}^{\cal I}   \right\} =
 \left\{{\cal L}_{1,i}^{{\cal I}c} \atop {\cal L}_{1,i}^{{\cal I}s} \right\}
\left\{\re  \atop \im \right\}  {\cal C}_{i}^{\cal I},
\quad \mbox{for}\quad
i=\{{\rm unp},  \cdots,{\rm TTP}-\}.
\end{eqnarray}
Since they linearly depend on the nine CFFs and three electromagnetic
form factors, we write these functions in terms of real valued $9\times
3$ matrices ${\cal \AM}$:
\begin{eqnarray}
\label{Def-C-Int}
{\cal C}_i^{\cal I} =
\left( {\cal H}_1   \cdots {\cal H}_5 \widetilde{{\cal H}}_1 \cdots
  \widetilde{{\cal H}}_4 \right)
 {\cal M}^{\cal I}_i
\left(
\begin{array}{c}
{G}_1\\ G_2 \\ G_3
\end{array}
\right),
\quad \mbox{for}\quad
i=\{{\rm unp},  \cdots,{\rm TTP}-\}.
\end{eqnarray}
In the case of a longitudinally polarized target the three matrices
${\cal M}_{\rm unp}^{\cal I} $, ${\cal M}_{\rm LP}^{\cal I},$ and
${\cal M}_{\rm LLP}^{\cal I} $ are presented in Appendix
\ref{App-IntTer} for small values of the momentum
transfer\footnote{\label{FooNot-1} The complete expressions for
all matrices of the interference and squared DVCS terms as well as
the complete squared BH term are available as TeX or MATHEMATICA
file from the authors via e-mail:
axel.kirchner@physik.uni-regensburg.de,
dieter.mueller@.physik.uni-regensburg.de
(dmueller@theorie.physik.uni-wuppertal.de). }, i.e., $-\Delta^2
\ll \AM^2$.

The squared DVCS amplitude is written in an analogous manner, namely,
\begin{eqnarray}
\label{Def-FC-DVCS}
c_{0,i}^{\rm DVCS}  =
{\cal L}_{0,i}^{\rm DVCS}
 {\cal C}_{i}^{\rm DVCS},
\quad \mbox{for}\
i=\{{\rm unp},  \cdots,{\rm TTP}-\}, 
\end{eqnarray}
where the nine functions 
\begin{eqnarray}
\label{Def-C-DVCS}
{\cal C}_i^{\rm DVCS} =
\left({\cal H}_1 \cdots \widetilde{{\cal H}}_4 \right) {\cal M}_i^{\rm DVCS}
\left(
\begin{array}{c}
{\cal H}_1^\ast\\ \vdots \\ \widetilde{{\cal H}}_4^\ast
\end{array}
\right)
\end{eqnarray}
are expressed by $9\times9$ hermitian matrices ${\cal M}_i^{\rm DVCS}$.
The leptonic factors ${\cal L}_{0,i}^{\rm DVCS}$ are given in Tab.\
\ref{Tab-LepFac} and the approximation of the matrices for
longitudinally polarized target can be found in Appendix \ref{App-DVCS}.
The analogous BH contributions are listed in Appendix \ref{App-BH}.

Let us summarize. For a given harmonic in $\phi$ we have altogether nine
possible observables. In principle, they can be measured by an
appropriate adjustment of the spin vector $S_\mu$ and Fourier analysis
with respect to the azimuthal angle $\varphi$. The interference term
linearly depends on the CFFs and is, thus, of special interest. In
facilities that have both kinds of leptons it can be separated by means
of the charge asymmetry. Combined with single spin-flip and unpolarized
as well as double spin-flip measurements, the $\sin/\cos$ harmonics can
be separated. This would provide access to the imaginary and real part
of the nine linear combinations ${\cal C}^{\cal I}_i$ and so to the nine
GPDs. To extract information from the nine GPDs one needs a realistic
model for them, where the model parameters have to be adjusted by
fitting the measured data. Certainly, this goal requires a deep
understanding of non-perturbative physics and a dedicated facility.

\section{Models for nucleus GPDs}
\label{Sec-ModGPD}

The main uncertainty in the estimate of observables is the lack of
knowledge on GPDs. Since GPDs are hybrids of exclusive and inclusive
quantities, we may expect that the constraints on models arising from
the reduction to parton densities in the forward limit or to elastic
electromagnetic form factors, appearing as the lowest moment of GPDs,
ensure the right order of magnitude for cross sections and asymmetries.
However, one should keep in mind that the predictions for DVCS in terms
of GPDs are rather complex and that only few quantities, like the sign
and the size of the single beam spin asymmetry or the size of the
unpolarized cross section, are ``predictable'' in a rather model
independent manner. We can also rephrase this statement in a positive
sentence, namely, most of the observables are sensitive to details of
the GPD models. One example is for instance the charge asymmetry for
DVCS on an unpolarized proton target. However, we should emphasize again
that the interpretation of experimental data requires a careful
consideration of possible kinematical effects as it has been stressed
in \cite{BelMueKir01}.

\subsection{Scalar target}

To start with the simplest case, we want to consider a scalar nucleus
target A with charge $Z e$. In this case only one GPD arises at
twist-two level and two of them at twist-three level
\cite{BelMue00a,BelMueKirSch00,AniPirTer00,RadWei00a}. The former one is
defined as expectation value of twist-two light--ray operators 
\begin{eqnarray}
\label{Def-GPD-sca}
H^{i/{\rm A}}(x,\eta,\Delta,\cQ^2) = \int \frac{d\kappa}{2\pi} e^{i \kappa P_+ x} 
\langle {\rm A}(P_2) | \bar\psi_i(-\kappa n) \gamma_+  \psi_i(\kappa n) | 
 {\rm A}(P_1) \rangle
\Big|_{\eta= \frac{\Delta_+}{P_+}}\, , 
\end{eqnarray}
where the gauge link has been omitted. Here the +--component of a four
vector is given by the contraction with the light--like vector $n$, e.g.,
$P_+ = n\cdot P$. Thus, $-1\le x \le 1$ can be interpreted as a momentum
fraction with respect to the light--cone component $P_+$, while the
skewedness parameter $\eta$ appears as a scaling variable. 
Lorentz invariance requires that the $n$th moment with
respect to $x$ of this functions is a polynomial in $\eta$ of order
$n+1$. This is ensured by the GPD support, which can be
implemented in a simple manner by the so-called double distribution (DD)
representation \cite{MueRobGeyDitHor94,Rad96} 
\begin{eqnarray}
\label{DD2GPD}
H^{i/{\rm A}}(x,\eta,\Delta^2,\cQ^2)=  \int_{-1}^1 dy
\int_{-1+|y|}^{1-|y|} dz\;
x\, \delta(x-y - \eta z) h^{i/{\rm A}}(y,z,\Delta^2,\cQ^2)\, .
\end{eqnarray} 
Note that, in contrast to the common definition, we included here an
extra factor $x$, which guarantees that the term of order $\eta^{n+1}$
is not absent in the $n$th $x$--moment. Moreover, time reversal
invariance combined with hermiticity requires that they are even in
$\eta$ and, thus, the double distribution $h$ is even in $z$. The
polynomiality condition is separately ensured for the two regions $y \ge
0$ and $y \le 0$. The so--called positivity constraints, mentioned
before, are not implemented in this representation.

 For $\Delta\to 0$ the GPD
(\ref{Def-GPD-sca}) reduces to the parton and anti--parton distributions 
\begin{eqnarray}
\left\{ { q^{i/{\rm A}} \atop \overline{q}^{i/{\rm A}} }\right\}(x,\cQ^2) 
= \left\{ {\, H^{i/{\rm A}}(x,\eta=0,\Delta^2=0,\cQ^2) \atop 
- H^{i/{\rm A}}(-x,\eta=0,\Delta^2=0,\cQ^2) } \right\} 
\end{eqnarray} 
in a nucleus with the familiar interpretation.
Both of them have the support $0 \le x \le 1$ and we applied
here the standard sign conventions for anti--quarks, so that
$\overline{q}^{i/{\rm A}}$ is positive. They satisfy the charge sum rule 
\begin{eqnarray}
\label{SumRul-Cha}
\int_0^1 dx  \sum_{i=u,d,s} Q_i
   \left[ q^{i/{\rm A}}(x,\cQ^2) - \overline{q}^{i/{\rm A}}(x,\cQ^2)\right] 
&\!\!\! =&\!\!\! \int_0^1 dx \left[ Q_u q^{u_{\rm val}/{\rm A}}(x,\cQ^2) +
Q_d q^{d_{\rm val}/{\rm A}}(x,\cQ^2)\right]
\nonumber\\
&\!\!\! =&\!\!\!  Z\, ,
\end{eqnarray}
and the Baryon number conservation sum rule 
\begin{eqnarray}
\label{SumRul-Bar}
\int_0^1 dx \frac{1}{3} \sum_{i=u,d,s} 
   \left[ q^{i/{\rm A}}(x,\cQ^2) - \overline{q}^{i/{\rm A}}(x,\cQ^2)\right] 
&\!\!\! =&\!\!\! \int_0^1 dx \frac{1}{3}\left[ q^{u_{\rm val}/{\rm A}}(x,\cQ^2) + 
q^{d_{\rm val}/{\rm A}}(x,\cQ^2)\right]
\nonumber\\
&\!\!\! =&\!\!\! A\, .
\end{eqnarray}
In both sum rules the sea quarks drop out and we can read off the 
number of valence quarks
\begin{eqnarray}
\label{Nor-ValGPD} 
&& \int_0^1 dx q^{u_{\rm val}/{\rm A}}(x,\cQ^2) = N_{u_{\rm val}} =  2 Z+ N \, ,
\nonumber\\
&& \int_0^1 dx q^{d_{\rm val}/{\rm A}}(x,\cQ^2) = N_{d_{\rm val}} =  Z+ 2 N \, ,
\qquad N= A-Z\, .
\end{eqnarray}
Furthermore, we have the momentum sum rule: 
\begin{eqnarray}
\label{SumRul-Mom}
\int_0^1 dx\, x \sum_{i=u,d,s} 
   \left[ q^{i/{\rm A}}(x,\cQ^2) + \overline{q}^{i/{\rm A}}(x,\cQ^2)\right] +
\int_0^1 dx\, x\, g^{\rm A}(x,\cQ^2)
=1\, ,
\end{eqnarray}
where $g^{\rm A}(x,\cQ^2) $ is the gluon density.

Now we like to connect the nucleus parton distributions to the
proton ones. Isospin symmetry breaking effects are rather small and,
thus, we employ this symmetry to connect their parton content:
\begin{eqnarray}
 &&\!\!\!\!\! q^{u/p} = q^{d/n}\equiv q^{u}\, , 
\ \ 
\overline{q}^{u/p} = \overline{q}^{d/n} \equiv 
\overline{q}^{u},
\qquad
 q^{d/p}= q^{u/n}  \equiv q^{d} \, ,
\ \ 
 \overline{q}^{d/p}  = \overline{q}^{u/n}  \equiv \overline{q}^{d}\, , 
\nonumber\\
&&\!\!\!\!\! q^{s/p} =\overline{q}^{s/p}   = q^{s/n} =\overline{q}^{s/n} \equiv q^{s}\, ,
\hspace{1.8cm}
g^{p}  = g^{n}  \equiv g\, .
\nonumber
\end{eqnarray}
Considering the nucleus as a system of almost free nucleons, we can
immediately express its parton distributions by those of the proton 
\begin{eqnarray}
\label{RelqA2qp}
&& q^{u/{\rm A}} = Z  q^{u} + N q^d \, , \qquad
q^{d/{\rm A}}= Z  q^{d} + N q^u\, , 
\nonumber \\
&& q^{s/{\rm A}} = (Z+N)  q^{s} , \qquad g^{\rm A} = (Z+N)  g\, .
\end{eqnarray}
Additionally, we have to relate the momentum fraction variable $x$ to that 
of the struck nucleon. Since the momentum of each nucleon is the $1/A$th part of the 
nucleus one, the appropriate scaling of the momentum fraction gives for instance
\begin{eqnarray}
\label{RelqA2qpSca}
q^{u/{\rm A}}(x)  =  
A \theta(1- |x A|)  \left[ Z  q^{u}(x A) + N q^d (x A) \right]\, 
\end{eqnarray}
and analogous for the other parton distribution.
Here the loss in phase-space due to the restriction $x \le 1/A$ is
compensated by the prefactor $A.$ Note that $q^{u/{\rm A}}(x)$ does in reality
not vanish at $x=1/A$, however, this simple prescription is for our
purpose justified as long as $\Bx= A x_A $ is in the valence quark region
or below. It is easy to convince ourself that all sum rules
(\ref{SumRul-Cha},\ref{SumRul-Bar},\ref{SumRul-Mom}) will be obeyed,
if they are satisfied for the nucleon. For the ratio of the structure
functions $F_2$, normalized per nucleon, in deep inelastic scattering
one finds for an isoscalar target $(N=Z)$
\begin{eqnarray}
\frac{F_2^{\rm A} (\Ax)}{F_2^{\rm N}(\Bx)} \simeq 
\frac{1}{A} \frac{\Ax\sum_i   Q_i^2 q^{i/{\rm A}}(\Ax)}
                                 {\Bx \sum_i   Q_i^2 q^{i/N}(\Bx) } = 1\, , 
\quad \mbox{with} \quad \Ax= \Bx/A.
\end{eqnarray}
Certainly, experimental measurements show that this is only true for
$\Bx \sim 0.25$ and in other regions one finds characteristic deviations
from this ratio, which reflect different aspects of the binding forces
between the nucleons. For the kinematics we are interested in, i.e., $\Bx
\sim 0.1$, it is a small difference of the order of a few percent, which
can safely be neglected for our purpose.

Inspired by the parton picture, we introduce a terminology for the GPDs.
However, we remind that  we deal now with a {\em generalization} of {\em
distribution amplitudes} and the probabilistic interpretation arises
only in the forward limit, because of the optical theorem. We uniquely
separate a GPD in valence-- and sea--like ones 
\begin{eqnarray}
\label{GPD-Sep-ValSea}
H^{i/{\rm A}}(x,\eta) = \theta(x \ge -|\eta|)  H^{i_{\rm val}/{\rm A}}(x,\eta) 
+  H^{i_{\rm sea}/{\rm A}}(x,\eta) \, .
\end{eqnarray}
The sea--like GPD is antisymmetric in $x$ and can be separated in quark
and anti--quark ones. The latter are obtained from Eq.\ (\ref{DD2GPD})
due to the restriction $y\le 0$  and  the whole sea might be expressed by 
the anti--quark contribution
\begin{eqnarray}
H^{i_{\rm sea}/{\rm A}}(x,\eta)&\!\!\! = &\!\!\! 
\overline{H}^{i/{\rm A}}(x,\eta) - \overline{H}^{i/{\rm A}}(-x,\eta)  .
\end{eqnarray} 
The valence--like part follows by  subtracting the sea quark GPD from 
the whole one:
\begin{eqnarray}
H^{i_{\rm val}/{\rm A}}(x,\eta) = \theta(x \ge -|\eta|)\left[H^{i/{\rm A}}(x,\eta)-
 H^{i_{\rm sea}/{\rm A}}(x,\eta) \right] \, .
\end{eqnarray}
Note that the valence--like GPDs have no symmetry property and that we
included the region $ -|\eta|\le x \le |\eta|$ in our definitions,
although their partonic interpretation belongs more to the excitation of
a mesonic like state. However, Lorentz invariance requires that the
`inclusive' region, which is closer to a probabilistically partonic
interpretation, is entirely determined by the `exclusive' one -- the
reverse is generally not true.

Let us now discuss the form factor aspects of GPDs. We define for any
parton species its own partonic form factor, which is given by the
lowest moment:
\begin{eqnarray}
\label{Def-SumRulGen}
\int_{-\xi}^{1} dx\, H^{i/{\rm A}}(x,\eta,\Delta^2,\cQ^2) = 
N_{i}  F^{i}( \Delta^2)\, .
\end{eqnarray}
The above given definition (\ref{GPD-Sep-ValSea}) induces the
equality of sea- and anti-quark form factors. Moreover, Lorentz
invariance and current conservation ensure that the r.h.s. is
independent on $\eta$ and $\cQ^2$, respectively. We normalize the partonic form
factors  to one
\begin{eqnarray}
F^{i}( \Delta^2=0)= 1\, .
\end{eqnarray} 
Thus, $N_{i}$ might be interpreted as the number of a given quark species $i$ 
inside the nucleus.

The above given sum rules for parton densities can now be generalized to
the non-forward kinematics. The charge sum rule (\ref{SumRul-Cha}) turns
into the relation between partonic and electromagnetic form factors: 
\begin{eqnarray}
\label{SumRul-ScaChaFF}
F( \Delta^2)  = Q_u N_{u_{\rm val}} F^{u_{\rm val}}( \Delta^2) + 
Q_d N_{d_{\rm val}} F^{d_{\rm val}}( \Delta^2)\, , \qquad 
F( \Delta^2=0)=Z\, ,
\end{eqnarray}
while the Baryon conserving ones (\ref{SumRul-Bar}) reads 
\begin{eqnarray}
\label{SumRul-ScaBarFF}
F^{\rm Bar} ( \Delta^2)  = 
\frac{1}{3} N_{u_{\rm val}} F^{u_{\rm val}}( \Delta^2) +
\frac{1}{3}  N_{d_{\rm val}} F^{d_{\rm val}}( \Delta^2)\, , \qquad 
F^{\rm Bar}( \Delta^2=0)=A\, ,
\end{eqnarray}
The sea--like GPDs are antisymmetric in $x$ by definition and so they can
not contribute to these sum rules. The first moment of the singlet
combination, including the gluonic component, 
\begin{eqnarray}
\label{SumRul-EneMomFF}
\int_{-1}^{1} dx\, x\left[\sum_{i=u,d,s} H^{i/{\rm A}}(x,\eta,\Delta^2,\cQ^2) +
 H^{g/{\rm A}}(x,\eta,\Delta^2,\cQ^2)  \right] = 
\frac{1}{P_+^2} \langle {\rm A} (P_2) | T_{++} |  {\rm A}(P_1) \rangle
\end{eqnarray}
is given by the expectation value of the +--components of the
energy-momentum tensor $T_{\mu\nu}$. Since of current conservation, it is 
also independent on the
scale $\cQ$. The r.h.s.\ is then a polynomial in $\eta$ of
the second order whose coefficients are given by the two gravitational form
factors, which appear in the matrix element 
$\langle {\rm A}(P_2) | T_{\mu\nu} |  {\rm A}(P_1) \rangle $: 
\begin{eqnarray}
\label{SumRul-GraForFac}
 \frac{1}{P_+^2}  \langle {\rm A}(P_2) | T_{++} |  {\rm A}(P_1) \rangle = 
T^{1}(\Delta^2) + \eta^2 T^{2}(\Delta^2), 
\quad \mbox{with}\quad  T^{1}(\Delta^2=0) =1\, .
\end{eqnarray}
The normalization of $T^{1}(\Delta^2=0)$ is consistent with the momentum sum
rule (\ref{SumRul-Mom}). It has been argued in Ref. \cite{Pol02} that
the second form factor is related to the spatial distribution of strong forces  
inside a nucleus and that its partonic content can be accessed via the GPDs.

\begin{figure}[t]
\begin{center}
\mbox{
\begin{picture}(550,110)(0,0)
\put(0,20){\insertfig{3.5}{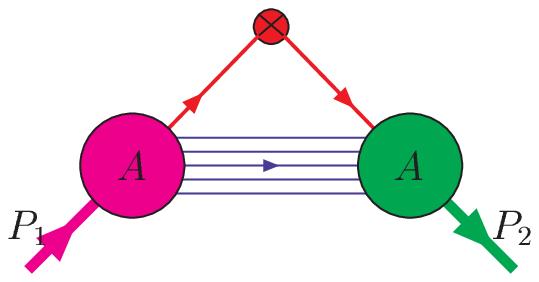}}
\put(42,0){\small (a)}
\put(115,15){\insertfig{3.5}{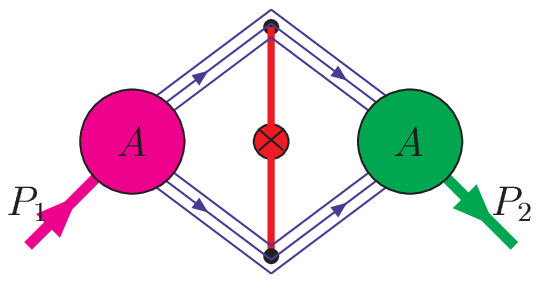}}
\put(158,0){\small (b)}
\put(230,20){\insertfig{3.5}{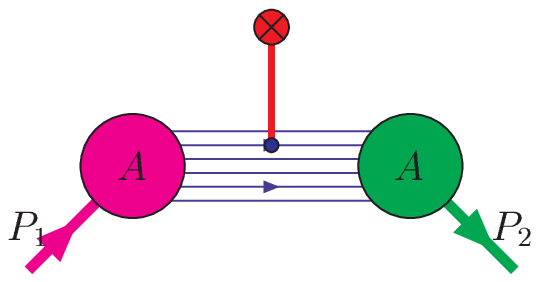}}
\put(276,0){\small (c)}
\put(345,20){\insertfig{3.5}{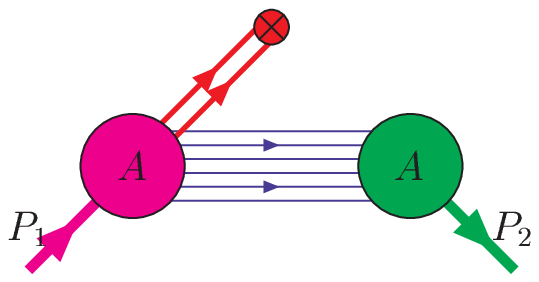}}
\put(390,0){\small (d)}
\end{picture}
}
\end{center}
\caption{
Different contributions to a nucleus GPD in which the light--ray operator couples 
to a nucleon (a), meson exchange in the $t$-channel (b),  mesonic contact 
term (c), and virtual nucleon pair (d). 
\label{Fig-GPDCon}
}
\end{figure}
To relate the nucleus GPDs to those of the proton, we rescale, as in the
forward case, the longitudinal momentum fractions, i.e., their
$x$-dependence and $\eta$-dependence. This is again motivated by
considering the nucleus as a system of almost free nucleons, which have
approximately the same momenta $P_1/A= (1-\eta)P_+/2A $. At leading
order in $1/Q^2$, the interaction with the two photons take then place
on one of them, which has the outgoing momentum $P_1/A+\Delta= (1-\eta+2
A\eta)P_+/2A + \Delta^\perp $. Thus,  instead of the GPD
definition (\ref{Def-GPD-sca}) we have to consider the matrix element, see Fig.\
\ref{Fig-GPDCon}a
\begin{eqnarray}
\int \frac{d\kappa}{2\pi} e^{i \kappa P_+ x} 
\langle N(P_1/A+ \Delta) | \bar\psi_i(-\kappa n) \gamma_+  
\psi_i(\kappa n) | N(P_1/A) \rangle
\Big|_{\eta= \frac{\Delta_+}{P_+}}\, .
\end{eqnarray}
To make contact with the nucleon GPD, we refer in this 
 matrix element to the light--cone component of the struck nucleon
$P^N_+ = [1-\eta +A\eta] P_+/A$. This implies the rescaling prescription
for the light--cone momentum fraction and skewedness parameter 
\begin{eqnarray}
x_N =    \frac{1-\eta_N}{1-\eta} A x  \qquad \mbox{and}\qquad 
\eta_N \equiv \frac{\Delta_+}{P_+^N } =  \frac{1-\eta_N}{1-\eta} A  \eta\, .
\end{eqnarray} 
From the second equality one easily recovers the scaling law for the
kinematical variable, namely, $\Bx = A \Ax$. However, this picture can not be
quite realistic, since transferring the whole transversal momentum
$\Delta^\perp$ to the struck nucleon, should, compared to a binding energy of
about 10 MeV per nucleon, induce target dissociation. On the
other hand, nuclei form factors are non-zero even for rather large
momentum transfer, which shows that there is a certain probability that
the transversal momentum transfer is almost equally distributed among
the constituents\footnote{Certainly, it remains an experimental problem
to ensure that the measured Compton scattering process is coherent and a
theoretical task to derive the predictions for target dissociation. }. 
{\em Assuming} that the $\Delta^2$-dependences factorize, we
can take care of this bound state effect by replacing  the partonic proton
form factors with the nucleus ones:
\begin{eqnarray}
\label{GPD-A-Sca}
 H^{u_{\rm val}/{\rm A}}(x,\eta,\Delta^2) \!\!\! &=& \!\!\!
F^{u_{\rm val}/{\rm A}}(\Delta^2) \Big|\frac{dx_N}{dx}\Big|
\theta(|x_N| \le 1)\left[ Z H^{u_{\rm val}}
+ N H^{d_{\rm val}} \right](x_N,\eta_N)\, , 
\nonumber\\
H^{d_{\rm val}/{\rm A}}(x,\eta,\Delta^2) \!\!\! &=& \!\!\! F^{d_{\rm val}/{\rm A}}(\Delta^2) 
\Big|\frac{dx_N}{dx}\Big| \theta(|x_N|  \le 1) \left[ Z H^{d_{\rm val}}
+ N H^{u_{\rm val}} \right](x_N,\eta_N)\, , 
\nonumber\\
H^{i_{\rm sea}/{\rm A}}(x,\eta,\Delta^2) \!\!\! &=& \!\!\!   F^{{\rm sea}/{\rm A}}(\Delta^2) 
 \Big|\frac{dx_N}{dx}\Big|\theta(|x_N|  \le 1)  A\, H^{i_{\rm sea}}(x_N,\eta_N)\, , 
\\
H^{g/{\rm A}}(x,\eta,\Delta^2) \!\!\! &=& \!\!\! F^{g/{\rm A}}(\Delta^2)  
\Big|\frac{dx_N}{dx}\Big| \theta(|x_N|  \le 1) A\, H^{g}(x_N,\eta_N)\, .
\nonumber
\end{eqnarray}
Here we introduced the reduced proton GPDs by $H^{i}(x,\eta)=
H^{i}(x,\eta,\Delta^2=0)$. The support of the proton GPDs induces the
restriction $ A |x| \le (1-\eta)/(1-\eta_N)$, which of course is only an
artifact of our approximation. It is easy to check that by construction
all sum rules remain valid. For an isoscalar target isospin symmetry
requires 
\begin{eqnarray}
H^{u/{\rm A}}(x,\eta,\Delta^2)= H^{d/{\rm A}}(x,\eta,\Delta^2)\, 
\end{eqnarray} 
and, thus, also the partonic $u$ and $d$ form factors are equal.

Obviously, what we presented here is the simplest version of a
convolution model for nuclei GPDs, which certainly can be improved by
incorporation of binding effects. Hopefully, they might provide only
small corrections to the scaling law (\ref{GPD-A-Sca}) and should
explain the distribution of transversal momentum transfer. Nevertheless,
in the case that such effects are responsible for symmetry breaking,
e.g., the breaking of spherical symmetry implies a non-vanishing
quadrupole moment for deuteron, they might become a leading contribution
to GPDs that decouples in the forward case. Thus, it is worth to have
a closer look on them.

First we consider the possibility that the light--ray operator insertion
couples to meson exchange currents in the $t$--channel, as it is
shown in Fig.\ \ref{Fig-GPDCon}b. The corresponding matrix element reads
\begin{eqnarray}
&&\hspace{-0.7cm}\int \frac{d\kappa}{2\pi} e^{i \kappa P_+ x} 
\langle n(P_1/A+ \Delta/2 )  p(P_1/A + \Delta/2 )| 
\bar\psi_i(-\kappa n) \gamma_+  \psi_i(\kappa n) | 
p(P_1/A)  n(P_1/A)\rangle
\Big|_{\eta= \frac{\Delta_+}{P_+}}^{t-\rm{channel}}\, 
\nonumber\\
&&\hspace{-0.7cm}\propto \int_{-1}^{1} du\,  \phi(u) 
\Big|\frac{u dx^{\rm ex}}{dx}\Big|
\int \frac{d\kappa}{2\pi} 
e^{i \kappa P_+^{\rm ex} x^{\rm ex}}\langle \pi([u-1]\Delta/2)| 
{\cal O}(\kappa,-\kappa)
|\pi([u+1]\Delta/2) \rangle
\Big|_{\eta^{\rm ex}= \frac{\Delta_+}{P_+^{\rm ex}}}\, .
\nonumber
\end{eqnarray}
Here the probability amplitude $\phi(u)$ with $|u| \le 1 $ is a  function
of the light--cone momentum fraction $u$, defined by
$P_+^{\rm ex} =u \Delta_+ = u \eta P_+$. The momentum fraction $x^{\rm
ex}$ and skewedness parameter $\eta^{\rm ex}$, entering in the GPD, are
\begin{eqnarray}
x^{\rm ex} =  \frac{x}{u\eta} \quad \mbox{and}\quad 
\eta^{\rm ex} \equiv \frac{\Delta_+}{P_+^{\rm ex}} 
=   \frac{1}{u}\, .
\end{eqnarray}
The support property of the resulting GPD 
\begin{eqnarray}
H^{t-{\rm cannel}}(x,\eta) \propto \frac{x}{|\eta|}  \int_{-1}^1\! du\, 
{\rm sign}(u \eta)\, \phi(u)   
\int_{-1}^1\! dy\!\!\int^{1-|y|}_{-1+|y|}\! dz\,
\delta (x- \eta[u y+ z] ) F_\pi^{t-{\rm cannel}}(y,z)\, 
\end{eqnarray}
can be deduced from the DD representation. It only contributes in the
`exclusive' region, since for the `inclusive' region we find $|x/\eta|
>1\ge |u y+ z| $. Time-reversal invariance combined with hermiticity
requires that $\phi$ and $F_\pi^{t-{\rm cannel}}(y,z)$ are symmetric in
$u$ and $z$, respectively. Consequently, the exchange GPD is also
symmetric in $x$ and the polynomiality condition is satisfied in
general. For the lowest moment we have the normalization:
\begin{eqnarray}
\int_{-1}^1dx  H^{t-{\rm cannel}}(x,\eta) \propto 
 \int_{-1}^1\! du\, |u| \phi(u)   
\int_{-1}^1\! dy y\!\!\int^{1-|y|}_{-1+|y|}\! dz\,  
F_\pi^{t-{\rm cannel}}(y,z)\, .
\end{eqnarray}
Remembering that the integral over $z$ gives the parton densities of
the pion, i.e., $q_\pi(y)/y$, we obviously, establish a link between
`exclusive' $t$-channel contributions and the parton picture. Again the
total sea quark contribution $q^{\rm sea}_\pi(y)$ is antisymmetric in
$y$ and, thus, it drops out in the sum rules
(\ref{SumRul-ScaChaFF},\ref{SumRul-ScaBarFF}) and also for higher
moments (which are even). Such an $I=1$ isospin exchange contribution is
`filtered' out for an isoscalar target, however, it might be important for
other targets or  dissociation processes. Note that
for an isoscalar target, resonance exchange contributions appear, like
the famous $\rho\pi\gamma$ exchange current. We should add that in such
a case also contact terms, depicted in Fig.\ \ref{Fig-GPDCon}c, have to be
included, which for instance are given by 
\begin{eqnarray}
&&\hspace{-0.7cm}\int \frac{d\kappa}{2\pi} e^{i \kappa P_+ x} 
\langle N(P_1/A+ \Delta/2 )  \pi(\Delta/2 )| 
\bar\psi_i(-\kappa n) \gamma_+  \psi_i(\kappa n) | 
N(P_1/A) \rangle
\Big|_{\eta= \frac{\Delta_+}{P_+}}\, .
\end{eqnarray}
Moreover, Lorentz invariance requires that virtual nucleus anti-nucleus
states contribute to the nucleon GPD. An example is depicted in Fig.\
\ref{Fig-GPDCon}d and its matrix element reads 
\begin{eqnarray}
\int \frac{d\kappa}{2\pi} e^{i \kappa P_+ x} 
\langle \Omega| 
\bar\psi_i(-\kappa n) \gamma_+  \psi_i(\kappa n) | 
N(p_1)  \overline{N}(p_2)\rangle
\Big|_{\eta= \frac{\Delta_+}{P_+}}\, .
\end{eqnarray}
It is far beyond the scope of this paper to establish a closer link
between the effective forces of nuclei and the fundamental degree of
freedom in QCD, however, our discussion shows that GPDs are an
appropriate tool for this task.

Based on a `popular' model for the reduced proton GPDs, we present now
our model for a scalar nucleus. To implement the support properties we
deal with the DD representation \cite{MueRobGeyDitHor94, Rad96} at a
given input scale $\cQ=\cQ_0$:
\begin{eqnarray}
\label{Mod-GPD-ScaTar}
H^{i}(x,\eta)=  \int_{-1}^1 dy
\int_{-1+|y|}^{1-|y|} dz\;
\delta(x-y - \eta z) q^{i}(y) \pi(|y|,z;b^i) + 
{\rm sign}(\eta) D^i(x/\eta)
\end{eqnarray}
and adopt Radyushkin's proposal for the factorization of the DD into a
forward parton density $q^i(y)$ given at the input scale $\cQ_0$ and a
profile function
\begin{eqnarray}
\pi(y,z;b) = \frac{\Gamma\left(b+\frac{3}{2}\right)}{\sqrt{\pi} \Gamma(b+1)}
\frac{\left[(1-y)^2 - z^2\right]^b}{(1-y)^{2b+1}}\, ,
\end{eqnarray}
which is normalized to 1:
\begin{eqnarray}
\int_{-1+|y|}^{1-|y|} dz\, \pi(|y|,z;b) = 1.
\end{eqnarray}
Here the free parameter $b$ models the skewedness effect that arises
from the $x$-shape of GPDs for a given value of $\eta$. In the limit
$b\to \infty$ the GPDs are independent of $\eta$ and are simply given by
$q^i(x)$ itself. For the unpolarized parton densities of the proton we
will take the MRS A' parameterization \cite{MarRobSti95}, given at the
input scale $\cQ^2_0=4\ \GeV^2$, with $\bar{u}= \bar{d}=\bar{s}/2$. For
the electromagnetic form factors we adopt here for simplicity the
following parameterization
\begin{eqnarray} 
F^{u_{\rm val}/{\rm A}}(\Delta^2) =
F^{d_{\rm val}/{\rm A}}(\Delta^2) =\frac{1}{Z} F^{\rm A}(\Delta^2)\, , \qquad
F^{{\rm sea}/{\rm A}}(\Delta^2) = e^{B^{\rm sea} A^{2/3} \Delta^2} \, ,
\end{eqnarray} 
where the slope  $A^{2/3} B^{\rm sea}$ of the sea quarks is
defined in such a way that $B^{\rm sea}$ only weakly depends on $A$.
This is motivated by the fact that the nucleus density is rather
independent on $A$ and, thus, we expect that  the radius of
the sea quark distribution should scale with $A^{1/3}$. For the
electromagnetic form factor we adopt the parameterization of Ref.\
\cite{Steetal75}:
\begin{eqnarray}
F^{\rm A}(\Delta^2) = Z \left(1-\frac{1}{6} c^2 \Delta^2\right)^{-1} 
\exp\left\{\frac{1}{6} \Delta^2  b^2  \right\}\, ,
\end{eqnarray}
with $b = 2.04 \times 5.07\ \GeV^{-1}$ and  $c = 1.07 A^{1/3} 
\times 5.07\ \GeV^{-1}$.

There are a few comments in order. First the factorization in $\Delta^2$
for medium values has no theoretical or experimental justifications. In
comparison with experimental data, which are presently measured at a
mean value of $<-\Delta^2> \sim 0.1...0.3\ \GeV^2$, the
$\Delta^2$-dependences effectively enters only the normalization of the
CFFs via the slope parameters. Second remark concerns the polynomiality
property of the moments. In comparison to Eq.\ (\ref{DD2GPD}) we choose
for simplicity another DD representation in Eq.\ (\ref{Mod-GPD-ScaTar}).
The prefactor $x$ is now neglected and the highest possible term
$\eta^{n+1}$ for the $n$th odd $x$--moment is restored by the so-called
$D$-term. For detailed discussions on this subject see Ref.\
\cite{PolWei99,BelMueKirSch00}. This $D$-term is antisymmetric in $x$
and can not spoil the sum rule (\ref{Def-SumRulGen}), however, it enter
the sum rule (\ref{SumRul-GraForFac}) and is discussed below. In our
language we count it as a sea-quark contribution. The third remark
concerns the factorization of the DDs ansatz, that is inspired by an
intuitive picture and provide in general an enhancement at the point
$x=\pm\eta$. Finally, we should comment on the meson exchange
contributions. For a scalar target we will not explicitly take them into
account. They effectively enter here the parameterization of the form
factors. Moreover, they have do vanish in the forward limit and,
therefore, they should be suppressed for valence like contributions at
small $\Delta^2$, while for sea--like contributions they can be included
in the antisymmetric $D$-term.

Let us finish this section by expressing the nucleus CFF ${\cal H}^{\rm
A}$ by the proton ones in the twist-two and twist-three sector. From the
ansatz (\ref{GPD-A-Sca}) and the convolution\footnote{Remove the index
$k$ for a scalar target .} (\ref{DefTw2}), one easily establish the
following scaling law for the twist-two CFF of a isoscalar nucleus
\begin{eqnarray}
{\cal H}^{\rm A} = A^2 \frac{1+\xi_N}{1+\xi}  
\left[ \frac{Q_u^2 + Q_d^2}{2 Z} 
F^{\rm A}(\Delta^2) 
\left(\frac{{\cal H}^{u_{\rm val}}}{Q_u^2}+ 
\frac{{\cal H}^{d_{\rm val}}}{Q_d^2} \right)(\xi_N) + 
 F^{{\rm sea}/{\rm A}}(\Delta^2) {\cal H}^{{\rm sea}}(\xi_N)
\right]\, .
\end{eqnarray}
Here ${\cal H}^{i}(\xi_N)$ denotes the contribution that appears in the
decomposition of the proton CFF ${\cal H} = {\cal H}^{u_{\rm val}} +
{\cal H}^{d_{\rm val}} +{\cal H}^{{\rm sea}}$ at $\Delta^2=0$. Note
that, as long we rely on the ansatz (\ref{GPD-A-Sca}), this scaling law
is valid beyond the LO approximation, since the hard-scattering
coefficients have the following functional dependence $1/\xi C(x/\xi)$.
It is interesting to note that in Ref.\ \cite{Pol02}, it was argued that
the $D$-term contribution to the Compton form factor, which is
independent of $\Bx$, scales with $A^{7/3}$. Its contribution to the
second gravitational form factor is
\begin{eqnarray}
T^{2}(\Delta^2)= \frac{1}{2} \frac{d^2}{d\eta^2} 
\sum_{i=u,d,s,g} \int_{-1}^{1}dx x H^{i/{\rm A}}(x,\eta,\Delta^2)  = 
\frac{4}{5} A^2 d^{\rm A}(\Delta^2)\, .
\end{eqnarray}
To study the $D$-term with respect to $A$ and in relation to the sea
quark contribution per nucleon, we have rescaled $d^{\rm A}$ by $A^2$. Based
on the estimate in the same reference, one finds that $d^{\rm
A}(\Delta^2=0)$ slightly increases with $A \ge 8$:
\begin{eqnarray}
d^{\rm A}(\Delta^2=0) & \!\!\!=\!\!\!& 
d^{Q/{\rm A}}(\Delta^2=0)+d^{G/{\rm A}}(\Delta^2=0)\		
\approx 
-0.2 A^{1/3}\left(1+ 3.8/A^{2/3}\right) \, ,
\\
d^{Q/N}(\Delta^2=0) & \!\!\!\approx\!\!\!& 
-4.0\quad \mbox{at} \quad \mu\approx 0.6\ \GeV
\, , 
\end{eqnarray}
where the second result for the nucleon $D$-term in the quark sector has
been predicted by a model calculation within the chiral soliton model
\cite{PetPobPolBoeGoeWei97,SchBofRad02}. As it will be discussed in
Section \ref{SubSec-EstAna}, a present measurement of the charge
asymmetry on the proton target does not allow for a definite conclusion
of the $D$-term contribution. Unfortunately, we might conclude that for
nucleus target, e.g., $d^{{\rm A}}\approx -1.3$ for $A=200$, this term
will not induce a significant contribution, too. Certainly, it is an
important task to confront this expectation with experimental
measurements.

Employing the equation of motion, the twist-tree GPDs are decomposed in
the so called Wandzura--Wilczek (WW) term, entirely expressed by the
twist-two GPD $H$, and the quark-gluon-quark GPD $H^{\rm qGq}$, carrying
new dynamical information. The two twist-three GPDs enter in the DVCS
amplitude always in the same linear combination \cite{BelMueKirSch00}:
\begin{eqnarray}
{\cal H}^{\rm eff} = {\cal H}^{\rm eff-WW} + {\cal H}^{\rm qGq}, \,
\end{eqnarray} 
where 
\begin{eqnarray}
{\cal H}^{A{\rm eff-WW}} &\!\!\! =\!\!\!& \frac{2}{1+\xi}{\cal H}^{\rm A} + 
2 \xi \frac{\partial}{\partial \xi} 
\sum_{i=u,d,s}\int_{-1}^1 dx \frac{Q_i^2}{\xi + x} \ln\frac{2\xi}{\xi-x-i 0} 
\\
&&\hspace{6cm}\times \left\{H^{i/{\rm A}}(x,\xi)- H^{i/{\rm A}}(-x,\xi)\right\} \, .
\nonumber
\end{eqnarray} 
Using the ansatz (\ref{GPD-A-Sca}), one easily derives the WW
piece in terms of the proton GPDs $H^i$. For instance, the total sea
quark contribution reads in terms of the SU(2) symmetric antiquark GPDs
\begin{eqnarray}
&&\!\!\!\!\!\! \frac{2}{1+\xi}{\cal H}^{{\rm sea}/{\rm A}}(\xi,\Delta^2) + A^2
 F^{{\rm sea}/{\rm A}}(\Delta^2) 
2\xi \frac{\partial}{\partial \xi} \frac{1+\xi_N}{1+\xi}  
\sum_{i=u,d,s}\int_{-1}^1 dx \frac{Q_i^2}{\xi_N + x} 
\ln\frac{2\xi_N}{\xi_N-x-i 0} 
\nonumber\\
&&\hspace{6cm}\times 
2\left\{\overline{H}^{i/{\rm A}}(x,\xi_N)- 
\overline{H}^{i/{\rm A}}(-x,\xi_N)\right\} \, ,
\end{eqnarray}
where $\xi_N = A\xi/(1 + \xi - A \xi)$ is to be considered as a function of
$\xi$. The analogous formula in the valence quark sector follows by an
appropriate replacement of form factor and GPDs together with the
corresponding charge factors.

\subsection{ GPDs of spin-1/2 nucleus target} 

Analogous models for proton GPDs have been used in Ref.
\cite{BelMueKir01}, for a detailed discussion of GPDs see Ref.\
\cite{GoePolVan01}, and they are consistent with all experimental data
available at present in the LO analysis. It has been stressed in Ref.\
\cite{FreMcD01abc} that this not necessarily true in NLO. 
The problem arises
from the employed model of the flavour singlet GPDs, especially from the
gluonic ones. We will skip this issue here and refer to the discussion
in Ref.\ \cite{BelMueKir01}. Due to the spin structure of the nucleon,
one has to model four different sets of GPDs: 
\begin{eqnarray}
F = \left\{H,E,\widetilde H, \widetilde E \right\}\, ,
\end{eqnarray}
corresponding to helicity conserved and non-conserved form factors with
even and odd parity. Note that for an unpolarized target the dominant
contribution arises from the helicity conserved and parity even GPDs 
$H^i$. They satisfy the constraints
\begin{eqnarray}
\lim_{\Delta\to 0} H^i(x,\eta,\Delta^2) = q^i(x) \quad \mbox{and}\quad 
\int_{-1}^{1} dx\, H^i(x,\eta,\Delta^2) = F_1^{i} (\Delta^2)
\end{eqnarray}
For longitudinally polarized target both $H$ and $\widetilde H$ are
important. The latter ones are related to the polarized sea quark
content and axial--vector form factors:
\begin{eqnarray}
\lim_{\Delta\to 0} \widetilde H^i(x,\eta,\Delta^2) = \Delta q^i(x) 
\quad \mbox{and}\quad 
\int_{-1}^{1} dx\, \widetilde  H^i(x,\eta,\Delta^2) = G_1^{i} (\Delta^2)\, ,
\end{eqnarray}
respectively. 
The helicity non-conserved GPDs $E$ and $\tilde E$, which decouple in
the forward limit, obey the sum rules
\begin{eqnarray}
\int_{-1}^{1} dx\, E^i(x,\eta,\Delta^2) = F_2^{i} (\Delta^2)
\quad \mbox{and}\quad 
\int_{-1}^{1} dx\, \widetilde  E^i(x,\eta,\Delta^2) = G_2^{i} (\Delta^2)\, .
\end{eqnarray}
Perhaps they can be accessed in transversally polarized target
experiments. An essential observation was that the ansatz
(\ref{Mod-GPD-ScaTar}) for the sea-quarks, measured in the small $\Bx$
region, has to be suppressed by choosing a $b^{\rm sea}\to \infty$ and a
large slope $B^{\rm sea} \sim 9\ {\rm GeV}^2$. This suppression then
ensures also the correct size of the single beam spin asymmetry measured
in fixed target experiments. For instance, the mean value at HERMES is
$\Bx \approx 0.1$ and, thus, the value of $\xi \approx 0.05$ appearing
in the argument of GPDs is rather small. As we said, the contribution of
sea quarks remains relatively small with respect to the valence like
ones. If this observation is implemented in the correct way within our GPD
model remains an open issue at present.

The helicity conserved GPDs $H^{i/{\rm A}}$ and $\widetilde H^{i/{\rm
A}}$ of a nucleus with spin-1/2 can be obtained from those of a proton
by means of Eq.\ (\ref{GPD-A-Sca}). Unfortunately, not too much is known
about the structure of $E^{i/{\rm A}}$ and $\widetilde E^{i/{\rm A}}$,
beside the pion-pole contribution of $\widetilde E^{i/{\rm A}}$. For the
parametrization of the twist-three GPDs see Refs.\
\cite{BelKirMueSch01,BelMueKir01}.

\subsection{GPDs of spin-one target}

In this section we list the properties that are known about the spin-one
GPDs and have been already given in Ref.\ \cite{BerCanDiePir01}. For the
matrix element of the twist-two operators we use the analogous
kinematical decomposition as for the CFFs in Eqs.\ (\ref{Def-V1-Sp1})
and (\ref{Def-A1-Sp1}). Combining hermiticity and time reversal
invariance tells us that the GPDs are real valued functions, which
respect the symmetry properties:
\begin{eqnarray}
H_k(x,\eta) \!\!\!&=&\!\!\! H_k(x,-\eta)
\mbox{\ \ for\ } k=\{1,2,3,5\}, \qquad
H_4(x,\eta)  = -H_4(x,-\eta),
\nonumber\\
\widetilde{H}_k(x,\eta) \!\!\!&=&\!\!\!
\widetilde{H}_k(x,-\eta)
\mbox{\ \ for\ } k=\{1,2,4\}, \hspace{1.2cm}
\widetilde{H}_3(x,\eta) =- \widetilde{H}_3(x,-\eta).
\end{eqnarray}
Note that $H_4(x,\eta)$ and $\widetilde{H}_3(x,\eta)$ are
antisymmetric with respect to $\eta$, and, consequently, their moments
are odd polynomials in $\eta$.

From the definition of GPDs as Fourier transforms of  light--ray operators
it follows that their lowest moment is given by the form factors
appearing in the vector or axial--vector current, respectively. In the
former case they are related to the electromagnetic form factors
\begin{eqnarray}
\label{Def-SumRulV1}
\sum Q_i \int_{-1}^1 dx\, H_k^{i/{\rm A}} (x,\eta,\Delta^2,\cQ^2) 
\!\!\!&=&\!\!\! G_k(\Delta^2)
\mbox{\ \ for\ } k=\{1,2,3\},
\\
\label{Def-SumRulV2}
\int_{-1}^1 dx\, H_k^{i/{\rm A}} (x,\eta,\Delta^2,\cQ^2)\!\!\!&=&\!\!\! 0
 \mbox{\ \ for\ } k=\{4,5\},
\end{eqnarray}
or they have to vanish. For the deuteron the electromagnetic form
factors are  known from experimental measurements and their
parameterization is given in Appendix \ref{App-DeuForFac}
\cite{KobSya95,Abbetal00}. For a target with charge $Z e$ we choose the
normalization
\begin{eqnarray}
G_1(\Delta^2=0)=Z\, ,\quad G_2(\Delta^2=0)= Z \mu_A\,  
,\quad G_3^i(\Delta^2=0)=  Z (\mu_A+{\cal Q}_A-1),
\end{eqnarray}
where $Z \mu_A$ is the magnetic moment and $Z {\cal Q}_A$ is the
electrical quadrupole moment. The latter is induced by non-central nuclear
forces. Its non-vanishing value, measured for the deuteron, has been
considered as an evidence for the role of pions in nuclear physics. More
precisely, the one pion exchange provides at larger distances the
dominant contribution to the potential, which is of the Yukawa type, and
induces a D-wave admixture to  the S-wave function.

The sum rules for parity odd GPDs read
\begin{eqnarray}
\label{Def-SumRulA1}
\int_{-1}^1 dx\, \widetilde H_k^i(x,\eta,\Delta^2,\cQ^2) \!\!\!&=&\!\!\!
 \widetilde{G}_k^i(\Delta^2)
\mbox{\ \ for\ } k=\{1,2\},
\\
\label{Def-SumRulA2}
\int_{-1}^1 dx\, \widetilde H_k^i(x,\eta,\Delta^2,\cQ^2)\!\!\!&=&\!\!\! 0
 \mbox{\ \ for\ } k=\{3,4\}.
\end{eqnarray}
The two axial form factors $\widetilde{G}_k^i(\Delta^2)$ with
$k=\{1,2\}$ are defined by the matrix element of the current
$j^{5,i}_{\mu}= \bar{\psi}_i(0)\gamma_\mu \gamma^5 \psi_i(0)$ with
flavour $i$:
\begin{eqnarray}
\label{Def-J5-Cur}
J^{5,i}_\mu =
i \epsilon_{\mu \epsilon_2^\ast  \epsilon_1 P } \widetilde G_1(\Delta^2)
- \frac{i\epsilon_{\mu \Delta P \epsilon_1}\, \epsilon_2^\ast\! \cdot\! P+
i \epsilon_{\mu \Delta P \epsilon_2^\ast}\, \epsilon_1\! \cdot\! P}{\AM^2}
 \widetilde  G_2(\Delta^2)\, ,
\end{eqnarray} 
In principle they, i.e., certain linear combinations with respect to the
flavour number, can be measured due to weak interaction, however, to
our best knowledge this has not be done yet.

First we consider the forward limit in which $H_2,H_3,H_4$ as well as
$\widetilde H_2,\widetilde H_3, \widetilde H_4$ decouple from the
Compton amplitude and, thus, are not measurable in deep-inelastic
scattering. This does not mean that the functions in question vanish by
themselves. The remaining three functions are expressed in terms of
parton densities
\begin{eqnarray}
q^{i/{\rm A}}(x) \!\!\!& \equiv &\!\!\! H_1^{i/{\rm A}}(x,\eta=0,\Delta^2=0) = 
\frac{1}{3}\left\{q^{+1}(x)+ q^{-1}(x) + q^{0}(x) \right\},
\nonumber\\
\delta q^{i/{\rm A}}(x) \!\!\!& \equiv &\!\!\! H_5(x,\eta=0,\Delta^2=0) = 
q^{0}(x)  - \frac{1}{2}\left\{q^{+1}(x)+ q^{-1}(x)\right\},
\\
\Delta q^{i/{\rm A}}(x)  \!\!\!& \equiv &\!\!\! 
\widetilde H_1^{i/{\rm A}}(x,\eta=0,\Delta^2=0) =
q^{+1}_\uparrow(x)- q^{-1}_\uparrow(x),
\nonumber
\end{eqnarray}
where $q^{\Lambda} = q^{\Lambda}_\uparrow + q^{\Lambda}_\downarrow $.
Here $q^{\Lambda}_\uparrow(x)$ is the probability to find a (anti)quark
with momentum fraction $x >0$ ($x<0$) and positive helicity in the
target of helicity $\Lambda$. Note that these definitions contain both
quark $(x\ge 0)$ and antiquark $(x\le 0)$ contributions with the
following sign conventions:
\begin{eqnarray}
&& {\overline q}(x) = - q(-x)\, ,\quad  \delta {\overline q}(x) = - \delta q(-x) 
\, ,\quad  
\Delta{\overline q}(x) = \Delta q(-x),\quad \mbox{for}\quad x\ge 0.
\end{eqnarray}
The combination of quark distributions in $H_5$ enter the structure
function $b_1$, measurable in deeply inelastic scattering on a polarized
spin-one target. The sum rule (\ref{Def-SumRulA2}) then induces
\begin{eqnarray}
\label{SumRul-ass-b1}
\int_{-1}^1 dx\, \delta q (x) =0 \qquad \Longrightarrow\qquad
\int_{0}^1 dx\, \delta q^{\rm val} (x) =0.
\end{eqnarray}
To obtain the second sum rule for valence quarks we employed the
antisymmetry property of the sea quarks. This relation then converts
into a sum rule for $b_1$
\begin{eqnarray}
\int_{0}^1 dx\,  b_1(x) = 
2\sum_{i=u,d,s} e_i^2 \int_{0}^1 \delta \overline q_i (x) dx\, ,
\end{eqnarray}
which vanishes for an unpolarized quark sea \cite{CloKum90}. Considering
the deuteron as composed of almost free nucleons induces a vanishing ratio $b_1/F_1$
\cite{HooJafMan89}. The first preliminary measurement from the HERMES
collaboration \cite{Conetal02} indicates a tensor asymmetry, $A_{zz} =
-2b_1/3F_1$, that strongly depends on $\Bx$. It is compatible with zero
in the valence quark region, but significantly positive and negative for
large and small $\Bx$, respectively. Since we are in the following
interested in the valence quark region, we set $\delta \overline q_i
(x)=0$. Consequently, this simplifies the spin content of the
unpolarized quark distributions 
\begin{eqnarray}
\label{Con-q0}
q^{0}(x) = \frac{1}{2}\left\{q^{+1}+q^{-1} \right\}(x) 
\qquad \Longrightarrow\qquad 
q(x) = \frac{1}{2}\left\{q^{+1}+q^{-1} \right\}(x)\, .
\end{eqnarray}

Let us now discuss the modelling of GPDs. In this paper we only consider
numerical estimates for unpolarized or longitudinally polarized target,
in which ${H}_2, H_4, \widetilde{H}_2, \widetilde{H}_3$, and
$\widetilde{H}_4$ are relatively suppressed by $\xi\sim (\Bx/2 A)$ or
$\tau = \Delta^2/4 A^2 M_N^2$ in the DVCS amplitude and will be
neglected. To satisfy the sum rule (\ref{Def-SumRulV2}) for
$H_5(x,\eta)$ we set it to zero. This choice is motivated by the
knowledge of parton densities. Alternatively, we can also take an
antisymmetric function in $x$, e.g., proportional to
$H_1(x,\eta)-H_1(-x,\eta)$. It is worth noting that both GPDs can be
related to the proton GPDs $E$ and $\widetilde E$, respectively. They
are accessible in experiments with transversally polarized targets,
which, however, is beyond the scope of this paper.

For the GPDs $H_1^{i/{\rm A}}$ and $\widetilde H_1^{i/{\rm A}}$ we
employ the scaling relation (\ref{GPD-A-Sca}) to connect them to the
proton GPDs $H^{i}$ and $\widetilde H^{i}$, cf.\ Eqs.\ (\ref{RelqA2qp})
and (\ref{RelqA2qpSca}). For the latter we will neglect the D-wave
admixture, which provides in the forward limit for the deuteron a few
percent effect. To implement the support properties of GPDs as well as
the reduction to the parton densities in a simple manner we again employ
the DD representation 
\begin{eqnarray}
\label{Mod-GPD-Tar-1}
\left\{ {H^{i}_1 \atop \widetilde H^{i}_1} \right\} =
\left\{ G^i_1 \atop \widetilde G^i_1\right\} (\Delta^2)   \int_{-1}^1 dy
\int_{-1+|y|}^{1-|y|} dz\;
\delta(x-y - \eta z) \left\{  q^{i} (y) \pi(|y|,z;b^i_1) \atop  \Delta q^{i}(y) 
\pi(|y|,z;\widetilde b^i_1) \right\} \, ,
\end{eqnarray}
where a possible $D$-term is neglected. The partonic form factors are
normalized to one and for an isoscalar target we choose 
\begin{eqnarray}
\label{DefFFval-Deu}
G_1^{u_{\rm val}}(\Delta^2) = G_1^{d_{\rm val}}(\Delta^2) = \frac{1}{Z}G_1(\Delta^2)
\, .
\end{eqnarray} 
Analogous to the scalar case, cf.\ Eq.\ (\ref{SumRul-ScaChaFF}), this
choice ensures that the sum rules (\ref{Def-SumRulV1}) are satisfied,
where again the sea quarks do not contribute. Inspired by the counting
rules, we take for the deuteron a partonic sea quark form factor with a
rather simple parameterization
\begin{eqnarray}
\label{DefFFsea-Deu}
G_1^{\rm sea}(\Delta^2) = 
\left(1 + \frac{2\tau}{3} {\cal Q}^{\rm sea}\right) 
\left(1-\frac{\Delta^2}{m^2_{\rm sea}} \right)^{-6},
\end{eqnarray}
where the free parameters $m^2_{\rm sea}=6/B^{\rm sea}$, expressed by
the slope $B^{\rm sea}$ of the `charge' form factor, and the
`quardropol' moment for sea quarks are specified below. Although
theoretical considerations predict also a rather complex shape for the
axial form factor, see Ref.\ \cite{CooMil01} and references therein, we
will use the simple parameterization 
\begin{eqnarray}
\widetilde G_1^i(\Delta^2) &\!\!\!=\!\!\!& 
\left(1-\frac{\widetilde B^i}{5} \Delta^2 \right)^{-5} \quad\mbox{for}\quad
i=\{u^{\rm val}, d^{\rm val}\},
\\
\nonumber
\widetilde G_1^i(\Delta^2) &\!\!\!=\!\!\!&
\left(1-\frac{\widetilde B^i}{6} \Delta^2\right)^{-6} \quad\mbox{for}\quad
i=\{u^{\rm sea}, d^{\rm sea}, s^{\rm sea}\},
\end{eqnarray}
for the partonic form factors in the small $-\Delta^2$ region. The
normalization $\widetilde G_1^i(\Delta^2=0)=1$ ensures the correct
reduction to the forward limit.

As mentioned above, the quadrupole moment arises from a pure bound state
effect, successfully explained by the one pion exchange, and essentially
given by the matrix element sandwiched between S- and D-wave state of
the deuteron. As stated before, exchange and contact $I=1$ contributions
drop out and so the dominant contribution should arise from the overlap 
\begin{eqnarray}
H_3^{i/d} \sim \int \frac{d\kappa}{2\pi} e^{i \kappa P_+ x} 
\langle d(P_2, {}^3\! D_1) | \bar\psi_i(-\kappa n) \gamma_+  \psi_i(\kappa n) 
|  d(P_1, {}^3\! S_1) \rangle
\Big|_{\eta= \frac{\Delta_+}{P_+}}\, . 
\end{eqnarray} 
Certainly, we have to include the interaction between both
nucleons. For the integrated GPD, i.e., the form factor $G_3$, one can
write the form factor as a product of the isoscalar nucleon form factor
and the so-called body form factor. The latter mainly arises due to the
overlap of $D$ and $S$ waves and alters the $\Delta^2$ dependence of the
isoscalar form factor. An extension of this approach to GPDs has been
proposed in \cite{CanPir02a}, where  $H_3$ is essentially given as
convolution of the isoscalar GPD, modelled as $H^{\rm iso}= (H^u +
H^d)/2$. On the other hand, $H_3$ is probing the binding force of the
deuteron, and as we know from  deep inelastic scattering this effect
rather depends on the value of $\Bx$. For GPDs the situation is rather
complex and at present we have no calculation or model available that
allows us to have a deeper insight in this problem. Thus, we consider
two extreme cases and discuss their consequences in the next section: 
\begin{itemize}
\item 
We assume that $H_3$ arises from the binding forces between
`partons' that carry either rather large or small momentum fractions.
Thus, one might expect that such effects only slightly influence DVCS
observables at present fixed target experiments. Therefore, we set for moderate
values of $\Bx$ the CFF to zero:
\begin{eqnarray}
{\cal H}_3=0 \qquad \mbox{for} \qquad  0.1 \le \Bx \le 0.3\, .
 \end{eqnarray}
\item 
In the convolution model  $H_1$ and $H_3$ are both essentially
determined by the isoscalar GPD $H^{\rm iso}$. So we equate the reduced
GPDs $H_1$ and $H_3$:
\begin{eqnarray}
H_3(x,\eta) = H_1(x,\eta)\, .
 \end{eqnarray}
\end{itemize}

\section{Estimates for observables}
\label{Sec-EstObs}

In Section \ref{SubSec-EstAna} and \ref{SubSec-EstNum} we give
analytical and numerical estimates for the size of the unpolarized cross
section, the beam spin asymmetry
\begin{eqnarray}
\label{Def-ALU}
A_{\rm LU}(\phi)
=
\frac{
d\sigma^\uparrow(\phi) - d\sigma^\downarrow(\phi)
}{
d\sigma^\uparrow (\phi)+ d\sigma^\downarrow(\phi)
}\, ,
\end{eqnarray}
the longitudinally polarized target spin asymmetry
\begin{eqnarray}
\label{Def-AUL}
A_{\rm UL}(\phi)
=
\frac{
d\sigma^\Uparrow(\phi) - d\sigma^\Downarrow(\phi)
}{
d\sigma^\Uparrow (\phi)+ d\sigma^\Downarrow(\phi)
}\, ,
\end{eqnarray}
as well as for the charge asymmetry of the unpolarized cross section
\begin{eqnarray}
\label{Def-AC}
A_{\rm C}(\phi)
=
\frac{
d\sigma^+(\phi) - d\sigma^-(\phi)
}{
d\sigma^+ (\phi)+ d\sigma^-(\phi)
}\, .
\end{eqnarray}
Moreover, for a spin-1 target we consider also the beam spin asymmetry
\begin{eqnarray}
\label{Def-ALpm}
A_{{\rm L}\pm}(\phi)
=
\frac{
d\sigma^{\uparrow\Uparrow}(\phi)+d\sigma^{\uparrow\Downarrow}(\phi) -
 d\sigma^{\downarrow\Uparrow}(\phi)-d\sigma^{\downarrow\Downarrow}(\phi)
}{
d\sigma^{\uparrow\Uparrow} (\phi)+d\sigma^{\uparrow\Downarrow}(\phi)+
 d\sigma^{\downarrow\Uparrow}(\phi)+ d\sigma^{\downarrow\Downarrow}(\phi)
}\, 
\end{eqnarray}
and the tensor polarization 
\begin{eqnarray}
\label{Def-Azz}
A_{zz}(\phi) = 
\frac{ d\sigma_{zz} }
{3d\sigma_{\rm unp}} \, , \quad
d\sigma_{zz} \equiv d\sigma^{\Uparrow} +  
d\sigma^{\Downarrow}-2  d\sigma^{\Rightarrow}\, 
,
d\sigma_{\rm unp} \equiv \frac{1}{3}
\left( d\sigma^{\Uparrow} +  d\sigma^{\Downarrow}+d\sigma^{\Rightarrow} \right),
\end{eqnarray}
where $\Uparrow$, $\Downarrow$, and $\Rightarrow$ denote the magnetic
quantum numbers $\Lambda= \{+1,-1,0\}$ for a longitudinally polarized
target. We give estimates for HERMES and (upgraded) JLAB kinematics for
a lepton beam of $E=27.6\ \GeV$ and $E=6\ (12)\ \GeV$, respectively.

The single spin and charge asymmetries are dominated by the first
harmonics of the interference term (\ref{InterferenceTerm}), which arise
at twist-two level, while for the tensor polarization the squared BH
term dominates in the charge even sector. All observables are
contaminated by $O(1/\cQ)$ suppressed effects, induced by other
harmonics in both the interference and squared DVCS term or from pure
kinematical effects: higher harmonics of the squared BH term
(\ref{Par-BH}) in the denominator and additional $\phi$ dependence due
to the BH propagators ${\cal P}_1 {\cal P}_2(\phi)$. We found that
higher twist-three harmonics are small in the WW approximation
\cite{BelMueKir01}. For the charge asymmetry also a `constant'
contribution arises in the interference term, which is completely
determined by the twist-two GPDs. It can be relatively large with
respect to the dominant $\cos(\phi)$ term, since in contrast to higher
harmonics it does not vanish at the kinematical boundaries. Certainly,
both the size of asymmetries and their power suppressed contamination
depend on the ratio of DVCS  to BH amplitude, which is estimated to
be\footnote{This is based on the assumption that the CFF scales like
$A^2/\Bx$ as it follows from Eq.\ (\ref{GPD-A-Sca}) together with the
small $\Bx$ behavior of the sea quarks. In the valence quark region we
expect even a larger suppression.}
\begin{eqnarray}
\label{Rat-DVCS2BH}
\frac{{\cal T}^{\rm DVCS}}{{\cal T}^{\rm BH}} \sim
 \frac{A}{Z} \sqrt{\frac{- (1-y)\Delta^2}{ y^2 \cQ^2 }}\, .
\end{eqnarray}
For HERMES kinematics with $|\Delta^2| < 0.3\ \GeV^2,\ \Bx\sim 0.1$, and
$\cQ^2\sim 2\ \GeV^2$, i.e., $y\sim 0.5$, the ratio is ${\cal T}^{\rm
DVCS}/{\cal T}^{\rm BH} < 0.5 A/Z$. For JLAB the value of $y$ is
typically larger. This results into a stronger suppression of the DVCS
amplitude, even for a smaller photon virtuality. We realize that for
HERMES kinematics the ratio of interference to squared BH term is not
necessary small, which is obviously consistent with the measurement of
sizable beam spin and charge asymmetries. However, we  have maybe
overestimated the ratio (\ref{Rat-DVCS2BH}). If the sea quark
 contribution is not yet dominant in this fixed target kinematics,
we might have an additional suppression by a factor $\sqrt{\Bx}$.
Indeed, the measurement of the charge asymmetry for a proton target
shows that the real part of the interference term is suppressed by a
factor of about ten. The unpolarized squared BH amplitude, which can be
taken into account exactly, is dominated by its `constant' term, while
higher harmonics are suppressed by the factor $K \sim \sqrt{-
(1-y)\Delta^2 /{ \cQ^2 }}$.

Since we expect, in the kinematics we are interested in, that the
unintegrated asymmetries (\ref{Def-ALU}, \ref{Def-AUL}, \ref{Def-AC})
are dominated by the zero harmonic of the squared BH and first ones of
the interference term, the $\phi$ dependence of the BH propagators will
almost cancel each other. Thus, their Fourier series  read 
\begin{eqnarray}
\label{Def-AsyForExp}
A_{\rm LU}(\phi) \!\!\!&=&\!\!\!  
\sin(\phi^\prime_\gamma) A^{(1)}_{\rm LU}  
		+\cdots\, , 
\quad
A_{\rm UL}(\phi) =\sin(\phi^\prime_\gamma) A^{(1)}_{\rm UL}  
			+\cdots\, , 
\nonumber\\
A_{\rm C}(\phi) \!\!\!&=&\!\!\! A_{\rm C}^{(0)} + 
	\cos(\phi^\prime_\gamma)\, A^{(1)}_{\rm C}+\cdots\, ,
\end{eqnarray}
where the  Fourier coefficients are proportional to certain linar combinations of 
twist-two CFFs, contaminated by $1/\cQ^2$ power
suppressed contributions. For the charge asymmetry $A_{\rm
C}(\phi)$ we included also the `constant' twist-three
contribution, discussed above. To coincide with the definitions, used by
the HERMES and CLAS collaborations, we wrote the Fourier expansion with
respect to the azimuthal angle of the real photon
$\phi^\prime_\gamma=\pi-\phi$ for a frame in which the $z$-axis points in
the direction of the virtual photon momentum. This convention affects only the sign of
$A^{(1)}_{\rm C}$. The asymmetries of
$\phi^\prime_\gamma$ integrated cross sections, e.g.,
\begin{eqnarray}
\label{Def-MOMasy}
A_{\rm LU}
 \!\!\!&=&\!\!\! 
	\frac{2\int_{0}^{2\pi} d\phi^\prime_\gamma\; \ \sin(\phi^\prime_\gamma)\; 
\left(d\sigma^\uparrow - d\sigma^\downarrow\right)}{
\int_{0}^{2\pi} d\phi^\prime_\gamma\;  
\left(d\sigma^\uparrow + d\sigma^\downarrow\right)} \eqsim 
\frac{2\int_{0}^{2\pi} d\phi^\prime_\gamma\;  
\sin^2(\phi^\prime_\gamma)/{\cal P}_1 {\cal P}_2(\phi^\prime_\gamma)}{
\int_{0}^{2\pi} d\phi^\prime_\gamma\;  1/{\cal P}_1 {\cal P}_2(\phi^\prime_\gamma) } 
A^{(1)}_{\rm LU} \, , 
\end{eqnarray}
is {\em influenced} by the {\em angular dependence} of the {\em BH
propagators} and is generally smaller as the lowest harmonic of the
Fourier expansion (\ref{Def-AsyForExp}). Moreover, the dependence on the
kinematical variables can be {\em altered}. Note also that for $\Bx$,
$\Delta^2$, or $\cQ^2$ integrated asymmetries, the mean value for
different parts of the leptoproduction cross section  can be {\em quite
different}.

\subsection{Analytical approximation in the valence quark region 
\label{SubSec-EstAna}}

The lengthy expressions for the Fourier coefficients make it rather hard
to interpret future measurements in terms of GPDs. Thus, it would be
instructive to have an approximation at hand that serves as a guide for
adjusting model parameters by using efficient numerical codes. This can
be done for the observables and kinematics, we discussed above.
For instance, the beam spin asymmetry (\ref{Def-ALU}) can be simplified to
\begin{eqnarray}
A_{\rm LU}(\phi)
\sim
\pm \frac{\Ax}{y}
\frac{s_{1,{\rm unp}}^{\cal I}}{c_{0,{\rm unp}}^{\rm BH}} \sin(\phi)
\quad\mbox{with}
\quad \left\{ {+ \mbox{\ for\ } e^- \atop - \mbox{\ for\ } e^+} \right. ,
\end{eqnarray}
where we neglected possible contamination of the squared DVCS term.
Moreover, for $-\Delta_{\rm min} \ll -\Delta^2 \ll M^2_A$ and $\Bx= A\Ax
\lesssim 0.3 $ the Fourier coefficients can be drastically simplified
due to a rough approximation of kinematical factors, i.e., we expand the
Fourier coefficients $c_{0,{\rm unp}}^{{\rm BH}}$ and $s_{1,{\rm
unp}}^{{\cal I}}$ (see Appendix \ref{App-Apr}) to zero order in $\Ax$
and $\tau$. However, we have to pay special attention on terms that
contain $G_3$ and ${\cal H}_3$, since $G_3(\Delta=0)$ is enhanced by one
order of magnitude compared to the normalization of the other two form
factors. Thus, we take also $\tau G_3$ and $\tau {\cal H}_3$ into
account and obtain  for the beam spin asymmetry:
\begin{eqnarray}
\label{App-ALU} 
A_{\rm LU}(\phi) &\!\!\!\sim\!\!\! & \pm 
 \frac{\Ax (2-y)\sqrt{\frac{-\Delta^2 (1-y)}{\cQ^2 }}}{2-2y + y^2} 
\\
&& \times 
\im  \frac{ 2 G_1  {\cal H}_1 + (G_1-2 \tau G_3)
 ( {\cal H}_1  - 2 \tau{\cal H}_3) + \frac{2}{3} \tau G_3  {\cal H}_5}{2 G_1^2 +
(G_1-2 \tau G_3)^2} \sin(\phi)\, .
\nonumber
\end{eqnarray}
To demonstrate that these considerations are more on a rough
quantitative level, we mention that the kinematical prefactor $1 \le
(2-y)/(2-2y + y^2)\le (1 + \sqrt{2})/2$ should for consistency also be
neglected, which can induce a reduction of $21\%$. In the following we
take it into account. One realizes that the single beam spin asymmetry
is proportional to a linear combination of nucleus GPDs multiplied with
$\Ax= \Bx/A$. Since for large $A$ the form factors scale with $Z$ and
the CFFs with $A^2$, this asymmetry does not scale with $A$, however,
besides bound state effects it only depends on the ratio $A/Z$. It is
rather surprising that bound state effects enter already the
interference term for the unpolarized target not only due to a possible
modification of the scaling relation for ${\cal H}_1$, but also directly
by the appearance of ${\cal H}_3$ and ${\cal H}_5$.

In the case that the beam spin asymmetry is measured on an {\em
incomplete} unpolarized target, where only over the polarization states
with $\Lambda=\pm 1$ is summed, the prediction differs from the
asymmetry (\ref{App-ALU}):
\begin{eqnarray}
\label{Def-App-ALpm} 
A_{\rm L\pm}(\phi) &\!\!\!\sim\!\!\! & \pm 
\frac{\Ax}{y}
\frac{ s_{1,{\rm unp}}^{\cal I}-\frac{1}{3}s_{1,{\rm LLP}}^{\cal I}}
{c_{0,{\rm unp}}^{\rm BH}-\frac{1}{3}c_{0,{\rm LLP}}^{\rm BH}} \sin(\phi)
\quad\mbox{with}
\quad \left\{ {+ \mbox{\ for\ } e^- \atop - \mbox{\ for\ } e^+} \right.  
\nonumber
\end{eqnarray}
 and reads  in our kinematical approximation 
\begin{eqnarray}
\label{App-ALpm} 
A_{\rm L\pm}(\phi) &\!\!\!\sim\!\!\! & \pm 
 \frac{\Ax (2-y)\sqrt{\frac{-\Delta^2 (1-y)}{\cQ^2 }}}{2-2y + y^2} 
\\
&& \times 
\im   \frac{ 2 G_1  ({\cal H}_1- \frac{1}{3}  {\cal H}_5) + 2(G_1-2 \tau G_3)
 ( {\cal H}_1  - 2 \tau{\cal H}_3 - \frac{1}{3}  {\cal H}_5) }{2 G_1^2 +
2 (G_1-2 \tau G_3)^2} \sin(\phi)\, .
\nonumber
\end{eqnarray}

The approximation of the unpolarized charge asymmetry (\ref{Def-AC})
\begin{eqnarray}
A_{\rm C}(\phi)
\sim \frac{\Ax}{y}
\frac{c_{0,{\rm unp}}^{\cal I} + c_{1,{\rm unp}}^{\cal I}\cos(\phi)}
{c_{0,{\rm unp}}^{\rm BH}} + \cdots 
\end{eqnarray}
is obtained in the analogous manner. The dominant twist-two harmonic is
given as the real part of the same linear combination of GPDs as in the
beam spin asymmetry (\ref{App-ALU}):
\begin{eqnarray}
\label{App-AC} A_{\rm C}(\phi) &\!\!\!\sim\!\!\! &
\frac{\Ax \sqrt{\frac{-\Delta^2}{\cQ^2}(1-y)}}{y}
\\
&&\times 
\re  \frac{2 G_1  {\cal H}_1 + (G_1-2 \tau G_3)
( {\cal H}_1  - 2 \tau{\cal H}_3)+ \frac{2}{3} \tau G_3  {\cal H}_5}{
2 G_1^2 + (G_1-2 \tau G_3)^2  } \cos(\phi) + \cdots\, .
\nonumber
\end{eqnarray}
Also this asymmetry is nearly independent on $A$. As noted before, the
constant term only depends on twist-two GPDs and does not vanish at the
kinematical boundaries. In our approximation it is 
\begin{eqnarray}
\label{PreCAc02c1}
\frac{c_{0,{\rm unp}}^{\cal I}}{c_{1,{\rm unp}}^{\cal I}} \simeq 
\frac{2-y}{\sqrt{1-y}} 
\sqrt{\frac{-\Delta^2}{\cQ^2}} \, .
\end{eqnarray} 
For a scalar target the exact relation does not depend on GPDs
\cite{BelMueKirSch00}, while for a spin-1/2 target only a $\Ax$
suppressed dependence appears \cite{BelKirMueSch01,BelMueKir01}. For a
spin-1 target the analogous relation is unknown at present.

The single spin asymmetry (\ref{Def-AUL}) for longitudinally polarized
target, 
\begin{eqnarray}
A_{\rm UL}(\phi) \simeq \pm \frac{\Ax}{y} \frac{s_{1,{\rm LP}}^{\cal
I}}{c_{0,{\rm unp}}^{\rm BH}} \sin(\phi) + \cdots \quad\mbox{with}
\quad \left\{ {+ \mbox{\ for\ } e^- \atop - \mbox{\ for\ } e^+}
\right. , 
\end{eqnarray}
is mainly governed by $\widetilde {\cal H}_1$:
\begin{eqnarray}
\label{App-UL} 
A_{\rm UL}(\phi) \sim \pm
\frac{\Ax \sqrt{\frac{-\Delta^2}{\cQ^2}(1-y)}}{y}
\im  \frac{3 (G_1-\tau G_3) \widetilde {\cal H}_1 + \frac{3 \Ax}{2}  G_2
({\cal H}_1-\tau {\cal H}_3 - {\cal H}_5)}{2 G_1^2 + (G_1-2 \tau G_3)^2}
\sin(\phi).
\end{eqnarray}
Since  the polarized quark contribution is relatively
small compared to the unpolarized ones and the latter increase
faster with growing $1/\Bx$ (in the small $\Bx$ region), we included
also a $\Ax$ suppressed term that is proportional to ${\cal H}_1-\tau
{\cal H}_3- {\cal H}_5 $.

For a spin-1 target another quite interesting observable is the tensor
polarization asymmetry (\ref{Def-Azz}). Their measurement requires a
longitudinally polarized target. For an unpolarized beam this asymmetry
is dominated by the squared BH contribution, which provides a dominant
constant term that is proportional to $\tau G_3$:
\begin{eqnarray}
\label{ResAna-Azz}
 A_{zz}(\phi)\!\!\! &=& \!\!\! \frac{-4\tau G_3 \left(G_1 - \tau G_3\right)}
  {2 G_1^2 + \left(G_1 - 2 \tau G_3 \right)^2} \pm
\frac{\Ax \sqrt{\frac{-\Delta^2}{\cQ^2}(1-y)}}{y} 
\\
&&\hspace{2cm}\times 3 G_1\left[
\cos(\phi) \times \dots +  
\lambda \frac{y (2-y)}{2-2y+y^2} \sin(\phi) \times \dots \right] + \cdots\, .
\nonumber
\end{eqnarray}
Here the ellipsises, proportional to the first harmonics, arise from a
Taylor expansion with respect to $-\Delta^2/\cQ^2$ and stand
for the real and imaginary part of a certain linear combination of CFFs:
\begin{eqnarray} 
\label{Ana-Azz-LinCom-CFF}
{\cal H}_1 + \frac{2}{3}{\cal H}_5   - 
    \frac{G_1}{ 2G_1^2 + \left(G_1 - 2\tau G_3  \right)^2}
    \left(\!\! 2 G_1 {\cal H}_1  + 
         \left(G_1 - 2\tau G_3  \right) \left({\cal H}_1  -
2 \tau {\cal H}_3\right)  + 
         \frac{2 \tau}{3}  G_3  {\cal H}_5\!\! \right)\! .
\end{eqnarray}
The third term comes from the denominator, while the $\cos(\phi)$ term
that stems from the BH Fourier coefficients $c_{1,{\rm unp}}^{\rm BH}$
and $c_{1,{\rm LLP}}^{\rm BH}$ do not contribute in our approximation.
Consequently, to extract the interference term the squared BH
contribution has to be subtracted. Alternatively, we can form the tensor
asymmetry from the charge odd part of the cross section
\begin{eqnarray}
A_{{\rm C}zz}(\phi) = 
\frac{d\sigma^{e^+}_{zz}- d\sigma_{zz}^{e^-}}
{3 d\sigma^{e^+}_{\rm unp} + 3 d\sigma^{e^-}_{\rm unp}} 
\simeq - 
\frac{\Ax}{y} \frac{c_{1,{\rm unp}}^{\cal I}-c_{1,{\rm LLP}}^{\cal
I}}{c_{0,{\rm unp}}^{\rm BH}} \cos(\phi) + \cdots\, , 
\end{eqnarray}
where the ellipsis include a constant term and higher harmonics.
Another possibility is to use a polarized beam and to form the asymmetry
\begin{eqnarray}
\label{Def-ALzz}
A_{{\rm L}zz}(\phi) &\!\!\! =\!\!\! &
 \frac{d\sigma_{zz}(\lambda=+1) - d\sigma_{zz}(\lambda=-1)}
{3 d\sigma_{\rm unp}(\lambda=+1) + 3d\sigma_{\rm unp}(\lambda=-1)} 
\\
&\!\!\! \simeq \!\!\! &
\pm \frac{\Ax}{y} \frac{s_{1,{\rm unp}}^{\cal I}-s_{1,{\rm LLP}}^{\cal
I}}{c_{0,{\rm unp}}^{\rm BH}} \sin(\phi) + \cdots
\quad\mbox{with}
\quad \left\{ {+ \mbox{\ for\ } e^- \atop - \mbox{\ for\ } e^+}
\right. ,
\nonumber
\end{eqnarray}
which is dominated by the $\sin(\phi)$ harmonics of the interference
term only. These two asymmetries allow  to explore the real and imaginary
part of the following linear combination of GPDs:
\begin{eqnarray}
\label{Res-ALzzACzz}
\left\{ {  A_{{\rm L}zz} \atop A_{{\rm C}zz} } \right\}(\phi)
&\!\!\! \sim\!\!\! &
\sqrt{ \frac{-\Delta^2}{\cQ^2}(1-y)}\, 2\Ax
\left\{ {\mp 1 \atop -1/y } \right\}\left\{ {\sin(\phi)\atop \cos(\phi)} \right\}
\\
&& \times
 \left\{ {\im \atop \re} \right\}
\frac{G_1\left(\tau {\cal H}_3 + {\cal H}_5\right) + \tau G_3\left( 
{\cal H}_1-2\tau {\cal H}_3 -\frac{1}{3}{\cal H}_5\right) }{2 G_1^2 +
 (G_1-2 \tau G_3)^2} 
\nonumber
\end{eqnarray}

It is now instructive to compare these results with those for a spin-1/2
or -0 target and to explore its dependence on GPDs, which are of course
model dependent. The analogous predictions simply follow by setting
$G_3$, ${\cal H}_3$, and ${\cal H}_5$ to zero and by the replacements:
\begin{eqnarray}
 G_1 \to \left\{{  F_1 \atop  F}\right\} \, , \  
G_2 \to \left\{{ F_1+F_2 \atop 0 }\right\}\, , \ 
{\cal H}_1 \to \left\{{ {\cal H} \atop {\cal H} }\right\}\, ,
\widetilde{\cal H}_1 \to \left\{{ \widetilde{\cal H}  \atop 0}\right\}\, 
\quad
\mbox{for spin-} \left\{{ 1/2 \atop 0 }\right.\, . 
\nonumber
\end{eqnarray}
First we like to remind that the sum rules, e.g., Eq.\
(\ref{Def-SumRulV1}), suggest that the $\Delta^2-$dependence of the
valence--like  GPDs is  given by $G_i$ and $F_1$ (or the
scalar form factor $F$), respectively. Furthermore, the analyses of the
H1, HERMES and CLAS DVCS data for a proton target in terms of the
oversimplified model, given in Section \ref{Sec-ModGPD}, at LO indicate
that the unpolarized sea quark contribution is compared to the forward
case additionally suppressed\footnote{We remind that $\xi \approx
\Bx/(2-\Bx)$ and already for $\xi\sim 0.05$ one finds in deep inelastic
scattering an almost equal contribution of valence and sea quarks.}. We
suppose that this qualitative property holds also true for the nucleus
GPDs. Therefore, we neglect in the valence region the sea quark
contribution and model the remaining GPDs as a product of form factor
and valence quark distributions\footnote{Here we neglect any skewedness
effect. For the DD model, introduced in Section
\ref{Sec-ModGPD}, skewedness effects would provide an enhancement of the
CFFs. Strictly spoken, it is not known what caused the observed
suppression of sea quark contributions and thus the simple ansatz for
the DD's together with a rather large slope parameter might be
questionable. }. In LO of perturbation theory we find for nuclei targets
\begin{eqnarray}
\label{App-CFFH1}
\frac{\Ax \im {\cal H}^A(\xi,\Delta^2)}{F^A(\Delta^2)} &\sim &
  \pi \Ax \left\{ Q_u^2  q^{u_{\rm val}/{\rm A}}(\xi)+
Q_d^2  q^{d_{\rm val}/{\rm A}}(\xi) \right\}
\\
&\sim & \pi   
\frac{Z Q_u^2 + N Q_d^2}{Z}\;  \Bx \left\{  
q^{u_{\rm val}}\!\left(\frac{\Bx}{2-\Bx}\right)+ 
q^{d_{\rm val}}\!\left(\frac{\Bx}{2-\Bx}\right)
 \right\}
\, ,
\nonumber
\end{eqnarray}
expressed in terms of the quark densities of the proton. The
analogous formula holds true for the spin-1/2 case and also for the
${\cal H}_1$ contribution of spin-1 targets.

The beam spin asymmetry for the positron scattering off a proton target is
\begin{eqnarray}
{A_{\rm LU}(\phi)} \sim -\sqrt{\frac{-\Delta^2 (1-y)}{\cQ^2 }} 
\pi \Bx \left\{
Q_u^2 q^{u_{\rm val}}(\xi_N) +Q_d^2 q^{d_{\rm val}}(\xi_N)
\right\} \sin(\phi)\Bigg|_{\xi_N=\frac{\Bx}{2-\Bx}} \, .
\end{eqnarray}
The ratio of beam spin asymmetries for a spin-0 or -1/2 nucleus with
$Z\simeq N$ to the proton is 
\begin{eqnarray}
\label{ALU2ALU0}
\frac{A_{\rm LU}^{\rm A}(\phi)}{A_{\rm LU}(\phi)}
\sim 
\frac{\left(Q_u^2 +Q_d^2\right) \left\{  
q^{u_{\rm val}}(\xi)+ q^{d_{\rm val}}(\xi_N)\right\}}
{Q_u^2 q^{u_{\rm val}}(\xi_N) +Q_d^2 q^{d_{\rm val}}(\xi_N)} 
\Bigg|_{\xi_N=\frac{\Bx}{2-\Bx}}\, 
\quad \mbox{for} \quad J= \{0, 1/2\}\, .
\end{eqnarray}
For a spin-1 nucleus we would have the same prediction, if ${\cal
H}_3^{\rm val} = G_3/G_1 {\cal H}_1^{\rm val}$ holds true. However, if
${\cal H}_3^{\rm val}$ is negligible in this kinematics, it follows from
Eq.\ (\ref{App-ALU}) an additional $\Delta^2$-dependence, which is
crucial:
\begin{eqnarray}
\label{rat-ALU-J1toJ2}
\frac{A_{\rm LU}^{\rm A}(\phi)}{A_{\rm LU}(\phi)}
&\!\!\!\sim\!\!\! & 
 \frac{ 3 -2 \tau G_3/G_1}{2  + (1-2 \tau G_3/G_1)^2} 
 \frac{ \left(Q_u^2 +Q_d^2\right) \left\{  
q^{u_{\rm val}}(\xi_N)+ q^{d_{\rm val}}(\xi_N)\right\}}
{Q_u^2 q^{u_{\rm val}}(\xi_N) +Q_d^2 q^{d_{\rm val}}(\xi_N)}
 \Bigg|_{\xi_N=\frac{\Bx}{2-\Bx}}\, 
\mbox{for} \quad J= 1\, ,
\nonumber\\
\frac{A_{{\rm L}\pm}^{\rm A}(\phi)}{A_{\rm LU}(\phi)}
&\!\!\!\sim\!\!\! & 
 \frac{ 3 -3 \tau G_3/G_1}{2  + 2(1- \tau G_3/G_1)^2} 
 \frac{ \left(Q_u^2 +Q_d^2\right) \left\{  
q^{u_{\rm val}}(\xi_N)+ q^{d_{\rm val}}(\xi_N)\right\}}
{Q_u^2 q^{u_{\rm val}}(\xi_N) +Q_d^2 q^{d_{\rm val}}(\xi)} 
\Bigg|_{\xi_N=\frac{\Bx}{2-\Bx}}
\, .
\end{eqnarray}
Taking $\Bx\sim 0.1$, we expect from the parameterization of parton
densities that  $d_{\rm val}/u_{\rm val} \sim
0.5$. For $\Delta^2=-0.2\ \GeV^2$, $\cQ^2= 2.5\ \GeV^2$, and $E=27.5\
\GeV$ we estimate the beam spin asymmetry for a proton target and the
ratios (\ref{ALU2ALU0}) and (\ref{rat-ALU-J1toJ2})  for  nuclei ($Z\simeq N$) to be:
\begin{eqnarray}
A_{{\rm LU}}(\phi) \sim - 0.29 \sin(\phi)\, , \qquad 
\frac{A_{\rm LU}^{\rm A}(\phi)}{A_{\rm LU}(\phi)}
\sim \left\{{ 5/3 \atop 1}\right\} \ \mbox{for}\   
J= \left\{{\{0,1/2\} \atop 1}\right\}\, .
\end{eqnarray}
Our estimate for the $\sin(\phi)$--weighted asymmetry (\ref{Def-MOMasy})
is $A_{{\rm LU}} \sim - 0.23$, while the ratios of beam spin asymmetries
for different targets are unchanged. So we roughly expect an enhancement
of the beam spin asymmetry for nuclei with spin-0 and -1/2, which
simply arises from the ratio of squared charges for an isoscalar to an
isodoublet state and the $u$ quark dominance. For a spin-1 target, one
would expect the same. However, in the case that ${\cal H}_3$ (and also
${\cal H}_5$) does not contribute the factor $(3 -2 \tau G_3/G_1)/(2 +
(1-2 \tau G_3/G_1)^2)$ gives a $\Delta^2$ dependent suppression.
It decreases with growing $|\Delta^2| < 0.5\ \GeV^2$, reaches its
minimum at $|\Delta^2| \approx 0.5\ \GeV^2$ with $\approx -0.1$ and then
increases with growing $|\Delta^2|$. Note the sign change of the beam
spin asymmetry, which occurs in the region $0.4\ \GeV^2 < -\Delta^2 <
0.7\ \GeV^2$.

In the following we confront our oversimplified estimates for the
$\phi$--integrated beam spin asymmetry (\ref{Def-MOMasy}) with
preliminary HERMES data from the 2000 run, which have been taken for
proton, deuteron, and neon target \cite{EllShaVol02}:
\begin{eqnarray}
\label{Exp-HERMES-2000}
A_{\rm LU} \!\!\! &=&\!\!\! -0.18 \pm 0.03  \pm 0.03 , 
 \langle \Bx \rangle = 0.12, \langle -\Delta^2\rangle =0.18\ \GeV^2,  
\langle \cQ^2 \rangle = 2.5\ \GeV^2\, ,
\nonumber\\
A^{\rm Ne}_{\rm LU} \!\!\! &=&\!\!\! -0.22 \pm 0.03  \pm 0.03 , 
\langle \Bx \rangle = 0.09,  \langle -\Delta^2\rangle =  0.13\ \GeV^2, 
\langle \cQ^2 \rangle = 2.2\ \GeV^2\, ,
\nonumber\\
A^d_{{\rm L}\pm {\rm U}} \!\!\! &=&\!\!\! -0.15 \pm 0.03  \pm 0.03, 
 \langle \Bx \rangle = 0.1,  \langle -\Delta^2\rangle = 0.2\ \GeV^2, 
\langle \cQ^2 \rangle = 2.5\ \GeV^2\, ,
\end{eqnarray} 
where the first (second) error denotes the statistical (systematical)
uncertainty. Here the notation $A^d_{{\rm L}\pm {\rm U}}$ refers to the fact that 
this value is obtained from two different data sets of polarized and unpolarized 
deuteron target. Our rather naive estimates give
\begin{eqnarray}
\label{EstAna-ALU}
A_{\rm LU} = -0.26\, , \quad A^{\rm Ne}_{\rm LU} = - 0.34\, ,\quad
A^d_{\rm LU} = - 0.23\, ,\  A^d_{{\rm L}\pm} = - 0.21\, ,
\end{eqnarray}
which are slightly higher than the experimental values. Note that these
numbers include uncertainties induced by the kinematical approximation
and, of course, due to the simplification of GPDs. Compared to the beam
spin asymmetry on a proton target from the 1996/97 run
\begin{eqnarray}
A_{\rm LU} = -0.23 \pm 0.04 \pm 0.03 , 
\langle \Bx \rangle = 0.11, 
\langle -\Delta^2\rangle = 0.27\ \GeV^2, 
\langle \cQ^2 \rangle = 2.6\ \GeV^2\, ,
\end{eqnarray}
our naive analytical and numerical estimates, within a certain model
that contains also sea quarks and twist-three corrections, give in both
cases nearly the same value $A_{\rm LU} = -0.26$ and $A_{\rm LU} =
-0.27$, respectively. The sea quark content gives an enhancement and the
exact kinematics at twist-tree level a dumping, which results finally in
a similar prediction. Nevertheless, these numbers are rather model
dependent and in fact by varying the slope $B_{\rm sea}$,
$b$-parameters, and adding dynamical twist-tree contributions by `hand'
our prediction covered the range $0.16 \le |A_{\rm LU}| \le 0.37$
\cite{BelMueKir01}.

For JLAB kinematics with $\Bx=0.3$ [$0.2$], $E=6\ [12]\ \GeV$,
$\Delta^2= -0.25\ \GeV^2$, and $\cQ^2= 2.5\ \GeV^2$ the quantitative
estimates for proton, deuteron, and a isoscalar nuclei $A$ target are
\begin{eqnarray}
A_{\rm LU} \sim 0.25\,  [0.3]\, ,\qquad A^d_{\rm LU} \sim 0.2\,  [0.25]
,\qquad A^{\rm A}_{\rm LU} \sim 0.35\,  [0.45]\, .
\end{eqnarray}

For the charge asymmetry, there are important differences with respect
to the beam spin asymmetry. First we recall the non negligible constant
term. Moreover, this asymmetry arises from the real part of the Compton
amplitude and so it is also sensitive to the $D$--term, which drops out
in the imaginary part\footnote{A first discussion of the imaginary and
real part of the Compton amplitude in the context of field theory can be
found for instance in Ref. \cite{CheTun70,BroCloGun72}. The constant
term in the real part, which appears here due to the $D$--term, has been
related to a fixed pole in the language of Reggeization.}. At smaller
values of $\Bx$ the ratio of real to imaginary part is 
\begin{eqnarray}
\label{Rat-Re2Im}
{\cal R} =\frac{\re\ {\cal F}}{\im\ {\cal F}} \simeq 
\tan\left([2\alpha^i-1-{\cal S}_{\cal F}]\frac{\pi}{4}\right) 
+ \mbox{pure `exclusive' contributions}/\im\ {\cal F}
\, ,
\end{eqnarray}
where the signature ${\cal S}_{\cal F} = 1$ for the CFFs $\{ {\cal H},\ {\cal
E},\ {\cal H}_1,\dots,{\cal H}_5$\} and ${\cal S}_{\cal F} = -1$ for the set
$\{ \widetilde{\cal H},\ \widetilde{\cal E},\ \widetilde{\cal
H}_1,\dots,\widetilde{\cal H}_4\}$. Here we used that the GPDs behave
like $H(\xi) \propto \xi^{-\alpha^i}$ with $\alpha^i >0$. Since
$\alpha^i\sim 1/2$ and $\alpha^i > 1$ for unpolarized valence and sea
quarks, respectively, we expect also for larger values of $\Bx$ that the
valence and sea quarks give a negative and positive correction,
respectively:
\begin{eqnarray}
R^i=\frac{\re\ {\cal H}^i}{\im\ {\cal H}^i}= 
\left\{ { < 0 \atop  > 0}  \right\}\quad\mbox{for}\quad 
\left\{ { \mbox{valence quarks} \atop  \mbox{sea quarks}} \right. \, .
\end{eqnarray}
The $D$-term contribution has been estimated to be negative, too, and will
partially cancel the positive sea quark contribution. All these terms enter 
the Fourier coefficient $A^{(1)}_{\rm C}$, defined in Eq.\ (\ref{Def-AsyForExp})  
with a reversed sign.  The  HERMES data
\begin{eqnarray}
\label{ChaAsy-HERMES}
A_{\rm C}(\phi) = - 0.05 \pm 0.03 (\rm{stat}) + 
\left[0.11 \pm 0.04 (\rm{stat}) \right]\cos(\phi_\gamma^\prime)\, , 
\end{eqnarray}
for $\langle \Bx \rangle = 0.12,\ \langle -\Delta^2\rangle = 0.27\
\GeV^2,\ \langle \cQ^2 \rangle = 2.7\ \GeV^2$, can be explained by 
two different scenarios:
\begin{enumerate}
\item small sea quark contribution and rather small or no $D$-term 
\item large sea quark contribution and large negative $D$-term.
\end{enumerate}
The second scenario predicts a beam spin asymmetry of $A_{\rm LU} \simeq
-0.37$, which is slightly larger as the beam spin asymmetry of the
1996/97 run. So a definite conclusion about the $D$-term can not be done
with present data. However, we state: if high precision data will be
available, the ratio of charge asymmetry to beam spin asymmetry allows to
pin down the $D$-term contribution. The measured ratio 
\begin{eqnarray}
\frac{A^{(0)}_{\rm C}}{A^{(1)}_{\rm C}} = - 
\frac{c_{0,{\rm unp}}^{\cal I}}{c_{1,{\rm unp}}^{\cal I}} 
-\sim 0.5
\end{eqnarray}
coincides with our estimate (\ref{PreCAc02c1}), which yields the value
$\approx 0.65$.

In both scenarios one expects that the valence quark contribution
dominates in the real part of the CFF $\cal H$ for HERMES kinematics.
The real part of this dominant CFF for a proton and a scalar nuclei
target are 
\begin{eqnarray}
\label{App-reCFFH1pr}
\frac{\Bx \re {\cal H}_1(\xi,\Delta^2)}{F_1(\Delta^2)} &\!\!\!\sim\!\!\! & 
\pi    \Bx \left\{  
Q_u^2 R^{u_{\rm val}}  q^{u_{\rm val}}+ 
Q_d^2  R^{d_{\rm val}} q^{u_{\rm val}}
\sum_i Q_i^2 \overline{q}^{i} 
 \right\}
\, ,
\\
\label{App-CFFH1re}
\frac{\Ax \re {\cal H}^{\rm A}(\xi,\Delta^2)}{F^{\rm A}(\Delta^2)} &\!\!\!\sim\!\!\! & 
\pi \Bx\left\{  
\frac{Z Q_u^2  + N Q_d^2 }{Z}\;     
\left[R^{u_{\rm val}}  q^{u_{\rm val}}+ R^{d_{\rm val}} q^{u_{\rm val}}\right]
 \right\}
\, ,
\nonumber
\end{eqnarray}
where $R^{i}$ and $q^{i}$ depend on $\xi_N =\Bx/(2-\Bx)$,
$\Delta^2$, and $\cQ^2$. As above in the case of the imaginary part a
similar estimate holds true for a spin-1/2 nucleus target. For larger
value of $\Bx$, e.g.\ $\Bx= 0.12$ the asymptotic formula
(\ref{Rat-Re2Im}) overestimates the size of the real part. In the MRS A'
parameterization we find $R^{u_{\rm val}}(\xi_N=0.065) \sim - 0.4$ and
$R^{d_{\rm val}}(\xi_N=0.065) \sim - 0.25$. So we expect, e.g., $\cQ^2 =
2.7\ \GeV^2$ the following ratio of charge and single beam
asymmetries\footnote{This ratio has the advantage that the BH
contribution and partly the kinematical prefactors of the interference
term drop out and so it is after restoration of the $y$ dependence exact
up to the neglected squared DVCS term and the approximation of the CFFs.
$\widetilde H$ and $E$ CFFs, which are kinematically suppressed, 
are still neglected.}
on a proton target 
\begin{eqnarray}
\frac{A^{(1)}_{\rm C}}{A^{(1)}_{\rm LU}} \sim  
\frac{2-2 y+y^2}{(2-y)y}\frac{
Q_u^2 R^{u_{\rm val}}  q^{u_{\rm val}}+ 
Q_d^2 R^{d_{\rm val}}   q^{d_{\rm val}}
}{
Q_u^2 q^{u_{\rm val}}+ Q_d^2 q^{d_{\rm val}} 
} 
\sim -0.8\, ,
\end{eqnarray} 
which is consistent with HERMES data. For an isoscalar nucleus we expect a similar 
value:
\begin{eqnarray}
\label{Rat-AC2ALU-Sin}
\frac{A^{{\rm A}(1)}_{\rm C}}{A^{{\rm A}(1)}_{\rm LU}} \sim  \frac{2-2y+y^2}{(2-y)y}
\frac{
R^{u_{\rm val}} q^{u_{\rm val}}+ R^{d_{\rm val}} q^{d_{\rm val}}
\sum_i Q_i^2 \overline{q}^{i} 
}{
q^{u_{\rm val}}+ q^{d_{\rm val}} 
} 
\sim -0.7
\, .
\end{eqnarray} 
However, these ratios should be smaller in reality, since the sea quarks
enter the numerator and denominator with negative and positive sign,
respectively.

The ratio of charge to beam spin asymmetry for a spin-1 target has a
rather complex model dependence. Even if we set ${\cal H}_3^{\rm val}$
and ${\cal H}_5$ to zero, we should worry about the unknown `quadrupole'
contribution ${\cal H}_3$, induced by the sea quarks:
\begin{eqnarray}
\frac{A^{{\rm A}(1)}_{\rm C}}{A^{{\rm A}(1)}_{\rm LU}} \sim  \frac{2-2y+y^2}{(2-y)y}
\frac{
R^{u_{\rm val}} q^{u_{\rm val}}+ R^{d_{\rm val}} q^{d_{\rm val}}
+ 2\frac{G_1^{\rm sea}}{G_1} \left[1
- 
\frac{2\tau G_3^{\rm sea}(G_1 -2 \tau G_3)}{G_1^{\rm sea} (3G_1 -2 \tau G_3) }
\right]R^{\rm sea} \frac{2 \sum_i Q_i^2 \overline{q}^{i}}{\sum_{i=u,d} Q_i^2}
}{ 
q^{u_{\rm val}}+ q^{d_{\rm val}} 
+  2\frac{G_1^{\rm sea}}{G_1} \left[1 
- 
\frac{2\tau G_3^{\rm sea}  (G_1 -2 \tau G_3)}{
G_1^{\rm sea} (3G_1 -2 \tau G_3) }
\right] \frac{2 \sum_{i} Q_i^2 \overline{q}^{i}}{\sum_{i=u,d} Q_i^2}
 } 
\, .
\end{eqnarray} 
Here we have set for simplicity ${\cal H}_3^{\rm sea}/{\cal H}_1^{\rm sea} =
G_3^{\rm sea} /G_1^{\rm sea} = 1- \mu_d^{\rm sea}+ {\cal Q}_d^{\rm sea}$
and assumed the same $R^{\rm sea}$ for all sea quark species. If the
sea quark contributions are small, we would have the same ratio
(\ref{Rat-AC2ALU-Sin}) as for isoscalar nucleus with spin-0 or -1/2.

It is interesting that for larger values of $\Bx$ the sign of
the real part of the CFFs might be changed. In our model that happens
for valence $u$[$d$] quarks at $\Bx\sim 0.3 [0.2]$, while the sign of
sea quarks and $D$-term remains the same. There is also a
significant difference between these contributions. While the real
part of the sea quarks is expected to drop rather fast with increasing
$\Bx$, the $D$-term gives a constant contribution. JLAB kinematics would
be suitable to explore this region and provide so more insight in the
structure of GPDs, unfortunately, no positron beam is available.

Now we come to the longitudinally polarized target spin asymmetry. For a
proton target the ratio of this target spin to the beam spin asymmetry
is 
\begin{eqnarray}
\frac{A_{\rm UL}(\phi)}{{\rm A}_{\rm LU}(\phi)} \sim  \frac{2-2y+y^2}{(2-y)y} \left[
\frac{
Q_u^2  \Delta q^{u_{\rm val}} + 
Q_d^2   \Delta  q^{d_{\rm val}}
}{ 
Q_u^2 q^{u_{\rm val}}+ Q_d^2 q^{d_{\rm val}} 
} 
 + \frac{\Bx}{2} \frac{F_2}{F_1}\right]\ \, .
\end{eqnarray}
Both terms in the squared bracket on the r.h.s.\ are positive and become
smaller\footnote{The ratio of form factors is for small value of
$|\Delta^2|$ given by the anomalous magnetic moment $F_2/F_1 \approx
1.793 (1 -\Delta^2/1.26) $, while the ratio of polarized to unpolarized
quarks is estimated from the measured ratio of polarized to unpolarized
structure functions in deep inelastic scattering which behave like
$\sqrt{\Bx}$ for increasing $\Bx$, e.g., $g_1/F_1\sim 0.2$ for $\Bx\sim
0.1$ .} with decreasing $\Bx$. So within our simplifications we expect
the same sign as for the beam spin asymmetry and a rather small value
for not so large $\Bx$ or small $y$. In the case of a isoscalar nuclei
spin-1/2 target 
\begin{eqnarray}
\frac{A^{\rm A}_{\rm UL}(\phi)}{A^{\rm A}_{\rm LU}(\phi)}  \sim  \frac{2-2y+y^2}{(2-y)y} 
\left[
\frac{
 \Delta q^{u_{\rm val}}+ \Delta  q^{d_{\rm val}}
}{ 
 q^{u_{\rm val}} +  q^{d_{\rm val}} 
}
 + \frac{\Bx}{2 A} \frac{F^{\rm A}_2}{F^{\rm A}_1}
\right] \, , 
\end{eqnarray}
the form factor dependent part will be suppressed by the atomic mass
number $A$. As long as this term is not enhanced due to a node in
$F^{\rm A}_1(\Delta^2_0)$ at given $\Delta^2_0$, the ratio of polarized to
unpolarized quark distribution dominates and we expect for $\Bx\approx
0.2$, $\cQ^2 \approx 2\ \GeV^2$ and HERMES [JLAB@12\ \GeV] kinematics
\begin{eqnarray}
\frac{A^{\rm A}_{\rm UL}(\phi)}{A^{\rm A}_{\rm LU}(\phi)}  \sim  \frac{2-2y+y^2}{(2-y)y} 
\frac{
 \Delta q^{u_{\rm val}} +   \Delta  q^{d_{\rm val}}
}{ 
 q^{u_{\rm val}} +  q^{d_{\rm val}} 
}  
\sim  0.5\,  [0.2]\, .
\end{eqnarray}
If we neglect ${\cal H}_3$ and ${\cal H}_5$ CFFs, we find for an
isoscalar spin-1 target a similar prediction: 
\begin{eqnarray}
\label{Res-AUL2ALU-Deu}
\frac{A^{\rm A}_{\rm UL}(\phi)}{A^{\rm A}_{\rm LU}(\phi)} \sim \frac{2-2y+y^2}{(2-y)y}  
\frac{G_1-\tau G_3}{G_1-2 \tau G_3/3}  \left[
\frac{
 \Delta q^{u_{\rm val}}+   \Delta  q^{d_{\rm val}}
}{ 
 q^{u_{\rm val}}+  q^{d_{\rm val}} 
}  + \frac{\Bx}{2 A} \frac{G_2}{G_1-\tau G_3}  \right]\, ,
\end{eqnarray} 
which differs from the previous cases by a $\Delta^2$ dependent
prefactor. Unfortunately, this ratio is also rather sensitive to the
details of $H_3$ and $H_5$ GPDs, compare Eqs.\ (\ref{App-ALU}) and
(\ref{App-UL}), and so no definite estimate can be given.

Finally, we discuss the tensor polarization asymmetries. In the case
that ${\cal H}_3$ and ${\cal H}_5$, are small we find that these
asymmetries can be expressed by the charge and beam spin asymmetries:
\begin{eqnarray}
\label{Res-ALzz2ALU}
\frac{A^{\rm A}_{{\rm L}zz}(\phi)}{A^{\rm A}_{{\rm LU}}(\phi)} = 
\frac{A^{{\rm A}(1)}_{{\rm C}zz}}{A^{{\rm A}(1)}_{{\rm C}}}
\sim  \frac{2}{3} 
\frac{(-\tau+4\xi \tau - 3 \xi^2)  G_3}{G_1- \frac{2}{3} \tau  G_3}\, .
\end{eqnarray}  
For small values of $-\Delta^2$ and $\xi$ these ratios are rather small,
e.g., $-\Delta^2= 0.2 [0.1]\ \GeV^2$ and $\Bx=0.1$ we find the number
$\sim 0.2 [0.08]$. A significant deviation from this prediction would
indicate that ${\cal H}_3$ and/or ${\cal H}_5$ are comparable in size to
${\cal H}_1$. For instance, if ${\cal H}_5$ would be small and assuming
that ${\cal H}_3 = G_3/G_1{\cal H}_1 $ this ratio is roughly double so
large.

\subsection{Numerical results
\label{SubSec-EstNum}}

In the previous section we tried to convince the reader that a rather
simple GPD ansatz explains the measured DVCS asymmetries. To avoid a
misunderstanding, we do not claim that our estimates are precise. Rather
we expect that more realistic GPD models yield large deviations from
these naive estimates. Nevertheless we like to give an impression of
this model dependence in the following. We also have to keep in mind
that the kinematical approximations are rather rough. In the following
we correct this and take the exact expressions for the BH cross section
and $K$ factor, defined in Eq.\ (\ref{Def-K}) and the full $\xi$ and
$\Delta^2$ dependences in the twist-two predictions of interference and
squared DVCS term.

\begin{table}[t]
\vspace{-0.5cm}
\begin{center}
\begin{tabular}{|c||c|c||c|c||c||c|c|}
\hline
\multicolumn{3}{|c||}{all targets} & \multicolumn{2}{c||}{proton} & neon & deuteron  
\\
\hline\hline
model  &  $b_{\rm val}$  & $b_{\rm sea}$ 
&  $B_{\rm sea}$ $[{\rm GeV}^{-2}]$  & $\kappa_{\rm sea}$
&  $B_{\rm sea}$ $[{\rm GeV}^{-2}]$  
&  $B_{\rm sea}$ $[{\rm GeV}^{-2}]$   \\
\hline\hline
A & 1 & $\infty$ & 9 & 0 & 9 & 20  
\\
\hline
B & $\infty$ & $\infty$ & 9 & -3 & 12 & 30 
\\
\hline
C & 1 & 1  & 5 & 0 & 6 & 15 
\\
\hline
\end{tabular}
\end{center}
\caption{Parameter sets for $H$ and $H_1$ GPDs of a spin-0 nucleus, proton, and
deuteron.}
\label{Tab-ModH1}
\end{table}

First we consider the beam spin asymmetry measured at HERMES. For the
proton target we rely on the GPD models A, B, and C, employed in Ref.\
\cite{BelMueKir01}. The parameters of these three models are listed in
Tab.\ \ref{Tab-ModH1}. Compared to model A, the sea quarks are
suppressed and enhanced in model B and C, respectively, by a larger
slope $B_{\rm sea}$ and $b_{\rm sea}\to \infty$. We also suppress in
model B the valence quarks by taking $b_{\rm val}\to \infty.$ At
twist-three level the model A and C are based on the WW approximation,
while in the model B quark-gluon-quark GPDs are added by `hand', see
Ref.\ \cite{BelMueKir01} for details. We neglect the $D$-term, which
affects the observable in question only sligthly at twist-three level.
Taking the mean values of the 2000 data (\ref{Exp-HERMES-2000}), the
$\phi$--integrated beam spin asymmetry (\ref{Def-MOMasy}) at twist-two
and twist-three level reads: 
\begin{eqnarray}
A_{\rm LU} = \left\{ 
{
[ -0.29,-0.25,-0.41] 
\atop
[-0.29,-0.20,-0.37] } 
\right\}\ \ \mbox{for}\ \left\{ { \mbox{twist-two} \atop \mbox{twist-three} } \right\}
\ \ \mbox{and\ models}\ \ [{\rm A,B,C}] \, .
\end{eqnarray}
As we realize the model B predictions are consistent with the
experimental data, while the model A gives a slightly too high value for
$|A_{\rm LU}|$. We remind that both model predictions are consistent
with the 96/97 data. The large value of the model C prediction is caused
by the sea quarks.

Now we present the predictions for the neon target at twist-two and
-three level. Here again the sea quark contributions are relatively
suppressed in model B and C, see Tab.\ \ref{Tab-ModH1}, and we again add
for the B model a quark-gluon-quark GPD contribution by `hand', in the
same way as it has been done for the proton. We find with these models: 
\begin{eqnarray}
A_{\rm LU}^{\rm Ne} = \left\{ 
{
[ -0.30,-0.28,-0.54] 
\atop
[-0.34,-0.24,-0.49] } 
\right\}\ \ \mbox{for}\ \left\{ { \mbox{twist-two} \atop \mbox{twist-three} } \right\}
\ \ \mbox{and\ models}\ \ [{\rm A,B,C}] \, .
\end{eqnarray} 
Again the model B prediction is on the one $\sigma$ level consistent with the experimental 
data (\ref{EstAna-ALU}).

Now we discuss estimates for an unpolarized and longitudinally
polarized deuterium target in more detail. The models for the reduced
deuteron GPD $H_1$ are listed in Tab.\ \ref{Tab-ModH1} and they share
the same qualitative features as described above. The $D$-term
contributions are neglected and we set for simplicity $\mu_{\rm sea}= 1$
and ${\cal Q}_{\rm sea}= 0$. To illustrate possible bound state effects,
we sometimes take $H_3$ into account, equating the reduced GPD with that
of $H_1$, or alternatively $H_5$, equating it with the antisymmetric
part of $H_1(x,\eta,\Delta^2)$. These two additional sets will be
denoted with a prime or a hat, respectively. As mentioned above, we
neglect the kinematically suppressed contributions $H_2$, $H_4$,
$\widetilde H_2$, and $\widetilde H_3$. The model of $\widetilde H_1$ is
based on the GS A parameterization \cite{GehSti95} with $\widetilde
b_{\rm sea}=1$. If not stated otherwise, we equate the slope
parameters $\widetilde B_{\rm val}=\widetilde B_{\rm sea}=20\ {\rm
GeV}^{-2}$, which is roughly the size of the slope parameter for the
charge form factor $G_Q$. 

\begin{figure}[t]
\vspace{-0.2cm}
\begin{center}
\mbox{
\begin{picture}(600,300)(0,0)
\put(-3,143){\rotate{\small $d \sigma \! /
(d\cQ^2 d\Delta^2 d\Ax)\, [{\rm nb}/{\rm GeV}^4]$}}
\put(0,140){\insertfig{16}{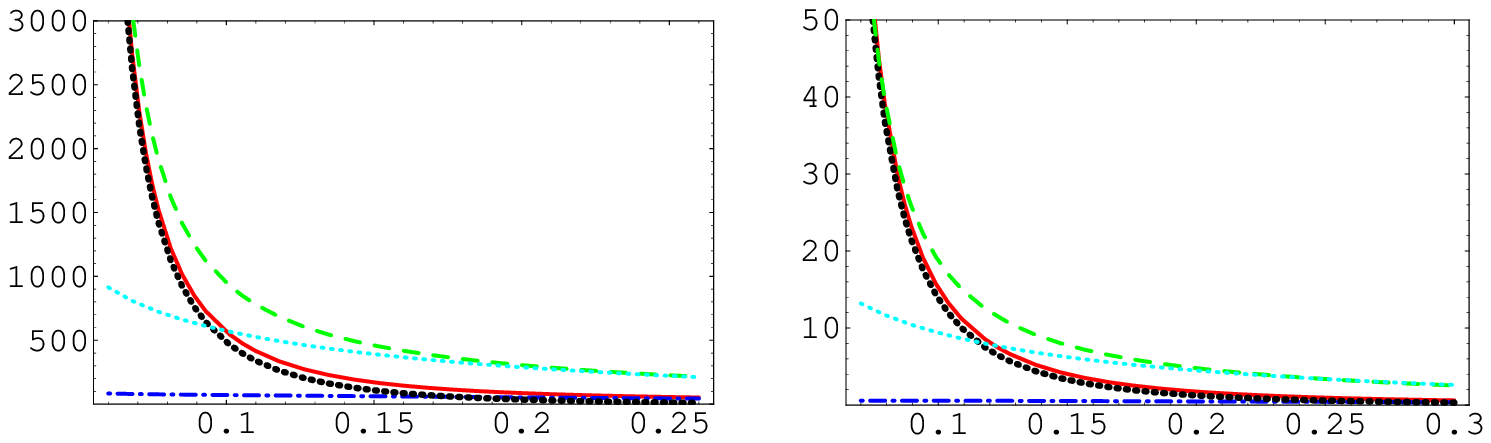}}
\put(180,250){(a)}
\put(405,250){(b)}
\put(-3,3){\rotate{\small $d \sigma \! /
(d\cQ^2 d\Delta^2 d\Ax)\, [{\rm nb}/{\rm GeV}^4]$}}
\put(0,0){\insertfig{16}{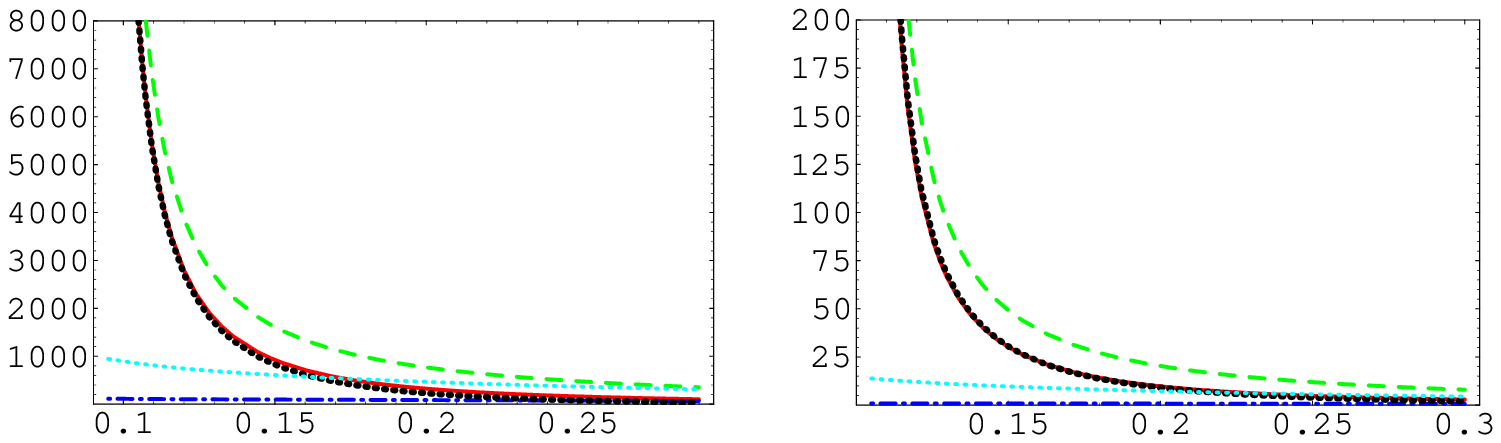}}
\put(180,110){(c)}
\put(405,110){(d)}
\put(210,-5){$\Bx$}
\put(430,-5){$\Bx$}
\end{picture}
}
\end{center}
\caption{\label{FigUnpCr}
Model dependent estimates for the cross section, integrated over the
azimuthal angle $\phi$, for the scattering of a positron [electron] beam
on a deuteron target at $E_e=27.6\ [12]\ \GeV$, $\cQ^2=2.5\ [2]\
\GeV^2$, $\Delta^2=-0.1\ \GeV^2$ (a) [(c)] and $\Delta^2=-0.3\ \GeV^2$
(b) [(d)]. The squared BH contribution is displayed as bold dotted line,
the DVCS cross section as dash-dotted (dotted) and the leptoproduction
cross section for positrons as solid (dashed) curves for model B (model
C).
}
\end{figure}
Let us first consider the size of the unpolarized cross sections for the 
positron and electron scattering off a deuteron target at HERMES and 
JLAB kinematics, respectively. It is demonstrated in Fig.\ \ref{FigUnpCr} that 
at larger value of $\Bx$, i.e., smaller value of $y$, the cross section 
is dominated by the DVCS one, however, its size is comparably small. 
With decreasing $\Bx$, the BH cross section starts to dominate over the model 
dependent DVCS one. In model C the sea quarks also induce  a growing DVCS cross 
section due to an associated large imaginary part in the DVCS amplitude. 
Typically, the ratio of BH to DVCS cross section and the total cross section is 
larger at JLAB compared to HERMES kinematics.

\begin{figure}[t]
\vspace{-0.2cm}
\begin{center}
\mbox{
\begin{picture}(600,300)(20,0)
\put(0,150){\insertfig{7.8}{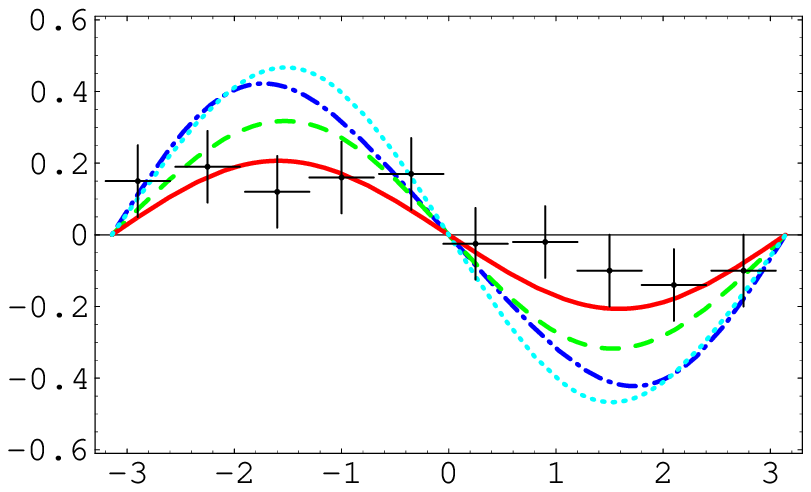}}
\put(240,152){\insertfig{7.8}{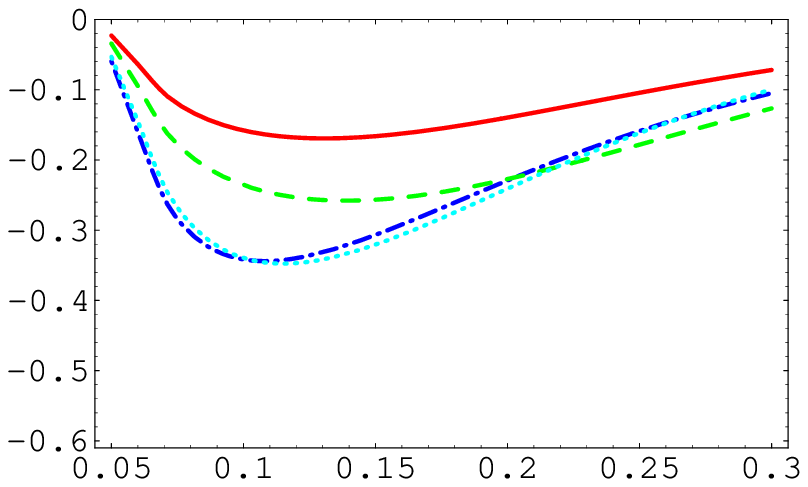}}
\put(185,260){(a)}
\put(425,190){(b)}
\put(200,147){$\phi$ [rad]}
\put(450,147){$\Bx$}
\put(0,0){\insertfig{7.9}{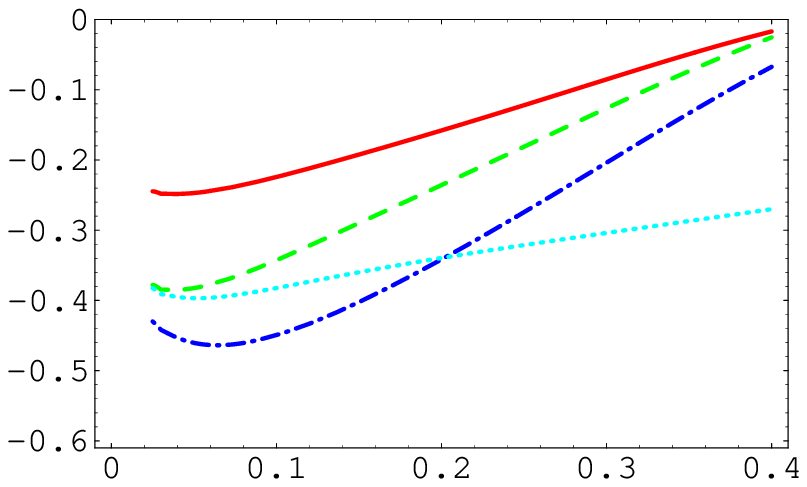}}
\put(240,2){\insertfig{7.7}{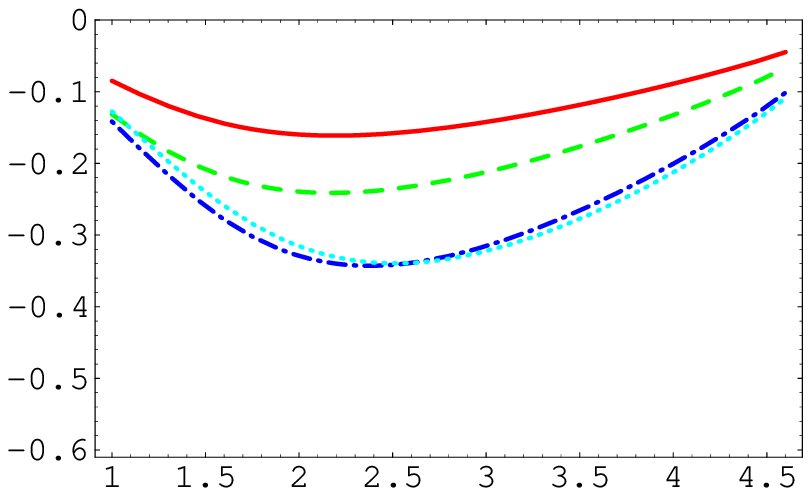}}
\put(185,40){(c)}
\put(425,40){(d)}
\put(175,-8){$-\Delta^2\ [\GeV^2]$}
\put(415,-8){$\cQ^2\ [\GeV^2]$}
\end{picture}
}
\end{center}
\caption{\label{FigALUHE}
The beam spin asymmetry ${\rm A}_{{\rm
L}\pm}(\phi)$ (a), ${\rm A}_{{\rm L}\pm}$ as function of $\Bx$ (b),
$-\Delta^2$ (c), and $\cQ^2$ (d) [without evolution] for the scattering
of a positron on a deuteron target at $E_e=27.6\ \GeV$. The fixed
kinematical variables are $\Bx=0.1$, $\cQ^2=2.5\ \GeV^2$, and
$\Delta^2=-0.2\ \GeV^2$, where the models A (dash--dotted), B (dashed),
$\hat {\rm B}$ (solid), and B$^\prime$ (dotted) are specified in the
text.
}
\end{figure}
In Fig.\ \ref{FigALUHE} we show the model dependence for the predictions
of the beam spin asymmetry $A_{{\rm L}\pm}$. In panel (a) we confront
them with the preliminary HERMES measurement \cite{EllShaVol02}.
However, we remind that the analysis is based on two different data sets
that, certainly, also contains incoherent scattering events. Model A
(dash--dotted) gives a larger asymmetry than model B (dashed), caused by
the larger sea quark content of the former model. Note that for the same
reason, model C predicts an even larger asymmetry, which is not
displayed. If we take into account the $H_3$ contribution via the
B$^\prime$ model (dotted), the beam spin asymmetry in panels (a,b,d) is
of the same size as the model A predictions. It is displayed in panel
(c) that model A and B$^\prime$ are distinguishable due to the
$\Delta^2$ dependence. In absence of $H_5$ we observe for the typical
HERMES kinematics only a rather small difference between the asymmetries
$A_{{\rm LU}}$ and $A_{{\rm L}\pm}$. The latter asymmetry is sensitive
to $H_5$ and we find a large reduction for the model $\hat{\rm B}$
(solid). We recall that the $\phi$--integrated asymmetries, displayed in
panels (b)--(d), are smaller than the non-integrated asymmetries at
$\phi=\pi/2$. This integration also changes the $\Delta^2$
dependence\footnote{The unintegrated asymmetries (\ref{Def-ALU}) and
(\ref{Def-ALpm}) are roughly spoken proportional to
$\sqrt{-\Delta^2/\cQ^2}$ and thus will constantly decrease with decreasing
$|\Delta^2|$. For the integrated asymmetry we rather observe a growing
that comes from the $\phi$ integration, indicated in Eq.\
(\ref{Def-MOMasy}). Of course, in any case the asymmetry vanishes in the
limit $\Delta^2\to \Delta^2_{\rm min}$.}. For JLAB kinematics we find
the same qualitative features, where of course for electrons the sign of
the asymmetry is reversed. Let us add that within the model B we have a
slightly larger value than in our result (\ref{EstAna-ALU}) of the
previous Section, caused by both kinematical and sea quark
contributions. The other models induce the typical effects, we have
already discussed:
\begin{eqnarray}
\label{Res-ALU-Num}
\left\{ {A_{{\rm LU}} \atop A_{{\rm L}\pm} } \right\}= 
\left\{{[-0.37, -0.26, -0.23, -0.34]  \atop [-0.34, -0.24, -0.16, -0.34]} \right\}\
 \ \mbox{for model}\ \ [{\rm  A, B,}\ \hat{\rm B},\ \rm{B'}]\, .
\end{eqnarray} 
These results will also be altered by twist-three contributions, i.e.,
change of the normalization and excitation of higher harmonics. A
quantitative estimate of such contributions is beyond the scope of this
paper. One might expect a similar change as in the case of a proton
target, discussed in great detail in Ref.\ \cite{BelMueKir01}.

\begin{figure}[t]
\vspace{-0.2cm}
\begin{center}
\mbox{
\begin{picture}(600,300)(10,0)
\put(0,150){\insertfig{7.7}{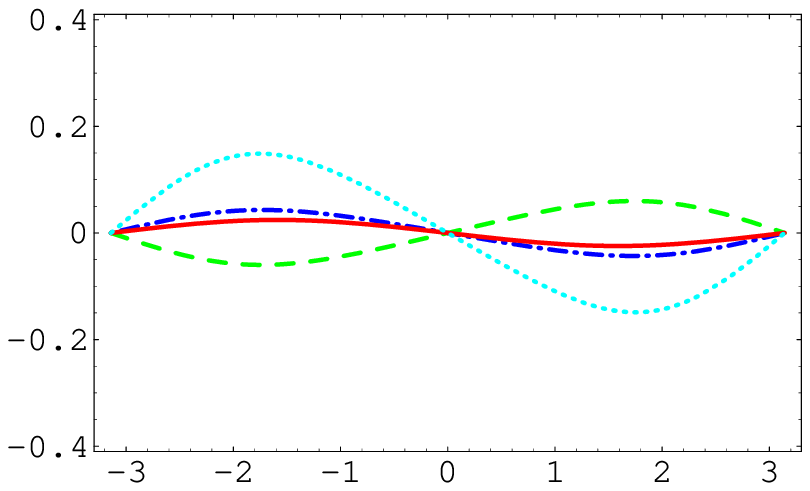}}
\put(235,150){\insertfig{7.9}{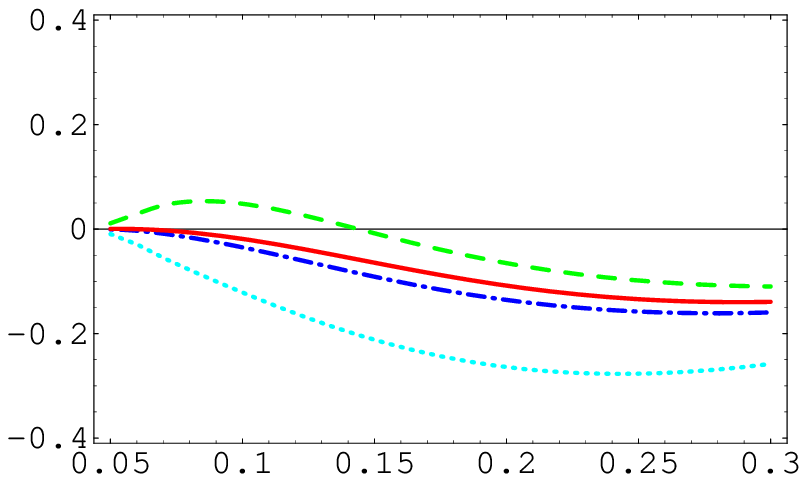}}
\put(180,260){(a)}
\put(415,260){(b)}
\put(0,0){\insertfig{7.9}{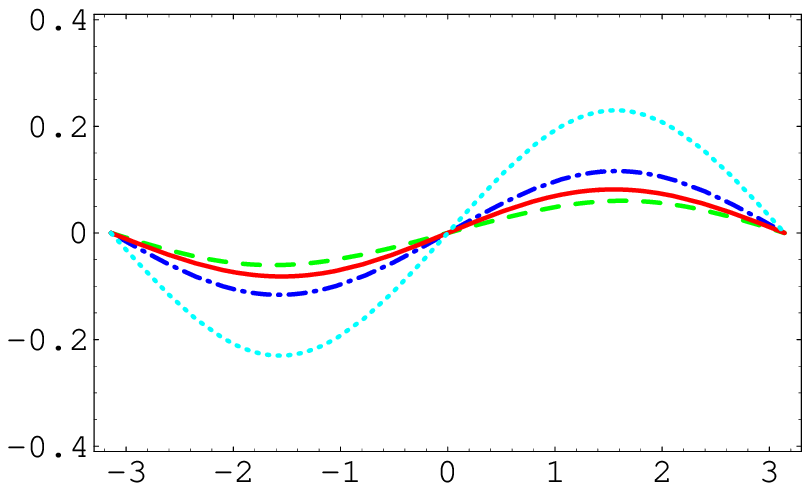}}
\put(240,0){\insertfig{7.6}{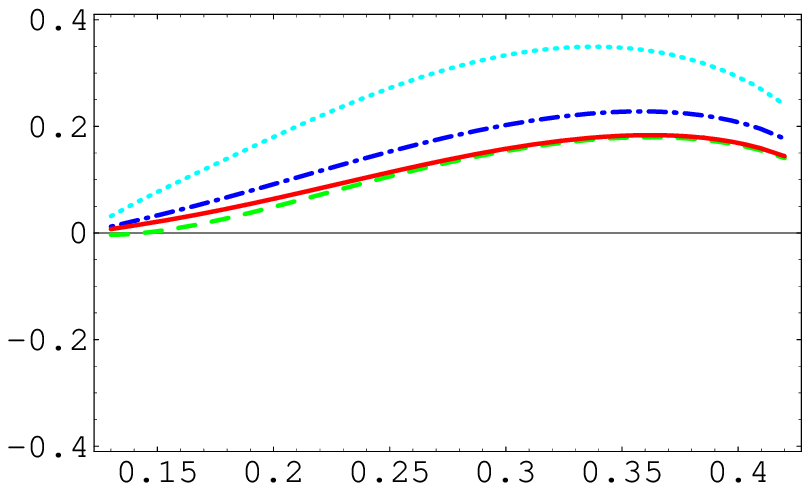}}
\put(180,40){(c)}
\put(415,40){(d)}
\put(190,-8){$\phi$ [rad]}
\put(445,-8){$\Bx$}
\end{picture}
}
\end{center}
\caption{\label{FigAUL}
The longitudinal target asymmetry $A_{\rm UL}(\phi)$ (a,c) 
and $A_{\rm UL}$ (b,d) versus 
$\phi$ and $\Bx$, respectively.  
For HERMES (a,b) [JLAB@12 GeV (c,d)] kinematics we take $\Bx=0.1$ [$\Bx=0.2$],
 $-\Delta^2 = 0.2\ \GeV^2$ and $\cQ^2 = 2.5\ \GeV^2$ as well as employ the 
model A with $\widetilde B_{\rm
val}=\widetilde B_{\rm sea}=20\ {\rm GeV}^{-2}$ (dash--dotted), 
$\widetilde B_{\rm val}=30\ {\rm GeV}^{-2}$, 
$\widetilde B_{\rm sea}=10\ {\rm GeV}^{-2}$  (dashed), 
$\widetilde B_{\rm val}=10\ {\rm GeV}^{-2}$, 
$\widetilde B_{\rm sea}=30\ {\rm GeV}^{-2}$ (dotted), and
$\hat{\rm B}$ with 
$\widetilde B_{\rm val}=\widetilde B_{\rm sea}=20\ {\rm GeV}^{-2}$ (solid). 
}
\end{figure}
As discussed above, see also Eq.\ (\ref{Res-AUL2ALU-Deu}), the size of the
longitudinal target spin asymmetry $A_{\rm UL}$ is rather model
dependent and allows to access the imaginary part of the parity odd CFF
$\widetilde{\cal H}_1$. This is demonstrated for HERMES and JLAB
kinematics in the panels (a,b) and (c,d), respectively, of Fig.\
\ref{FigAUL}. Here we set $\Bx=0.1$ and $\Bx=0.2$, respectively, as well
as $-\Delta^2 = 0.2\ \GeV^2$ and $\cQ^2 = 2.5\ \GeV^2$. For models A
(dash--dotted) and B (solid) with same $\widetilde{\cal H}_1$, taking
$\widetilde B_{\rm val} = \widetilde B_{\rm sea}=20\ {\rm GeV}^{-2}$, we
find a small asymmetry. Within this model parameter the contribution of
$\widetilde {\cal H}_1$ is negligible, since its valence and sea quark
contributions partly cancel each other. Thus, we observe in agreement
with Eq.\ (\ref{Res-AUL2ALU-Deu}) that the asymmetry has the same sign
as the beam spin asymmetry and generally increases with growing $\Bx$,
as it is displayed in panels (b) and (c). This is mainly caused by the
fact that the asymmetry is proportional to $1/y$. Thus, comparing panel
(a) and (c), we realize that for the typical JLAB kinematics (larger
$\Bx$) our estimates for $|A_{\rm UL}(\phi=\pi/2)|$ are slightly larger
than for the HERMES one. If the polarized valence quarks are relatively
suppressed with respect to the sea quarks, displayed by the dashed line,
the negative sea quark contribution takes over and induces the sign
change in panel (a). For larger value of $\Bx$, see panel (b), the
asymmetry turns out to be positive. For the contrary case that the valence
quark contribution is enhanced (dotted line), the target spin asymmetry
becomes sizably large.

The sign of the charge asymmetry $A_{\rm C}$ is not
predictable. For the mean values used above, we find that the model A
gives an asymmetry that is negligibly small, the B models provide a
positive and the C model a large negative contribution, where no $D$-term
has been taken into account. This observation is in qualitative
agreement with our results for the proton target and affirms once more
that an access to the $D$-term contribution is intricate.

\begin{figure}[t]
\vspace{-0.2cm}
\begin{center}
\mbox{
\begin{picture}(600,300)(10,0)
\put(0,150){\insertfig{7.7}{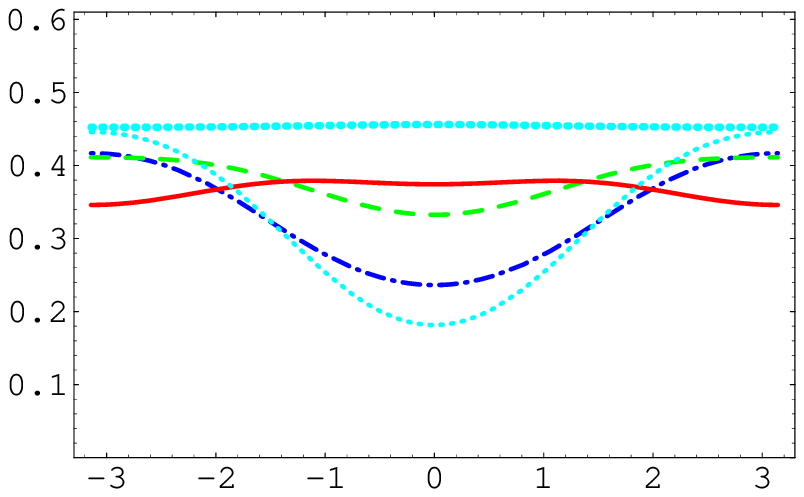}}
\put(235,150){\insertfig{7.9}{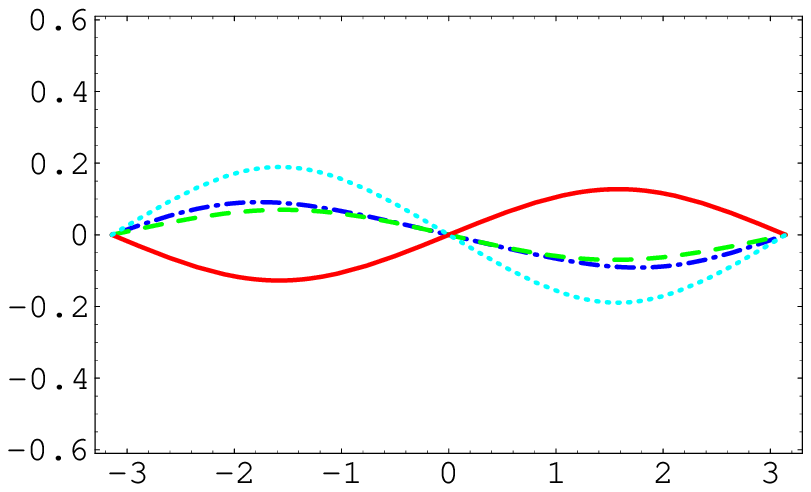}}
\put(180,190){(a)}
\put(415,190){(b)}
\put(0,0){\insertfig{7.9}{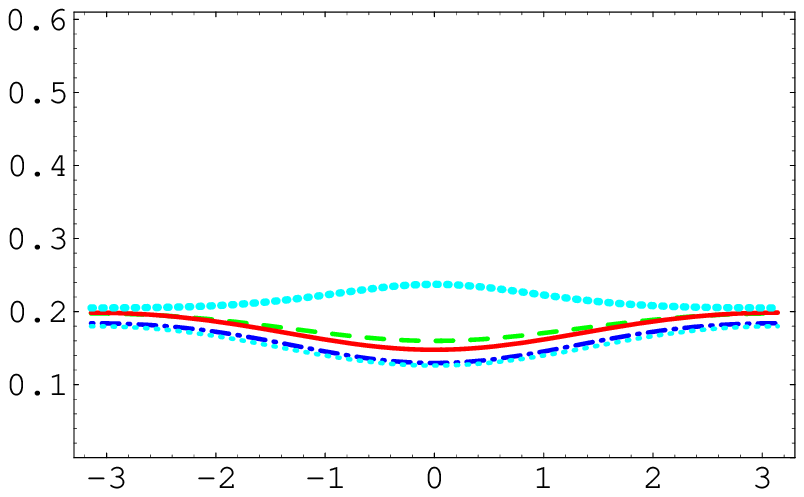}}
\put(240,0){\insertfig{7.9}{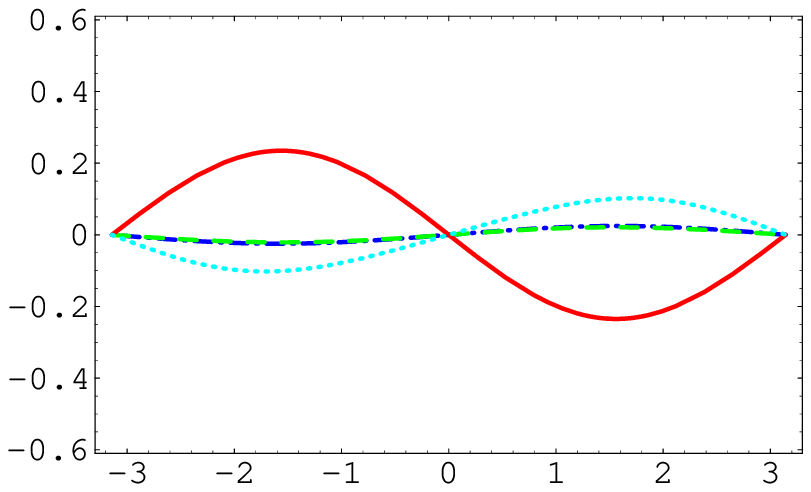}}
\put(180,115){(c)}
\put(415,115){(d)}
\put(190,-8){$\phi$ [rad]}
\put(430,-8){$\phi$ [rad]}
\end{picture}
}
\end{center}
\caption{\label{FigTenAS}
Tensor asymmetries $A_{zz}(\phi)$ (a,c) and $A_{{\rm
L}zz}(\phi)$ (b,d)  for HERMES (a,b) and JLAB@12 GeV (c,d) kinematics
within the same GPD models as in Fig.\ \ref{FigALUHE} and mean values as in
Fig.\ \ref{FigAUL}. The thick dotted line in panels (a,c) shows only the
BH cross section.
}
\end{figure}
The tensor asymmetry $A_{zz}(\phi)$, cf.\ Eq.\ (\ref{Def-Azz}), of
an unpolarized positron and electron beam is displayed in Fig.\
\ref{FigTenAS}(a) and (c) for the same HERMES and JLAB@12 GeV kinematics
as in Fig.\ \ref{FigAUL}, respectively. The thick dotted line shows the
pure BH contribution, which is in correspondence with Eq.\
(\ref{ResAna-Azz}) almost flat for HERMES kinematics (a). In the case of
JLAB kinematics (b) its contribution is much smaller compared to the
approximation (\ref{ResAna-Azz}), which is caused by higher order terms
in $\xi$ and $\tau$ that have not been  taken into account [see analogous
discussion of Eq.\ (\ref{Res-AUL2ALU-Deu})]. Subtraction of this
contribution gives a dominant $\cos(\phi)$ contribution that is
proportional to the real part of the linear combination
(\ref{Ana-Azz-LinCom-CFF}). As in the case of the charge asymmetry the sign
depends on the details of GPDs. In panels (b,d) we have ploted the tensor
asymmetry $A_{{\rm L}zz}(\phi)$, cf.\ Eq.\ (\ref{Def-ALzz}), for a
polarized lepton beam, which is essentially given by a $\sin(\phi)$
harmonic. Its amplitude is proportional to the imaginary part of CFFs
combination that enters Eq.\ (\ref{Res-ALzzACzz}). If both CFFs ${\cal
H}_3$ and ${\cal H}_5$ are small, as it is realized in models A
(dash-dotted) and B (dashed), we find only a small asymmetry, which has
the same sign as the beam spin asymmetries. This is in line with
approximation (\ref{Res-ALzz2ALU}). This tensor asymmetry is rather
sensitive to bound state effects. For instance, in the case that there
is an important $\im {\cal H}_3$ contribution the asymmetry becomes
sizable. On the other hand, if we assume a large positive $\im {\cal
H}_5$ and a negligible $\im {\cal H}_3$ contribution, i.e., taking model
B' (solid), we find a sizable asymmetry that has a reversed sign.

\section{Summary}

In this paper we considered the leptoproduction of a photon on nuclei up
to spin-1. For spin-0 and spin-1/2 targets the theoretical predictions
in terms of nuclei GPDs simply follow  from the known results, presented
in Refs.\ \cite{BelMueKirSch00,BelKirMueSch01,BelMueKir01} at the
twist-three level, by appropriate replacements of form factors, GPDs,
and kinematical variables. As a new result we added to this collection
the azimuthal angular dependence of the leptoproduction cross section of
a photon on a spin-1 target at twist-two level to leading order in
perturbation theory. Since the hard-scattering part is for all targets
unique, the result for the considered harmonics can immediately be
extended to NLO. At this order also a gluon tensor contribution appears,
which induces a $\cos(3 \phi)$ and $\cos(2\phi)$ harmonic in the
interference and squared DVCS term, respectively. We should add that the
elaboration of the twist-three sector, providing the zero and second
harmonics in the interference term and the first ones in the DVCS cross
section, is  straightforward in our covariant formalism.

For a spin-1 target nine CFFs enter the twist-two sector, which are
given as a convolution with nine GPDs. From the theoretical point of view,
the measurement of the imaginary and real part of all these CFFs is
possible for the case of a polarized lepton beam and target, with an
adjustable quantization direction, for both kinds of leptons is
available. Moreover, an appropriate Fourier analysis allows to eliminate
the twist-three contamination. In this way one has the maximal access to
the twist-two deuteron GPDs, given at the diagonal $x=\xi$ and as
convolution.

An important issue is the contamination of the leading twist-two
prediction by power suppressed contributions. Naively, one would expect
that such contributions scale with $\Delta^2/\cQ^2$ and $M_A^2/\cQ^2$.
Fortunately, within our model we showed that the so-called target mass
corrections have the same $A$ scaling as the leading twist-two
contributions. A more detailed investigation of power suppressed
contributions would be very desired.

Certainly, a dedicated experiment is required to measure all CFFs.
Having on hand only an unpolarized target one can access  the beam
spin and charge asymmetry via the Fourier coefficients $s^{\cal I}_{1,\rm
unp}$ and $c^{\cal I}_{1,\rm unp}$, given as a certain linear
combinations of ${\cal H}_1$, ${\cal H}_3$, and ${\cal H}_5$ CFFs. A
longitudinally polarized target allows to measure four further
independent observables. Two of them, e.g., the tensor polarization
combined with beam spin or charge asymmetry, are predicted by another
linear combination of the afore mentioned three CFFs, entering the
Fourier coefficients $s^{\cal I}_{1,\rm LLP}$ and $c^{\cal I}_{1,\rm
LLP}$. Unfortunately, such experiments do not allow to access all three
GPDs in question. The remaining two observables are related to the
target spin flip odd asymmetries, given by the FCs $s^{\cal I}_{1,\rm
LP}$ and $c^{\cal I}_{1,\rm LP}$. In a single target spin flip
experiment one can measure $s^{\cal I}_{1,\rm LP}$ and get access to
the imaginary part of $\widetilde {\cal H}_1$. On the other hand,
combining the data for polarized target with charge asymmetry or the
flip of the beam helicity allow then to measure $c^{\cal I}_{1,\rm LP}$
and so also the real part of $\widetilde {\cal H}_1$. In all these
observables, we discussed so far, the remaining five CFFs are
kinematically suppressed or do not enter the theoretical predictions.

To estimate the size of different observables, we proposed a simple
model for nuclei GPDs that is based on the $A$--scaling of the
longitudinal degrees of freedom and a $\Delta^2$ factorized ansatz of
the nucleon GPDs. For a spin-0 or -1/2 nucleus target, this ansatz is
sufficient to fix all reduced GPDs in terms of the proton ones (of
course, they are also unknown). The remaining degree of freedom concerns
mainly the form factors, especially, in the sea quark sector. Of course,
binding effects should alter this oversimplified ansatz, e.g., we expect
that t-channel exchange forces can induce an important contribution in
the `exclusive' region of GPDs. Such effects might be visible by a
precise measurement of single spin and charge asymmetries. For a spin-1
target the large number of GPDs opens a new window to study nuclear
binding effects. Compared to deep inelastic scattering or elastic
lepton-deuteron scattering, they provide additional information,
contained in the CFFs. Beside the modifaction of the the scaling law for
the CFFs ${\cal H}_1$ and $\widetilde {\cal H}_1$, they induce
non-vanishing CFFs ${\cal H}_3$ and ${\cal H}_5$. We conclude that
nucleus GPDs give a new testing ground for non-perturbative methods that
have been used in nuclear physics for the calculation of form factors, 
see reviews \cite{GilGro01,GarOrd01}.

Finally, we estimated for the typical kinematics of present fixed target
experiments, different asymmetries and showed that a qualitative
understanding of them is possible by means of analytical formulae, which
are obtained by a simplification of kinematics and GPD models. For a
deuteron target such an approximation allows to discuss the contribution
of the CFFs ${\cal H}_3$ and ${\cal H}_5$. The LO analyses of the
pioneering measurements of DVCS \cite{Adletal01,Airetal01,Steetal01} on
the proton suggest for $-\Delta^2 \sim 0.3\ \GeV^2$ the dominance of
valence-quark GPDs in the valence quark region with no essential
enhancement by the skewedness effect. Assuming the same features also
for the nuclei GPDs, our estimates agree with the
$\phi$-integrated beam spin asymmetries measured on neon and deuteron
targets. In the latter case this asymmetry is already sensitive to the
CFFs ${\cal H}_3$ and ${\cal H}_5$. A measurement of the tensor
polarization would certainly provide a new constraint for these CFFs.

\section*{Acknowledgements}
We are very grateful to M.~Diehl, A.~Freund, and A.~Sch\"afer for useful
discussions on different theoretical issues. We are indebted to
M.~Amarian, B.~Krauss, and R.~Shanidze for discussions of the
experimental settings and to J.~Ball for providing us the parameters of
deuteron form factor parameterization. One of us (A.K.) would like to
thank the Studienstiftung des deutschen Volkes for financial support. He
also wants to thank the CTP (MIT), especially, Prof.\ Jaffe and Prof.
Negele for kind hospitality and many useful discussions. 

{\ }
\\

\noindent
{\bf Note added:} Recently there appeared a note \cite{GuzStr03} where
DVCS on a spin-0 nucleus has been considered in the impuls
approximation. This result overlaps with a part of our analysis in
Sections \ref{Sec-ModGPD} and \ref{Sec-EstObs}.

\begin{appendix}

\section{Parameterization of the electromagnetic deuteron form factors
\label{App-DeuForFac}
}

The set of form factors $G_1,\ G_2,$ and $G_3$ may be expressed by
the charge monopole, magnetic dipole, and charge quadrupole form
factors:
\begin{eqnarray}
\label{Rel-among-FF}
G_C(\Delta^2) &\!\!\!=\!\!\!&
\left(1 -  \frac{2\tau}{3}\right) G_1(\Delta^2) +  \frac{2\tau}{3}
\left[ G_2(\Delta^2)-(1-\tau) G_3(\Delta^2)\right],
\nonumber\\
G_M(\Delta^2) &\!\!\!=\!\!\!&  G_2(\Delta^2),
\\
G_Q(\Delta^2) &\!\!\!=\!\!\!&  G_1(\Delta^2) - G_2(\Delta^2) +(1-\tau)
G_3(\Delta^2),
\nonumber
\end{eqnarray}
with $\tau = \frac{\Delta^2}{4 \AM^2}$. The normalizations of these
form factors read
\begin{eqnarray}
\label{Def-StaPro}
G_C(0)=1,\quad  G_M(0) = \mu_d = 1.714,\quad G_Q(0)= Q_d = 25.83.
\end{eqnarray}

Their parameterization, which is inspired by the counting rules
for large $-\Delta^2$ was taken from Ref. \cite{KobSya95}:
\begin{eqnarray}
G_C(\Delta^2) &\!\!\!=\!\!\!& \frac{G_D^2\left(\frac{\Delta^2}{4}\right)}{1-2\tau}
\left[\left(1+\frac{2}{3}\tau\right) g_{00}^+  + \frac{8}{3} \sqrt{-2\tau}
g_{+0}^+ - \frac{2}{3}(1+2\tau) g_{+-}^+  \right]
\nonumber\\
G_M(\Delta^2) &\!\!\!=\!\!\!& \frac{G_D^2\left(\frac{\Delta^2}{4}\right)}{1-2\tau}
\left[2 g_{00}^+  -\frac{2(1+2\tau)}{\sqrt{-2\tau}}g_{+0}^+ -2 g_{+-}^+ \right]
\\
G_Q(\Delta^2) &\!\!\!=\!\!\!& \frac{G_D^2\left(\frac{\Delta^2}{4}\right)}{1-2\tau}
\left[-g_{00}^+  +\sqrt{-\frac{2}{\tau}}g_{+0}^+ +\frac{1-\tau}{\tau} g_{+-}^+
\right].
\nonumber
\end{eqnarray}
Here we use the standard dipole parameterization for the nucleon
form factor
\begin{eqnarray}
G_D\left(\Delta^2\right) =  \left(1- \frac{\Delta^2}{0.71\ \GeV^2}
\right)^{-2}
\nonumber
\end{eqnarray}
and the helicity transition amplitudes in the infinite momentum frame
\cite{BroHil92} are given by
\begin{eqnarray}
g_{00}^+ = \sum_{i=1}^4 \frac{a_i}{\alpha_i^2-\Delta^2}, \quad
g_{+0}^+ = \sqrt{-\Delta^2} \sum_{i=1}^4 \frac{b_i}{\beta_i^2-\Delta^2}, \quad
g_{+-}^+ = -\Delta^2 \sum_{i=1}^4 \frac{c_i}{\gamma_i^2-\Delta^2}.
\end{eqnarray}
Counting rules predict for large $-\Delta^2$ the following
behavior
\begin{eqnarray}
g_{00}^+ \sim (-\Delta)^{-2}, \quad g_{+0}^+ \sim (-\Delta)^{-3},
\quad
g_{+-}^+ \sim (-\Delta)^{-4},
\end{eqnarray}
which gives together with the static properties (\ref{Def-StaPro}) six constraints
for twentyfour fitting parameters:
\begin{eqnarray}
\label{Con-FitParDFF0}
\sum_{i=1}^4 \frac{a_i}{\alpha_i^2} =1, \
\sum_{i=1}^4 \frac{b_i}{\beta_i^2} =\frac{2 -\mu_d}{2\sqrt{2}M_d}, \
\sum_{i=1}^4 \frac{c_i}{\gamma_i^2} =\frac{1-\mu_d - Q_d}{4M_d^2},\
\sum_{i=1}^4 b_i= \sum_{i=1}^4 c_i=\sum_{i=1}^4 c_i \gamma_i^2  =0.
\nonumber\\
\end{eqnarray}
\begin{table}[t]
\vspace{-0.5cm}
\begin{center}
\begin{tabular}{|l|c|c|c|c|}
\hline
\hspace{1cm}$i=$ & 1 & 2 & 3 & 4
\\ \hline\hline
$a_i\ [{\rm fm}^{-2}]$ & 1.57057 & 12.23792 & -42.04576 & Eq.\ (\ref{Con-FitParDFF0})
\\ \hline
$b_i\ [{\rm fm}^{-1}]$  & 0.07043 & 0.14443 & Eq.\ (\ref{Con-FitParDFF0}) & Eq.\
(\ref{Con-FitParDFF0})
\\ \hline
$c_i$  & -0.16577 & Eq.\ (\ref{Con-FitParDFF0}) & Eq.\ (\ref{Con-FitParDFF0}) &
Eq.\ (\ref{Con-FitParDFF0})
\\ \hline
$\alpha_i^2\ [{\rm fm}^{-2}]$  & 1.52501
& Eq.\ (\ref{Con-FitParDFF}) & Eq.\ (\ref{Con-FitParDFF}) & 23.20415
\\ \hline
$\beta_i^2\ [{\rm fm}^{-2}]$  & 43.67795
& Eq.\ (\ref{Con-FitParDFF}) & Eq.\ (\ref{Con-FitParDFF}) & 2.80716
\\ \hline
$\gamma_i^2\ [{\rm fm}^{-2}]$  & 1.87055
& Eq.\ (\ref{Con-FitParDFF}) & Eq.\ (\ref{Con-FitParDFF}) & 41.1294
\\ \hline
\end{tabular}
\end{center}
\caption{Fitting parameter sets for the electromagnetic Deuteron form factors.}
\label{Tab-DFF}
\end{table}
To reduce this set to twelve parameters, one may introduce for each group
${\alpha_i}$, ${\beta_i}$, and ${\gamma_i}$ the algebraic relations:
\begin{eqnarray}
\label{Con-FitParDFF}
\alpha_i^2 = \alpha_1^2 + \left(\alpha_4^2-\alpha_1^2\right)
\frac{i-1}{3}\quad \mbox{for}
\quad i=1,\dots,4.
\end{eqnarray}
The fitting parameters are taken from Ref.\ \cite{Abbetal00} and are given
in Tab.\ \ref{Tab-DFF}.

\section{Results for the Fourier coefficients\label{App-Apr}}

Below the twist-two results for unpolarized and longitudinally polarized
target are listed. All results have been expanded for small $\tau=
\frac{\Delta^2}{4 M^2}$, i.e., for $\Delta^2 \ll M^2$. Terms proportional
to ${\cal H}_3, G_3$ are given up to order ${\cal O}(\tau)$, because
$G_3$ is at $\Delta^2=0$ roughly $20$ times larger than the other form
factors. The BH amplitude squared has been exactly calculated. However,
the results are very cumbersome and, thus, they are only presented in
leading order of $1/\cQ^2$. We note that the rather lengthy results for
${\cal M}^{{\cal I},{\rm DVCS}}_{0,{\rm LLP}}$ and $C^{\rm BH}_{0,{\rm
LLP}}$ can be obtained through the much shorter results for unpolarized
and transverse--transverse coefficients via equations (\ref{TTPAbh}).

\subsection{Interference term\label{App-IntTer}}

Here we give the matrices ${\cal M}^{\cal I}_{1,{\rm unp}}$, 
${\cal M}^{\cal I}_{1,{\rm LP}}$, and ${\cal M}^{\cal I}_{1,{\rm TTP}\Sigma}$, 
which appear in Eq.\ (\ref{Def-C-Int}).

\begin{eqnarray}
{\cal M}^{\cal I}_{1,{\rm unp}}&=& 
\footnotesize
\frac{1}{9}\left(
\matrix{ 9 & 0 & -6\,\tau \cr 0 & 0 & 0 \cr -6\,\tau  & 0 &
12\,{\tau }^2 \cr -6\,\xi  & 0 & 12\,\xi \,\tau  \cr 9\,
   {\xi }^2 & -6\,{\xi }^2 & 2\,\left( {\xi }^2\,\left( 3 - 6\,\tau  \right)  
+ \tau  \right)  \cr 0 & 6\,
   \xi  & 0 \cr 0 & 0 & 0 \cr 0 & 0 & 0 \cr 0 & 0 & 0 \cr  }
\right) \nonumber\\
\nonumber\\
\nonumber\\
{\cal M}^{\cal I}_{1,{\rm LP}}&=& \footnotesize\frac{1}{6
(1+\xi)}\left( \matrix{ 0 & 6\,\xi & 0 \cr 0 & 3\,{\xi }^2 & 0 \cr
0 & -6\,\xi \,\tau  & 0 \cr 0 & -3\,{\xi }^2 & 0 \cr 0 & -2\,
   \xi  & 0 \cr 6\,\left( 1 + \xi  - {\xi }^2 \right)  & 6\,{\xi }^2 
& -6\,{\xi }^2\,\left( 1 - 2\,\tau  \right)  -
   6\,\tau  \cr -24\,{\xi }^2 & 0 & 24\,{\xi }^2\,\tau  \cr 0 & 0 
 24\,{\xi }^2\,\tau  \cr -6\,\left( 1 - \xi  \right) \,
   {\xi }^2 & -6\,{\xi }^3 & 6\,\xi \,\left( {\xi }^2\,\left(1 - 2\,\tau\right)
+ \tau  + \xi \,\tau  \right)  \cr  }
\right)
\nonumber\\
\nonumber\\
\nonumber\\
{\cal M}^{\cal I}_{1,{\rm TTP}\Sigma}&=& \footnotesize\frac{1}{6
(1+\xi)^2}\left( \matrix{ 6\,\left( 1 + \left( 2 - \xi  \right)
\,\xi  \right)  & 6\,{\xi }^2 & -6\,{\xi }^2\,\left( 1 - 3\,\tau
\right)  -
   6\,\tau  \cr 6\,\left( 1 - \xi  \right) \,{\xi }^2 & 6\,{\xi }^3 & 6\,\xi \,
   \left( -\left( {\xi }^2\,\left( 1 - 2\,\tau  \right)  \right)  
- \tau  - \xi \,\tau  \right)  \cr -6\,{\xi }^2\,
    \left( 1 - 3\,\tau  \right)  - 6\,\tau  & -6\,\xi \,
\left( 1 + 2\,\xi  \right) \,\tau  & 12\,\tau \,
   \left( {\xi }^2\,\left( 1 - 2\,\tau  \right)  + \tau  \right)  \cr
 -6\,\left( 1 - \xi  \right) \,{\xi }^2 & -6\,
   {\xi }^3 & 6\,\xi \,\left( {\xi }^2\,\left( 1 - 2\,\tau  \right)  
+ \tau  + \xi \,\tau  \right)  \cr -2\,
   \left( 1 + \left( 2 - \xi  \right) \,\xi  \right)  & -2\,{\xi }^2 & 2\,
   \left( {\xi }^2\,\left( 1 - 3\,\tau  \right)  + \tau  \right)  \cr 0 
& 3\,\xi \,\left( 1 + \xi  \right) \,
   \left( 2 + \xi  \right)  & 0 \cr 0 & -12\,{\xi }^3 & 0 \cr
 0 & 12\,{\xi }^3 & 0 \cr 0 & -3\,{\xi }^2\,
   {\left( 1 + \xi  \right) }^2 & 0 \cr  }
\right) \nonumber
\end{eqnarray}

\subsection{DVCS amplitude squared\label{App-DVCS}}

The matrices ${\cal M}^{\rm DVCS}_{0,i}$, entering Eq.\ (\ref{Def-C-DVCS}),
for $i \in \{{\rm unp},{\rm TP}-,{\rm LLP},{\rm LTP}+,{\rm TTP}\Sigma,{\rm
TTP}\Delta\}$ are of the form
\[
{\cal M}^{\rm DVCS}_{0,i}=\left\{ \matrix{ A^{\rm DVCS}_{0,i} & 0
\cr 0 & B^{\rm DVCS}_{0,i} \cr } \right\},
\]
with $A^{\rm DVCS}_{0,i}$ being a $5 \times 5$  and $B^{\rm
DVCS}_{0,i}$ being a $4 \times 4$ matrix. The 
matrices ${\cal M}^{\rm DVCS}_{0,j}$ for $j \in
\{{\rm LP},{\rm TP}+,{\rm TTP}\pm\}$ have the building blocks 
\[
{\cal M}^{\rm DVCS}_{0,j}=\left\{ \matrix{0 & C^{\rm DVCS}_{0,j}
\cr D^{\rm DVCS}_{0,j} & 0 \cr } \right\},
\]
where $C^{\rm DVCS}_{0,j}$ is a $5 \times 4$ and $D^{\rm
DVCS}_{0,j}$ is a $4 \times 5$ matrix.

We list only the submatrices $A^{\rm DVCS}_{0,i}, B^{\rm DVCS}_{0,i}, C^{\rm
DVCS}_{0,j}, D^{\rm DVCS}_{0,j}$ that are needed for an unpolarized
and longitudinally polarized target:
\\
\\
\begin{eqnarray}
A^{\rm DVCS}_{0,{\rm unp}}&=&\footnotesize\frac{1}{9}\left(
\matrix{ 18 & 0 & -12\,\tau & -12\,\xi  & 18\,{\xi }^2 \cr 0 &
-12\,{\xi }^2 & 0 & 0 & -24\,{\xi }^2 \cr -12\,\tau  & 0 & 24\,
   {\tau }^2 & 24\,\xi \,\tau  & 4\,\left( {\xi }^2\,\left( 3 - 6\,\tau \right)
 + \tau  \right)  \cr -12\,\xi  & 0 & 24\,\xi \,
   \tau  & 12\,{\xi }^2 & -4\,\xi \,\left( 2 + 3\,{\xi }^2 \right)  \cr 
18\,{\xi }^2 & -24\,{\xi }^2 & 4\,
   \left( {\xi }^2\,\left( 3 - 6\,\tau  \right)  + \tau  \right)  
& -4\,\xi \,\left( 2 + 3\,{\xi }^2 \right)  & 4 -
   24\,{\xi }^2 + 6\,{\xi }^4 \cr  }
       \right)\nonumber\\
       \nonumber\\
       \nonumber\\
B^{\rm DVCS}_{0,{\rm unp}}&=&\footnotesize\frac{1}{9}\left(
\matrix{ 12\,\left( 1 - {\xi }^2 \right)  & -48\,{\xi }^2 & 0 &
-12\,{\xi }^2 \cr -48\,{\xi }^2 & 0 & 0 & 0 \cr 0 & 0 & 0 & 48\,
   {\xi }^3 \cr -12\,{\xi }^2 & 0 & 48\,{\xi }^3 & -12\,{\xi }^2\,
\left( 1 + {\xi }^2 \right)  \cr  }
\right)\nonumber\\
\nonumber\\
\nonumber\\
\nonumber\\
 C^{\rm DVCS}_{0,{\rm LP}}&=&\footnotesize\frac{1}{3
(1+\xi)}\left(\matrix{ 6\,\left( 1 + \xi - {\xi }^2 \right)  &
-24\,{\xi }^2 & 0 & -6\,\left( 1 - \xi \right) \,{\xi }^2 \cr 3\,
   \left( 1 - \xi  \right) \,{\xi }^2 & -12\,{\xi }^3 & -12\,{\xi }^3 
& 3\,{\left( 1 - \xi  \right) }^2\,{\xi }^2 \cr -6\,
    {\xi }^2\,\left( 1 - 2\,\tau  \right)  - 6\,\tau  & 24\,{\xi }^2\,\tau  
& 24\,{\xi }^2\,\tau  & 6\,\xi \,
   \left( \tau  + \xi \,\left( \xi  + \tau  - 2\,\xi \,\tau  \right)\right)\cr
 -3\,\left( 1 - \xi  \right) \,{\xi }^2 & 12\,
   {\xi }^3 & 12\,{\xi }^3 & -3\,{\left( 1 - \xi  \right) }^2\,{\xi }^2 \cr
 -2\,
   \left( 1 + \left( 1 - \xi  \right) \,\xi  \right)  & 8\,{\xi }^2 & 0 &
 2\,\left( 1 - \xi  \right) \,{\xi }^2 \cr  }\right)\nonumber\\
   \nonumber\\
   \nonumber\\
D^{\rm DVCS}_{0,{\rm LP}}&=&\scriptsize\frac{1}{3
(1+\xi)}\left(\matrix{ 6\,\left( 1 + \xi - {\xi }^2 \right)  &
3\,\left( 1 - \xi  \right) \,{\xi }^2 & -6\,{\xi }^2\,
    \left( 1 - 2\,\tau  \right)  - 6\,\tau  & -3\,\left( 1 - \xi  \right) \,
{\xi }^2 & -2\,
   \left( 1 + \left( 1 - \xi  \right) \,\xi  \right)  \cr -24\,{\xi }^2 & -12\,
{\xi }^3 & 24\,{\xi }^2\,\tau  & 12\,
   {\xi }^3 & 8\,{\xi }^2 \cr 0 & -12\,{\xi }^3 & 24\,{\xi }^2\,\tau  
& 12\,{\xi }^3 & 0 \cr -6\,\left( 1 - \xi  \right) \,
   {\xi }^2 & 3\,{\left( 1 - \xi  \right) }^2\,{\xi }^2 & 6\,\xi \,
   \left( \tau  + \xi \,\left( \xi  + \tau  - 2\,\xi \,\tau  \right)  \right)
& -3\,{\left( 1 - \xi  \right) }^2\,
   {\xi }^2 & 2\,\left( 1 - \xi  \right) \,{\xi }^2 \cr  }
\right)\nonumber\\
\nonumber\\
\nonumber
\end{eqnarray}
\newpage \hspace{-1cm}\rotatebox{90}{
\begin{minipage}{23.5cm}
\begin{eqnarray}
A^{\rm DVCS}_{0,{\rm TTP}\Sigma}&=&\scriptsize\frac{1}{9
(1+\xi)^2} \left(\matrix{ 9\,\left( 2 + 2\,\left( 2 - \xi  \right)
\,\xi \right)  & 18\,\left( 1 - \xi \right) \,{\xi }^2 & 18\,
   \left( -\left( {\xi }^2\,\left( 1 - 3\,\tau  \right)  \right)  -\tau\right)
& -18\,\left( 1 - \xi  \right) \,
   {\xi }^2 & -6\,\left( 1 + \left( 2 - \xi  \right) \,\xi  \right)\cr
 18\,\left( 1 - \xi  \right) \,{\xi }^2 & -9\,
   {\left( 1 - \xi  \right) }^2\,{\xi }^2 & 18\,\xi \,
   \left( -\left( {\xi }^2\,\left( 1 - 2\,\tau  \right)  \right)
  - \tau  - \xi \,\tau  \right)  & 9\,
   {\left( 1 - \xi  \right) }^2\,{\xi }^2 & -6\,\left( 1 - \xi  \right)
 \,{\xi }^2 \cr 18\,
   \left( -\left( {\xi }^2\,\left( 1 - 3\,\tau  \right)  \right)- \tau\right)
  & 18\,\xi \,
   \left( -\left( {\xi }^2\,\left( 1 - 2\,\tau  \right)  \right)  - \tau  - 
\xi \,\tau  \right)  & 36\,\tau \,
   \left( {\xi }^2\,\left( 1 - 2\,\tau  \right)  + \tau  \right)  & 18\,\xi \,
   \left( \tau  + \xi \,\left( \xi  + \tau  - 2\,\xi \,\tau  \right)  \right)
& 6\,
   \left( {\xi }^2\,\left( 1 - 3\,\tau  \right)  + \tau  \right)  \cr 
-18\,\left( 1 - \xi  \right) \,{\xi }^2 & 9\,
   {\left( 1 - \xi  \right) }^2\,{\xi }^2 & 18\,\xi \,
   \left( \tau  + \xi \,\left( \xi  + \tau  - 2\,\xi \,\tau  \right)  \right)  
& -9\,{\left( 1 - \xi  \right) }^2\,
   {\xi }^2 & 6\,\left( 1 - \xi  \right) \,{\xi }^2 \cr 
-6\,\left( 1 + \left( 2 - \xi  \right) \,\xi  \right)  & -6\,
   \left( 1 - \xi  \right) \,{\xi }^2 & 
6\,\left( {\xi }^2\,\left( 1 - 3\,\tau  \right)  + \tau  \right)  & 6\,
   \left( 1 - \xi  \right) \,{\xi }^2 & 
2 + 2\,\left( 2 - \xi  \right) \,\xi  \cr  }\right)
\nonumber\\
\nonumber\\
\nonumber\\
B^{\rm DVCS}_{0,{\rm TTP}\Sigma}&=&\footnotesize\frac{1}{9
(1+\xi)^2} \left(\matrix{ 9\,{\left( 1 + \xi \right) }^2\,\left( 2
- {\xi }^2 \right)  & -36\,{\xi }^2\,\left( 1 + \xi  \right) \,
   \left( 2 + \xi  \right)  & -36\,{\xi }^3\,\left( 1 + \xi  \right)  & 
-9\,\left( 1 - \xi  \right) \,{\xi }^2\,
   {\left( 1 + \xi  \right) }^2 \cr -36\,{\xi }^2\,\left( 1 + \xi  \right) \,
\left( 2 + \xi  \right)  & 144\,{\xi }^4 & -144\,
   {\xi }^4 & 36\,{\xi }^3\,{\left( 1 + \xi  \right) }^2 \cr
 -36\,{\xi }^3\,\left( 1 + \xi  \right)  & -144\,{\xi }^4 & 144\,
   {\xi }^4 & 36\,{\xi }^3\,{\left( 1 + \xi  \right) }^2 \cr 
-9\,\left( 1 - \xi  \right) \,{\xi }^2\,
   {\left( 1 + \xi  \right) }^2 & 36\,{\xi }^3\,{\left( 1 + \xi  \right) }^2 &
 36\,{\xi }^3\,{\left( 1 + \xi  \right) }^2 & -9\,
   {\xi }^2\,{\left( 1 - {\xi }^2 \right) }^2 \cr  }
\right) \nonumber
\end{eqnarray}
\\
\subsection{BH amplitude squared \label{App-BH}}

\begin{eqnarray}
C^{\rm BH}_{0,{\rm unp}}&=&\frac{8\,\left( 2 - 2\,y +
      y^2 \right)}{3\,
    {\left( 1 + \xi  \right) }^2} \, \left( 2\,
       {\xi }^2 \,G_2^2+
      \left(1-\frac{\Delta_{\rm min}^2}{\Delta^2}\right)\,
       \left( 1 - {\xi }^2 \right)
         \,\left( 3\,
          G_1^2 -
         4\,\tau\,G_1\,
          G_3  +
         4\,
          {\tau }^2\,G_3^2 \right)
      \right)\nonumber\\
       \nonumber\\
      \nonumber\\
C^{\rm BH}_{0,{\rm LP}}&=&\frac{-8\,
    \left( 2 - y \right) \,y\,
    \lambda \,\xi}{{\left
       ( 1 + \xi  \right) }^3} \,G_2\, \left( \xi \,
       \left( 1 + 2\,\xi  \right)\,G_2\,
       + 2\,\left(1-\frac{\Delta_{\rm min}^2}{\Delta^2}\right)\,
       \left( 1 - \xi^2  \right) \,
       \left( G_1 -
         \tau\,G_3
         \right)  \right)\nonumber\\
\nonumber\\
         \nonumber\\
C^{\rm BH}_{0,{\rm TTP}\Sigma}&=& \frac{8\,
    \left( 2 - 2\,y + y^2 \right)
      \,\left( 1 - \xi  \right) }{{\left
       ( 1 + \xi  \right) }^3}\,
    \left(1-\frac{\Delta_{\rm min}^2}{\Delta^2}\right)\,\left(
   {\left( 1 + \xi  \right) }^2\,{G_1}^2 -
  2\,\xi \,
   \left( 1 + 2\,\xi  \right)\,
   \tau\,G_2\,
   G_3 - 2  \left(1-\frac{\Delta_{\rm min}^2}{\Delta^2}\right)\,
  \left( 1 - {\xi }^2 \right) \,
  \tau\,G_3\,\left( G_1 -
  \tau\,G_3  \right)\right)\nonumber
\end{eqnarray}

\end{minipage}}

\section{Estimate of target mass corrections \label{App-TarMasCor}}

Target mass corrections for a scalar and spin-1/2 target have been
evaluated in Ref. \cite{BelMue01}, where $\Delta^2/\cQ^2$ corrections
have been neglected. This result will serve us to derive the scaling law
of the target mass corrections with respect to the atomic mass number
$A$ for a general Compton process in the light-cone dominated region. It
is sufficient to consider the simplest case of a scalar target. Then
the target mass corrections in the parity even sector read for the
Compton amplitude 
\begin{eqnarray}
\label{F1-MasCorPro}
{\cal F}_1(\xi,\eta) \!\!\! &=&\!\!\! \int_{{\mit\Omega}} dy \, dz \, h(y, z)
\Bigg[ C_1^{(0)}\!\!\left(\frac{y+ \eta z}{\xi}-i0\right) +
 \frac{\xi^2 M^2 y^2}{Q^2 (y+ \eta z)^2}
C_1^{(1)}\!\!\left(\frac{y+ \eta z}{\xi}-i0\right)
\nonumber\\
&&\hspace{5cm} + {\cal O}\left(\frac{\xi^4 M^4}{Q^4} \right) +
\left\{ {y \to - y \atop z \to -z}\right\}
\Bigg]\, .
\end{eqnarray}
An analogous formula for the other leading twist amplitude ${\cal
F}_2$ holds true. Here $h(y, z)$ denotes the DD and the coefficients 
\begin{eqnarray}
C_1^{(0)}({\mit\Xi^{-1}})= -\frac{1}{1+\mit\Xi}\, ,\quad \mbox{and} \quad
C_1^{(1)}({\mit\Xi^{-1}})
= \frac{1}{\left( 1 + {\mit\Xi} \right)^2}
+ 2  \ln \left( \frac{{\mit\Xi}}{1 + {\mit\Xi}}\right)\, ,
\end{eqnarray}
depend on the variable $\mit\Xi= \xi/(y + \eta z)$. Let us set the mass
$M$ and the scaling variables in Eq.\ (\ref{F1-MasCorPro}) equal to the
nucleon ones: $M_N, \xi_N, \eta_N$. The Compton amplitude for a nucleus
A at smaller values of the Bjorken variable\footnote{We only use this
condition to simplify the discussion. For larger values of $\Bx$ one
should take the relation $\xi = (1+\xi) \xi_N/(1+\xi_N) A$.} follows
from the replacements: 
\begin{eqnarray}
M_N \to M = A M_N\, , \ \xi_N \to \xi \sim \xi_N/A\, ,\ 
\eta_N \to \eta\sim \eta_N/A\, ,
\ \ \mbox{and}\ \ h\to h^A\, .
\end{eqnarray}
Corresponding to our nuclei GPD model (\ref{GPD-A-Sca}) and the
GPD definition (\ref{DD2GPD}) in terms of a DD, we read off the
scaling law for the nucleus DD:
\begin{eqnarray}
h^A(y,z) \propto \theta(|A y \pm z| \le 1) h(A y,z) \, .
\end{eqnarray}
Changing the integration variable $y= y^\prime/A$ and renaming
$y^\prime \to y$, we find that the Compton amplitude for a nucleus
target reads
\begin{eqnarray}
F_1^A (\xi,\eta) \!\!\! &\propto&\!\!\! \int_{{\mit\Omega}} dy \, dz \, h (y, z)
\Bigg[ C_1^{(0)}\!\!\left(\frac{y+ \eta_N z}{\xi_N}-i0\right) +
 \frac{\xi_N^2 M_N^2 y^2}{Q^2 (y+ \eta_N z)^2}
C_1^{(1)}\!\!\left(\frac{y+ \eta_N z}{\xi_N}-i0\right)
\nonumber\\
&&\hspace{6cm} + {\cal O}\left(\frac{\xi^4_N M_N^4}{Q^4} \right) +
\left\{ {y \to - y \atop z \to -z}\right\}
\Bigg]\, .
\end{eqnarray}
As we realize by comparison with Eq.\ (\ref{F1-MasCorPro}), setting $M=M_N$,
$\xi=\xi_N$, and $\eta = \eta_N$, the target mass corrections do not
scale with $A$ in respect to the leading twist-two term.
 It can be shown in the same way from Eqs.\ (30) and (31)
in Ref. \cite{BelMue01} that also the resummed target mass corrections
possess  the same property. From the same representation it also
follows that this statement is true for a target with non--zero spin
content, including the parity odd sector.

Beside this kind of corrections also dynamical ones, which are expressed
in terms of multi--parton correlation functions, appear in the CFFs.
Unfortunately, only little is known about them  and it remains an
open task to consider the full twist-four sector.

\end{appendix}


\end{document}